%% file: main.tex
\documentclass[11pt,a4paper]{article}
\usepackage[utf8]{inputenc}
\usepackage[T1]{fontenc}
\usepackage[english]{babel}
\usepackage{lmodern}
\usepackage{multicol}
\usepackage{calc}
\usepackage{caption}
\usepackage{subcaption}
\usepackage{chngcntr}
\usepackage{enumitem}
\usepackage{setspace}
\usepackage{gensymb}
\usepackage{graphicx} 
\usepackage{tabularx}

\usepackage{authblk}

\usepackage{layout}
\usepackage{multicol}
\usepackage{multirow}
\usepackage{amsmath,amsfonts,amssymb,amsthm}
\usepackage{mathrsfs}
\usepackage{float}

\usepackage[left=2.5cm,right=2.5cm,top=2.5cm,bottom=2.5cm]{geometry}
\usepackage[final]{pdfpages}
\usepackage{pstricks}
\usepackage{siunitx}

\usepackage{soul}
\usepackage{titlesec}
\usepackage{ulem}
\usepackage{url}

\usepackage{dsfont}
\usepackage{array}

\usepackage{wrapfig}
\usepackage{fancyhdr}
\usepackage[lastpage,user]{zref}

\usepackage{authblk}

\usepackage{xcolor}
\usepackage{physics}
\usepackage{hyperref}
\usepackage{sectsty}
\definecolor{myblue}{rgb}{0,0.1,0.5}
\definecolor{blueplot}{rgb}{0,0,1}
\definecolor{redplot}{rgb}{0.84,0.15,0.16}
\definecolor{greenplot}{rgb}{0,0.5,0}

\usepackage[style=numeric,sorting=none]{biblatex} 
\usepackage{lineno} 

\hypersetup{ 
colorlinks=true,
breaklinks=true, 
urlcolor=myblue,
linkcolor=black,
citecolor=myblue,
pdfauthor={Nicolas De Angelis}
}

\usepackage{varwidth}
\DeclareCaptionFormat{myformat}{
  \begin{varwidth}{\linewidth}
    \centering
    #1#2#3
  \end{varwidth}
}

\graphicspath{{./Pictures/}}

\begin{document}

\title{Temperature dependence of radiation damage annealing of Silicon Photomultipliers}

\input{authors.tex}

\renewcommand{\thefootnote}{\arabic{footnote}}

\date{\today}

\maketitle

\section*{Abstract}
The last decade has increasingly seen the use of silicon photomultipliers (SiPMs) instead of photomultiplier tubes (PMTs). This is due to various advantages of the former on the latter like its smaller size, lower operating voltage, higher detection efficiency, insensitivity to magnetic fields and mechanical robustness to launch vibrations. All these features make SiPMs ideal for use on space based experiments where the detectors require to be compact, lightweight and capable of surviving launch conditions. A downside with the use of this novel type of detector in space conditions is its susceptibility to radiation damage. In order to understand the lifetime of SiPMs in space, both the damage sustained due to radiation as well as the subsequent recovery, or annealing, from this damage have to be studied. Here we present these studies for three different types of SiPMs from the Hamamatsu S13360 series. Both their behaviour after sustaining radiation equivalent to 2 years in low earth orbit in a typical mission is presented, as well as the recovery of these detectors while stored in different conditions. The storage conditions varied in temperature as well as in operating voltage. The study found that the annealing depends significantly on the temperature of the detectors with those stored at high temperatures recovering significantly faster and at recovering closer to the original performance. Additionally, no significant effect from a reasonable bias voltage on the annealing was observed. Finally the annealing rate as a function of temperature is presented along with various operating strategies for the future SiPM based astrophysical detector POLAR-2 as well as for future SiPM based space borne missions.\newline

\noindent\textit{Keywords: SiPM, Photodetectors, Annealing, radiation, space environment, LEO} 

\newpage
\tableofcontents

\newpage
\section{Introduction} \label{sec:intro}

Silicon Photo-Multipliers (SiPMs) are semiconductor photodetectors. They consist of a matrix of pixels, all operating in parallel at the same operation voltage, thereby functioning in avalanche mode. As a result, a single incident optical photon which produces a photo-electron can induce an avalanche in a pixel with a typical gain of $10^6$, thereby producing a signal detectable using standard electronics. The fine pixelization of the detector furthermore allows for a linear sensitivity to optical photons typically up to several thousands of photons. This makes SiPMs comparable to photo-multiplier tubes (PMTs) in sensitivity while having several significant advantages. These advantages over PMTs are their insensitivity to magnetic fields, compact size, lower operating voltages (30-60~V instead of 1-2~kV), as well as mechanical robustness. A more detailed overview on SiPMs can be found in \cite{ACERBI201916}.

The advantages listed above are important for using SiPMs in a range of different physical experiments. On first inspection, one of the most clear implementations where SiPMs are preferable over PMTs is for usage in space based detectors. Their compact size is preferable as space based detectors require to be lightweight and compact, while their relative mechanical robustness, compared to PMTs, which contain an entrance glass window, makes it easier to implement them in a design capable of surviving launch conditions. The lower operating voltage means that no power supplies capable of providing typical voltages of kV are required. Such power supplies are complex, heavy and prone to failure while those required to operate SiPMs are small and radiation tolerant Commercial Off The Shelf (COTS) versions are available \cite{HVPS_TN, FEEradpaper, GRID-02results}. Finally, their immunity to magnetic fields simplifies their usage in orbit where magnetic fields are present and variable. For these reasons SiPMs appear to be an ideal candidate for replacing PMTs when, for example, reading out scintillator based detectors in space. As a result a range of upcoming astrophysical experiments use SiPMs in their design (BurstCube \cite{Smith:2019zra}, EIRSAT-1 \cite{Murphy:2022ldf}, CAMELOT \cite{CAMELOT} to name a few) and the first group of detectors making use of SiPMs has already been launched in recent years, the most complex one being GECAM \cite{GECAM}. 

The downside of SiPMs over PMTs is their susceptibility to radiation damage \cite{MITCHELL2021164798, Mianowski_2020, GARUTTI201969}. Several studies have shown bulk defects in the silicon structure appearing due to radiation damage \cite{GARUTTI201969, VANLINT1987453, Honniger2007}. Ionizing radiation has been shown to produce point defects and cluster related defects in the sensitive volume, while neutrons scattering in the silicon lattice can induce displacements of the atoms resulting in cluster damage. In space environments it is the damage induced by ionizing particles, mainly protons, which is of interest as the neutron flux is relatively low \cite{SiPMpaper}. The damage induced is furthermore highly dependent on the orbit \cite{SiPMpaper} and varies with time, for example, more damage is expected to be induced in the South Atlantic Anomaly (SAA) than in other parts of the orbit. Exposing the SiPMs to radiation equivalent to several years in space within several minutes or even seconds is not sufficient in order to model the performance of the SiPM after a fixed amount of time in orbit. Indeed, one needs to know the induced radiation damage, but also the potential recovery from this damage with time. While the effects of radiation are discussed in detail in several publications \cite{Mianowski_2020, HiroTaka, MITCHELL2021164798} (and studied in-orbit by several experiments like SIRI \cite{SIRIinstr, SIRIresults} and SIRI-2 \cite{SIRI-2}, or GRID-01 \cite{GRID} and GRID-02 \cite{GRID-02results}) and are therefore well known, details on the recovery of the damage with time and its dependence on temperature are lacking\footnote{The detailed study of the temperature related silicon lattice recovery after irradiation is lacking for SiPMs, but has already been performed for other types of silicon detectors \cite{Moll1999}}, preventing current missions from predicting the performance of the SiPMs over time.

Here we present a detailed study of the annealing process of 3 types of SiPMs as a function of temperature as well as its dependence on the applied bias voltage. The three types of SiPMs selected were the S13360-6025PE, S13360-6050PE and S13360-6075PE by Hamamatsu\footnote{Datasheet to be found here: \url{https://www.hamamatsu.com/content/dam/hamamatsu-photonics/sites/documents/99_SALES_LIBRARY/ssd/s13360_series_kapd1052e.pdf}}. These SiPMs only differ in the microcell pitch, which is 25 $\mu\mathrm{m}$ for the S13360-6025PE, 50 $\mu\mathrm{m}$ for the S13360-6050PE and 75 $\mu\mathrm{m}$ for the S13360-6075PE. They are later referred as the 25, 50, and 75$\mu m$ SiPMs, respectively. The effects of radiation damage on each of these has been previously discussed in detail in \cite{SiPMpaper}. The details of the application of the radiation damage to the SiPMs are discussed first, followed by a discussion on the setup and method used to study the annealing for different environments. The results of the tests are discussed thereafter along with a discussion on the implications for future space missions, particularly for the POLAR-2 mission which aims to have a total of 6400 SiPM channels onboard \cite{ICRC21_POLAR-2}. 

\newpage
\section{Experimental setups} \label{sec:meas_setup}
\subsection{Irradiation setup}

The irradiation of 21 SiPM sensors has been performed at the Institute of Nuclear Physics of Krakow, Poland (IFJ-PAN). All sensors have been irradiated with a 58~MeV proton beam from an isochronous cyclotron \cite{IFJcyclotron}, whose intensity profile is shown in Figure \ref{fig:IFJ_beam}.

\begin{center}
\begin{figure}[H]
\captionsetup{justification=centering}
\begin{centering}
\includegraphics[height=.38\textwidth]{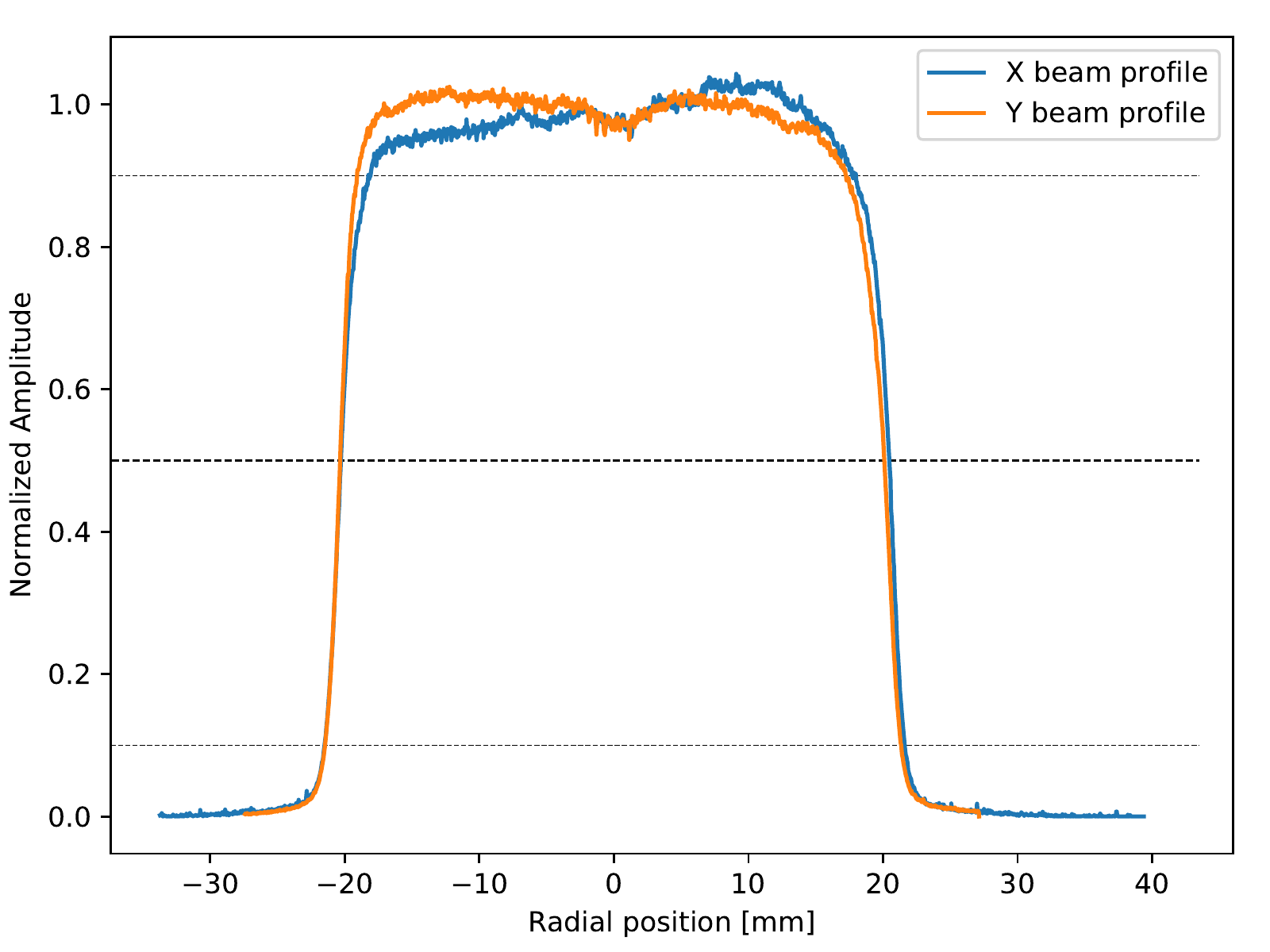}\includegraphics[height=.4\textwidth]{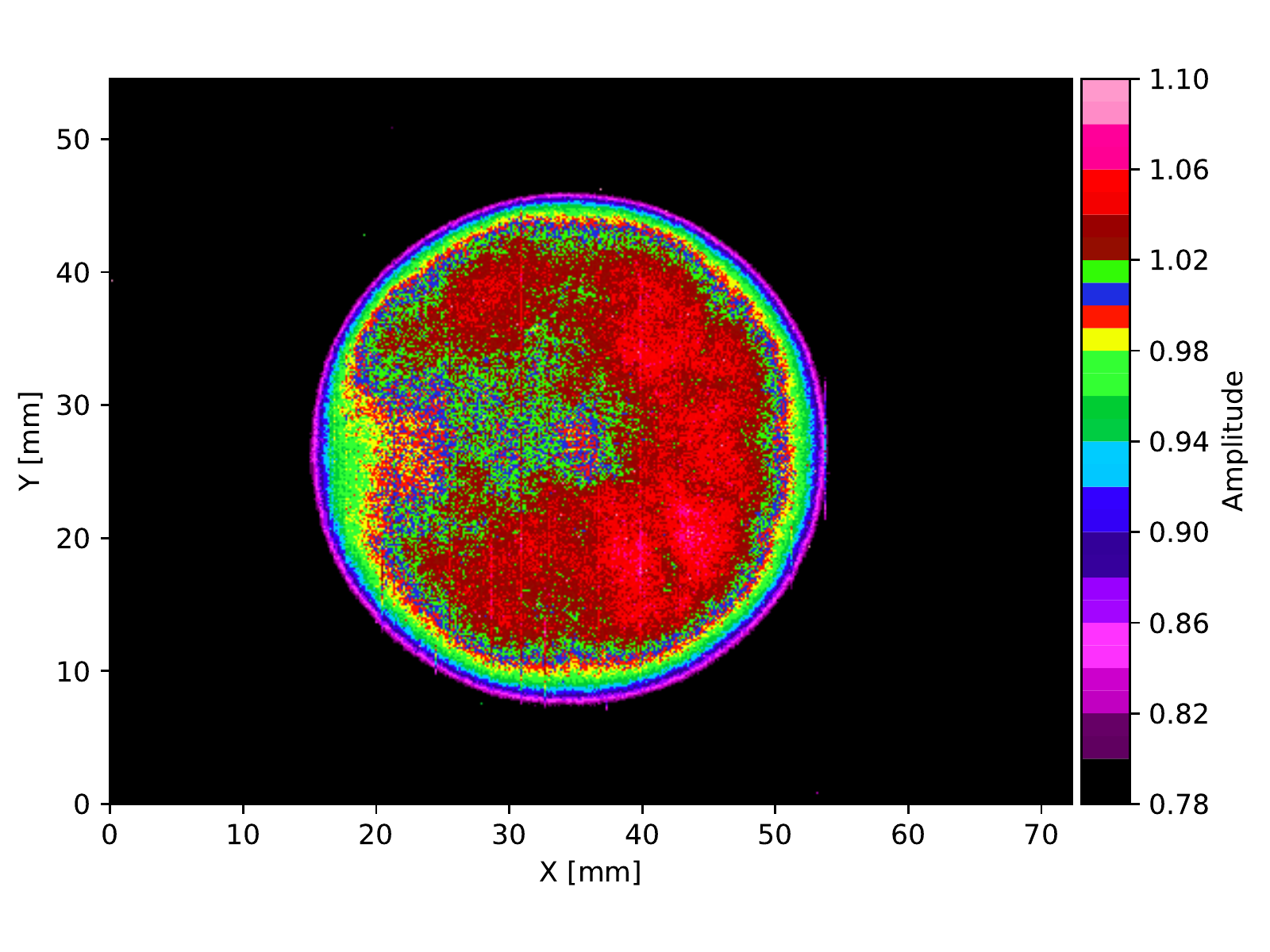}
  \par\end{centering}
  \protect\caption{Normalized intensity of the proton beam along the X and Y directions (left) and 2D beam profile (right)}
    \label{fig:IFJ_beam}
\end{figure}
\par\end{center}

GEANT4 \cite{GEANT4} simulations were performed in order to calculate the required beam exposure time in the sensitive volume of the SiPMs, as well as to determine the equivalence between dose and time spent in the low earth orbit (LEO) environment \cite{SiPMpaper}. Each sample has been irradiated with a fluence\footnote{The non-uniformity of the inner part of the circular beam, where the SiPMs are placed during irradiation, as well as the precision of the irradiation time and the flux variation with time are contributing to the uncertainty on the total fluence.} of ($1.00\pm$ 0.08)$\cdot10^8$~protons/cm$^2$ over a time of 10~s, which corresponds to a dose of 0.134~Gy, or 1.7 years in LEO for SiPMs inside of the the POLAR-2 instrument. Figure \ref{fig:IFJ_setup} shows the irradiation room as well as a few SiPMs mounted in front of the beam output. 

\begin{center}
\begin{figure}[H]
\captionsetup{justification=centering}
\begin{centering}
\includegraphics[height=.38\textwidth]{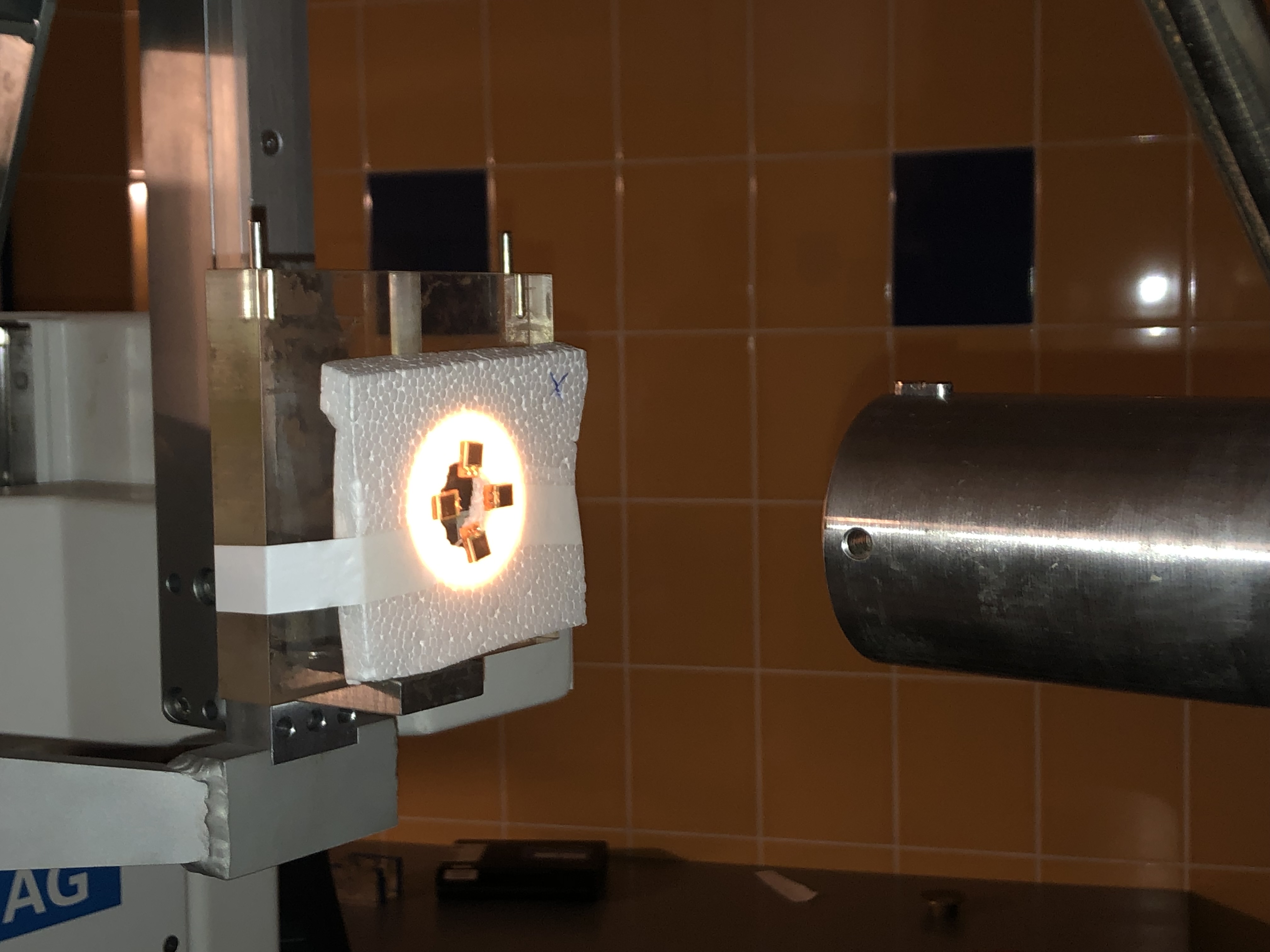}
  \par\end{centering}
  \protect\caption{Four single channel SiPMs placed on the beam axis at IFJ-PAN}
    \label{fig:IFJ_setup}
\end{figure}
\par\end{center}

\subsection{SiPM Storage Environment}

With the aim of studying the silicon lattice recovery through annealing effect, SiPMs have been stored in different thermal and electrical bias conditions. SiPMs of each of the 3 types were stored within 24 hours \footnote{The irradiation took place in Krakow, Poland, while the storage and characterization of the SiPMs post-irradiation were performed in Geneva, Switzerland. The samples therefore spent about 24~h at ambient temperature before being stored in controlled conditions. This does not affect the results since we study the temperature dependence of the annealing effect relative to the first measurement point (after transportation) and all samples were stored under the same condition in this 24~h period.} after irradiation at 6 different temperatures ranging from -22.8$\pm$1.8 to 48.7$\pm$3.3$^\circ$C. Several 50~$\mu$m SiPMs were also stored at room temperature\footnote{All the setups where placed in a clean room, where the temperature is controlled at 20$^\circ$C.} with different bias voltages (2 with 3~V overvoltage, 1 with 8~V and 1 with 12~V). The number of SiPMs stored at each condition is summarized in Table \ref{tab:list_sipms}, while the setups for biasing the SiPMs and storing them at the highest temperatures are shown in Figure \ref{fig:temp_storage_setup}.

\begin{table}[ht]
\centering
\hspace*{-0.5cm}\begin{tabular}{|l||c|c|c|c|c|c|}
\hline
Chamber ID & 1 & 2 & 3 & 4 & 5 & 6 \\ 
\hline
Chamber type & Freezer & Fridge & Room temperature & \multicolumn{3}{|c|}{Polystyrene + Power resistor} \\ 
\hline
\begin{tabular}{@{}l@{}}Storage \\ Temperature [$^\circ$C]\end{tabular} & -22.8$\pm$1.8 & 6.3$\pm$0.9 & 20.5$\pm$0.6 & 29.7$\pm$0.6 & 38.7$\pm$1.6 & 48.7$\pm$3.3 \\ 
\hline
\# of 25$\mu m$ SiPMs & 1 & 1 & 1 & 1 & 1 & 1 \\
\hline
\# of 50$\mu m$ SiPMs & 1 & 1 & 1+4 biased & 1 & 0 & 1 \\
\hline
\# of 75$\mu m$ SiPMs & 1 & 1 & 1 & 1 & 1 & 1 \\
\hline
\end{tabular}
\caption{List of storage conditions  for the SiPMs used in this study. One SiPM of each cell size (25, 50, and 75 $\mu m$) is passively stored at each of the 6 temperatures. For the 50$\mu m$, the effect of different over-voltages (3, 8, and 12V) is studied at room temperature. The SiPMs in chamber \#6 were later placed in a climatic chamber in order to characterize the annealing at higher temperatures (75 and 100$^\circ$C)}
\label{tab:list_sipms}
\end{table}

As illustrated in Table \ref{tab:list_sipms}, the coldest storage temperatures were reached using a freezer and a fridge, while the hottest ones were reached using power resistors placed in purpose build polystyrene chambers (see Figure \ref{fig:temp_storage_setup}) and tuned to reach specific temperatures. The temperatures were monitored every few days to ensure their stability and calculate their uncertainties. The annealing effect on even higher temperatures have been studied as well using a climatic chamber to reach temperatures of 75 and 100$^\circ$C. The biased SiPMs were stored in a dark box and biased using an in house developed power supply board based on the LT3482 DC/DC converter from Linear Technology. For these biased SiPMs great care was taken in the design of the dark box in order for the SiPMs to remain in good thermal contact with the box. Thereby, the temperature of the SiPMs, despite the significant heat dissipation inside of the SiPM, was at $20.5\pm0.6^\circ$C throughout the test.

\begin{center}
\begin{figure}[H]
\captionsetup{justification=centering}
\begin{centering}
\includegraphics[height=.5\textwidth]{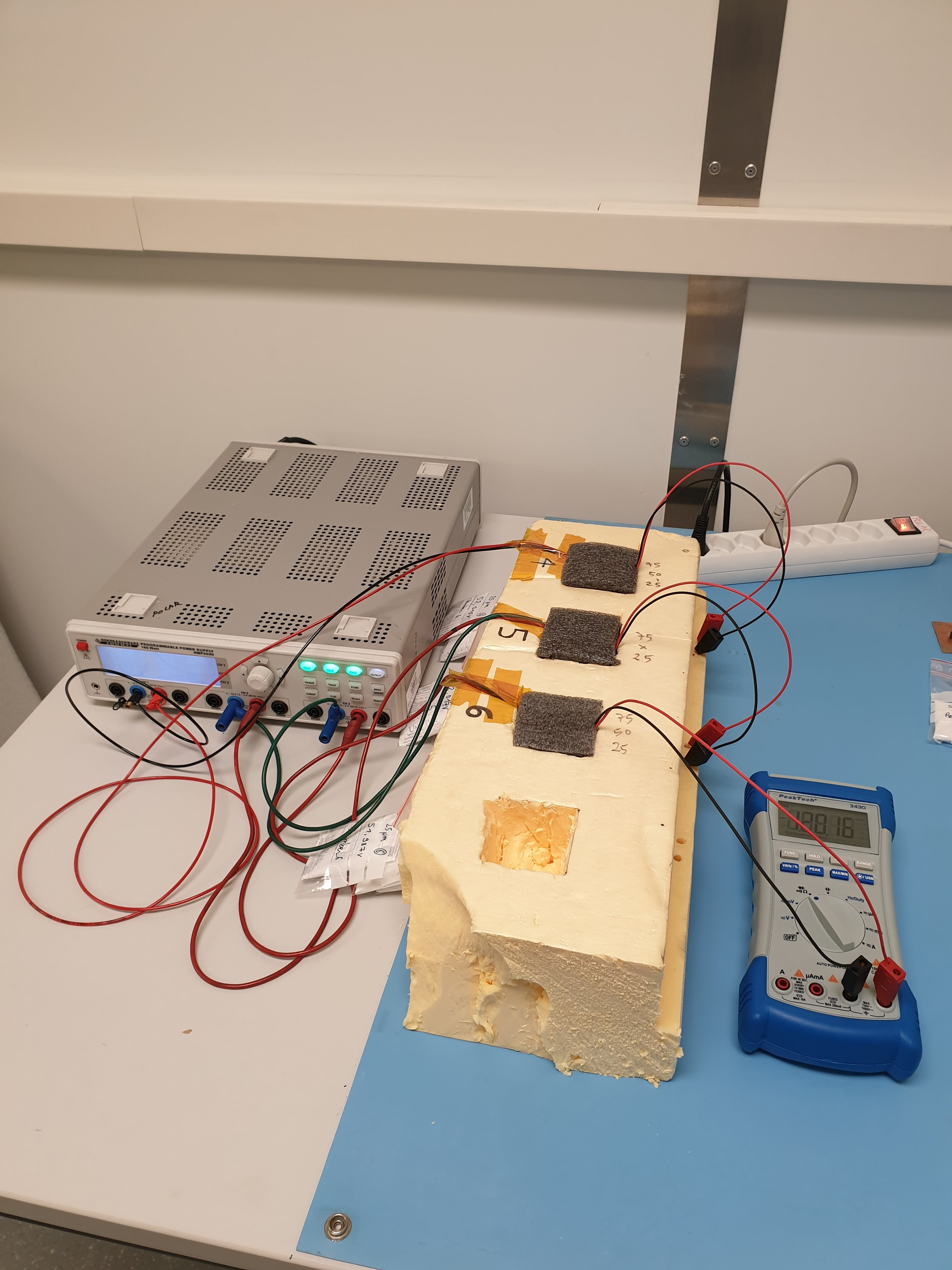}\hspace*{0.1cm}\includegraphics[height=.5\textwidth]{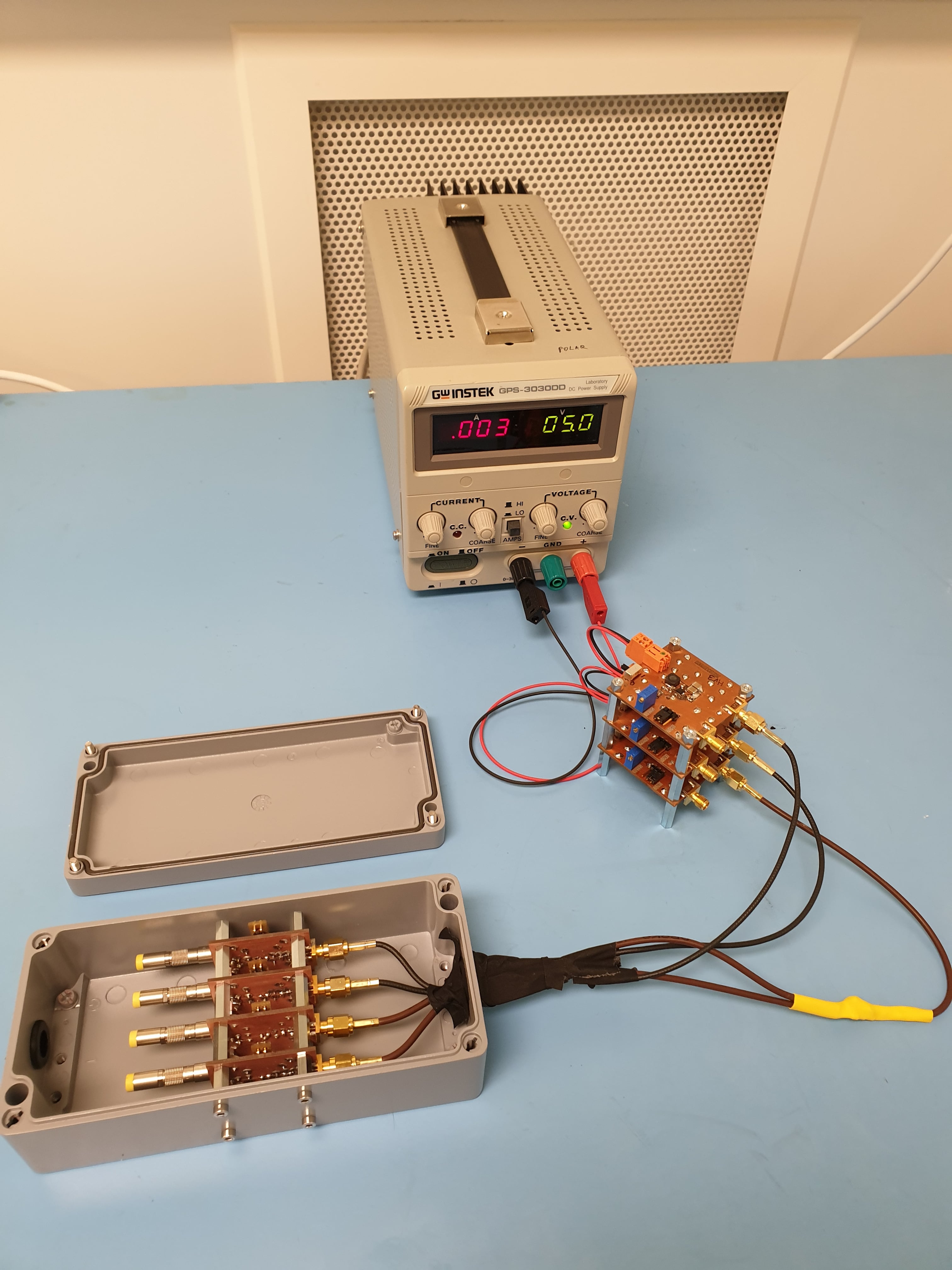}
  \par\end{centering}
  \protect\caption{\textbf{\emph{Left:}} SiPM storage setup for high temperatures. A block of polystyrene (yellow) is used in which cavities of $50\times50\times50\,\mathrm{mm^3}$ are cut. The SiPMs, along with a power resistor are stored in the various cavities. The power resistors are powered using the power supply which provides a current calibrated to reach the required stable temperature in the cavities. The cavities are closed off with a lid of foam while the temperature is occasionally monitored using a PT100. \textbf{\emph{Right:}} The SiPMs stored at different bias voltages. The SiPMs are stored in a dark box, seen opened in the figure, which allows to run the SiPMs in dark conditions with a good thermal conductions towards the aluminium box. As a result the SiPMs remain thermalized with the room. The bias voltages are provided through several in house developed power supply boards based on the LT3482 DC/DC converter from Linear Technology.}
    \label{fig:temp_storage_setup}
\end{figure}
\par\end{center}

\subsection{Characterization setups}
\label{subsec:char_setup}

The two main features measured to study the annealing effect are the dark current and the dark count rate (DCR) of the SiPMs with respect to the time after irradiation. The former is actually measured versus voltage through an I-V characterization at room temperature all along the 2 months of storage, while the latter is measured for a 5~V overvoltage at both 0 and 20$^\circ$C at the end of the 2 months storage. Great care was taken to minimize the time for each SiPM to be taken out of their thermal storage unit, while at the same time it was ensured that the SiPMs reached room or chamber temperature when taking the measurement. It was found that removing the SiPMs from their storage for 10 minutes would allow for complete thermalization with the room. This way the total time outside of the thermal storage unit was negligible (15 minutes per measurement: 10 minutes for thermalizing and 5 for measuring the I-V characteristics) while ensuring that all I-V measurements are performed with the SiPM thermalized to 20$^\circ$C.

\begin{center}
\begin{figure}[H]
\captionsetup{justification=centering}
\begin{centering}
\hspace*{-0.2cm}\includegraphics[height=.5\textwidth]{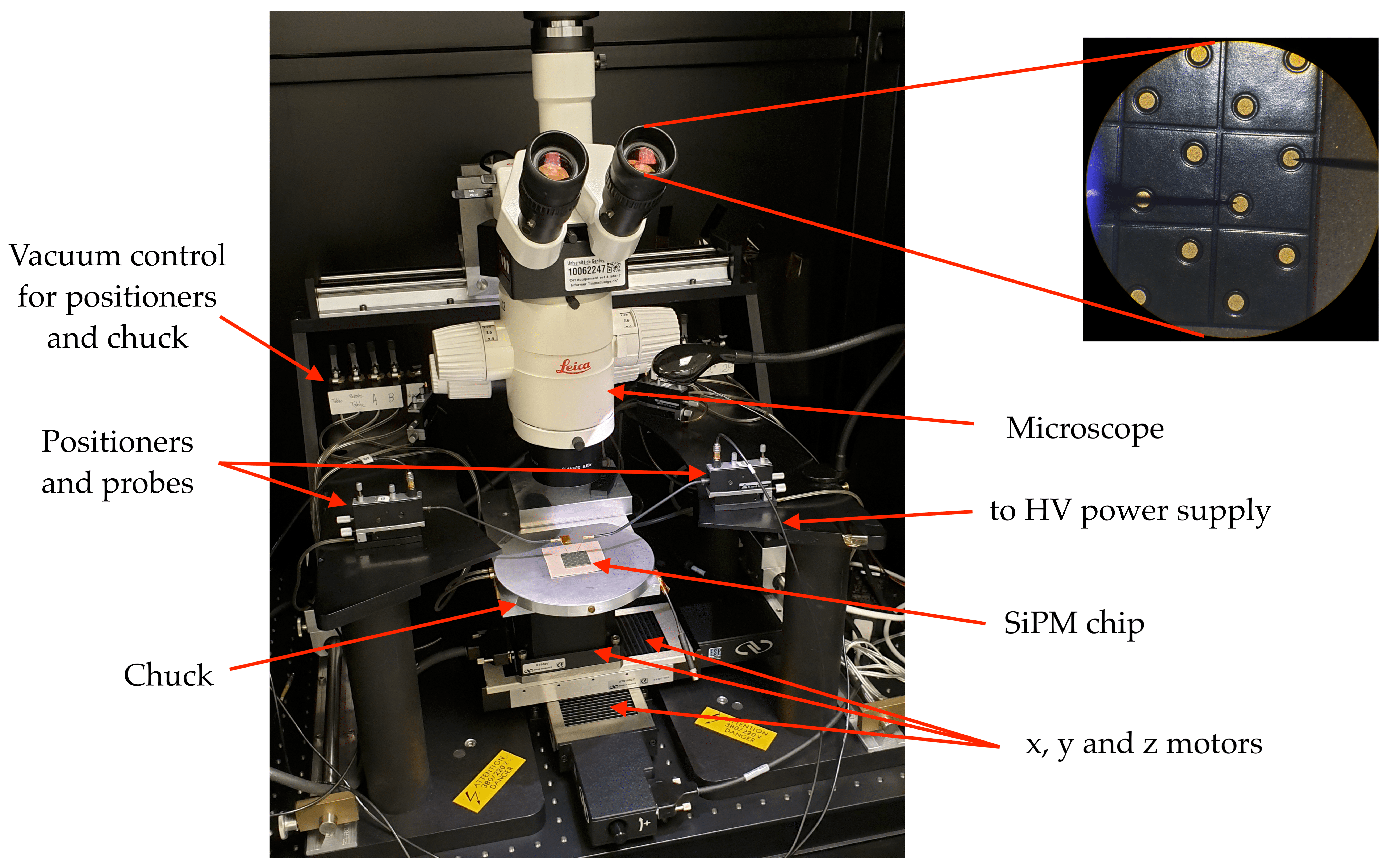}\hspace*{0.5cm}\includegraphics[height=.5\textwidth]{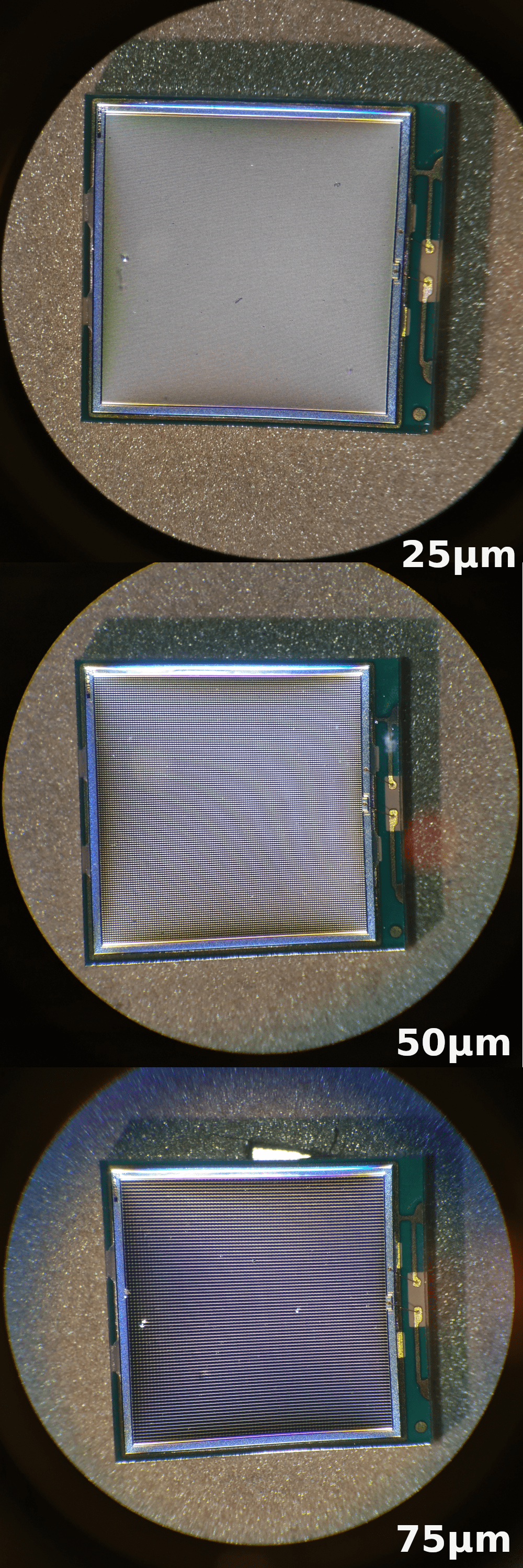}
  \par\end{centering}
  \protect\caption{Probe station setup used for I-V characterization (left) and microscope view of the 25/50/75~$\mu$m SiPMs (right)}
    \label{fig:probe_station_herd}
\end{figure}
\par\end{center}

The I-V characterization is made using an in-house designed probe station shown in Figure \ref{fig:probe_station_herd}, based on Keithley electrometers and controlled through a LabView\textsuperscript{TM} program. The DCR is measured in a climatic chamber, shown in Figure \ref{fig:tia_setup}, where the SiPM waveforms are readout with a Teledyne Lecroy Wavesurfer 510 oscilloscope through a trans-impedance amplifier (TIA) board based on ADA4817 and AD8000 operational amplifiers. The use of a climatic chamber ensures a stable temperature all along the measurements, and allows to acquire dark spectra at different temperatures, namely 0 and 20$^\circ$C.

\begin{center}
\begin{figure}[H]
\captionsetup{justification=centering}
\begin{centering}
\includegraphics[height=.5\textwidth]{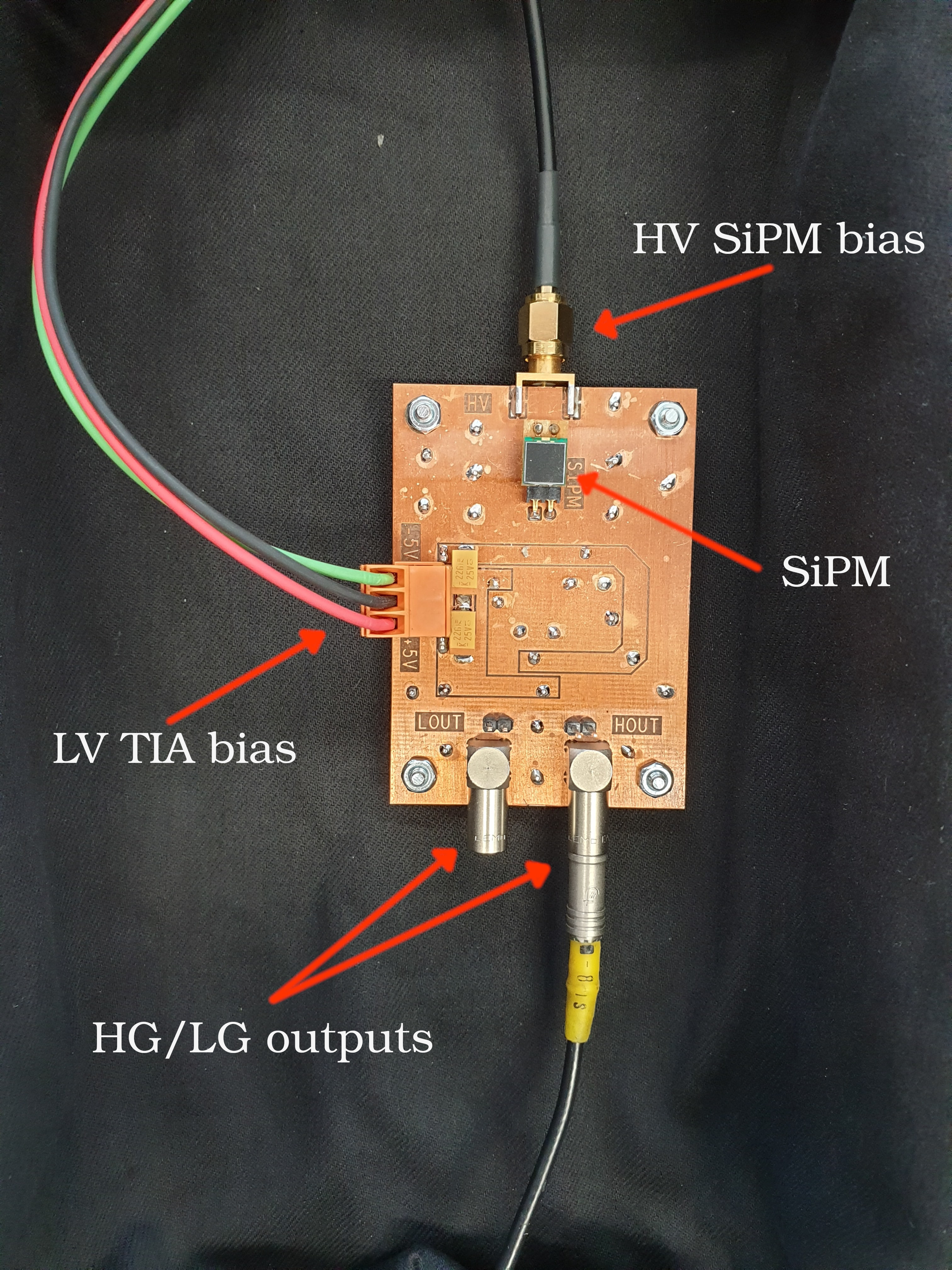}
  \par\end{centering}
  \protect\caption{Transimpedance amplifier board used to characterize the SiPMs dark spectra}
    \label{fig:tia_setup}
\end{figure}
\par\end{center}

For each dark spectrum, 20 waveforms of 10~ms are acquired with a threshold of about 0.5 photoelectron (p.e.). An analysis similar to that described in \cite{SiPMpaper} is applied to the waveforms. A Savitzky-Golay algorithm is applied twice to clean the data. The waveform is then subtracted to a time-delayed copy of itself in order to ease the peak finding process. Finally, an amplitude threshold of 20~mV for the 50~$\mu$m SiPM, 25~mV for the 75~$\mu$m SiPM, and a deadtime threshold\footnote{This inter-pulse time threshold is used to get rid of delayed crosstalk and afterpulsing.} of 120~ns are applied (see Figure \ref{fig:initial_darkspectrum}). Note that the SiPMs are operated at 5~V overvoltage to have a higher prompt crosstalk (compared to that at 3~V overvoltage, that is the Hamamatsu recommended operation point) and therefore more peak to fit in the dark spectrum. This make the analysis easier and does not affect its quality since we are interested in the relative comparison of the photoelectron resolution for different annealing conditions. Each dark spectra is fitted using a sum of Gaussians.

\newpage
\section{Results}

\subsection{Temperature dependence of the annealing effect on the SiPM current}
\label{subsec:annealing_current}

The measured I-V characteristics for 75~$\mu$m SiPMs are shown in Figure \ref{fig:i-v_curves} for both the SiPMs stored at the lowest and room temperatures. The curves for other microcell sizes and annealing temperatures are provided in appendix \ref{sec:appendix_IV}. A zoomed-in version is also displayed as a corner plot in log-scale in order to highlight the evolution of the I-V shape around the breakdown region. From those plots it is clearly seen that there is almost no effect on the post-irradiation dark current on the SiPMs stored at -22.8$\pm$1.8$^\circ$C, while for those stored at room temperature almost half of the dark current is recovered after about 2 months.

\begin{center}
\begin{figure}[H]
\captionsetup{justification=centering}
\begin{centering}
\includegraphics[width=.5\textwidth]{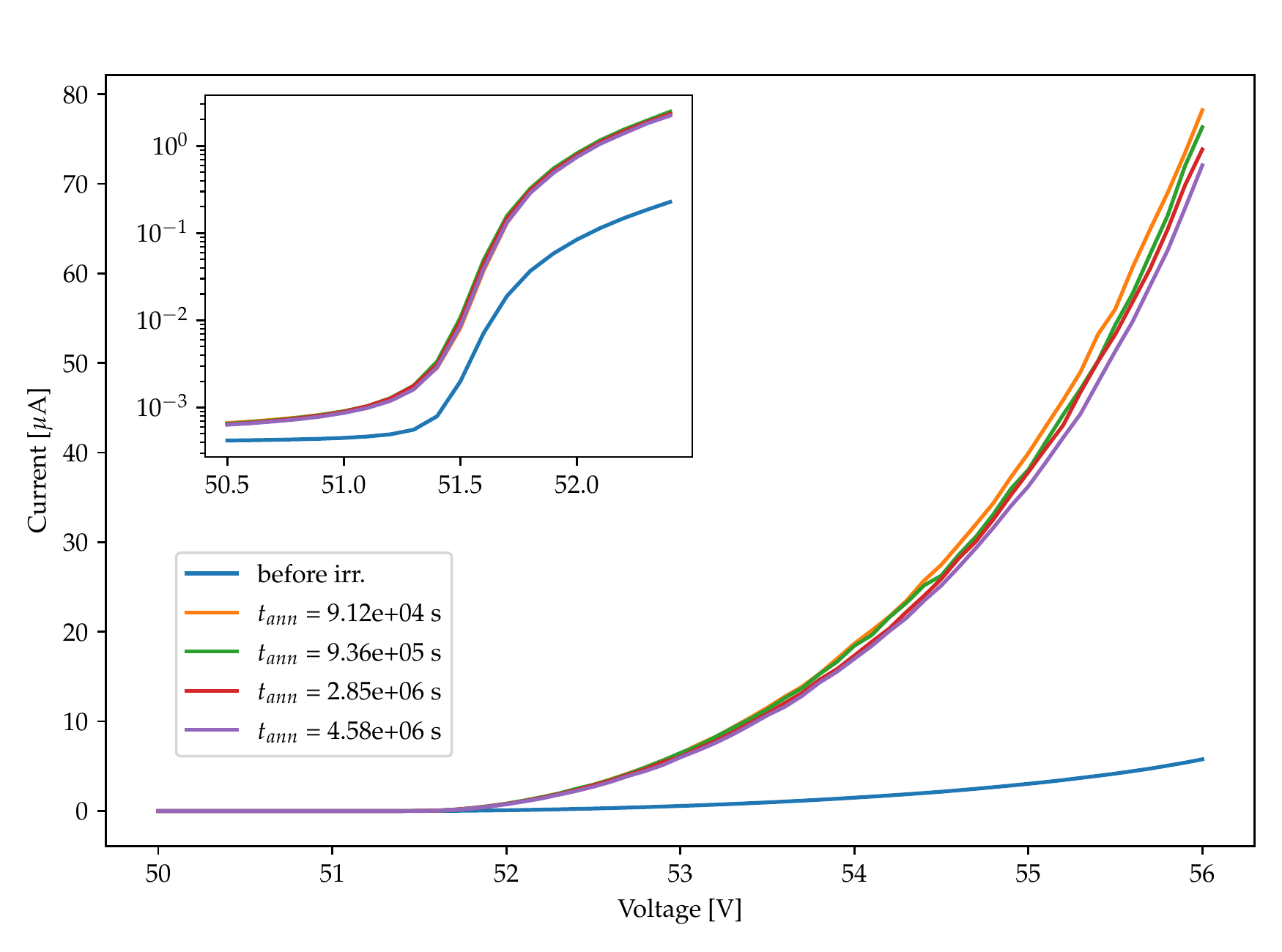}\includegraphics[width=.5\textwidth]{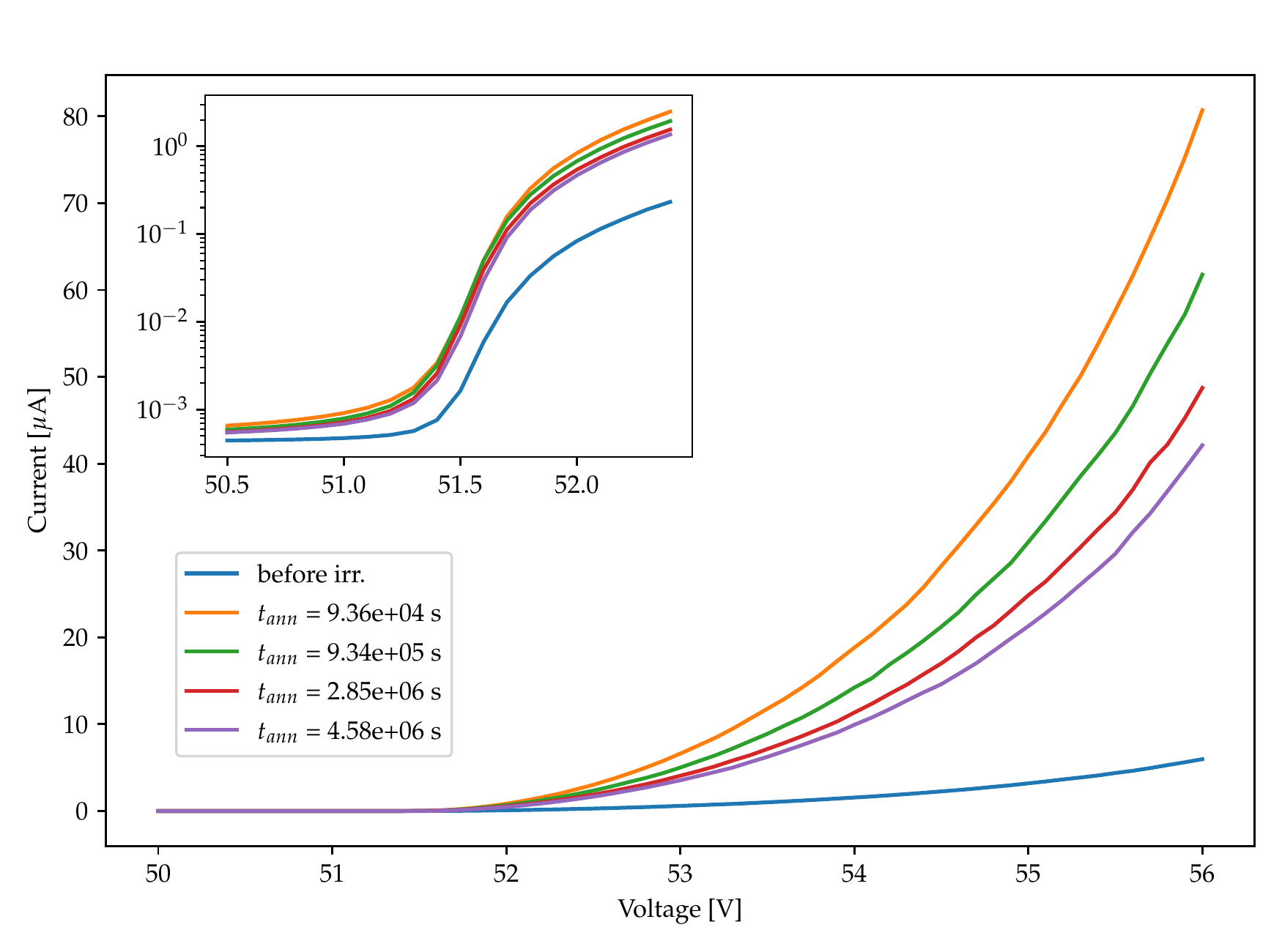}
  \par\end{centering}
  \protect\caption{Post-irradiation time evolution of the I-V characteristics of 75$\mu m$ SiPMs at $-22.8\pm1.8^\circ$C (left) and at $20.5\pm0.6^\circ$C (right). We only show a few I-V curves here even though a lot more curves were measured for the convenience of the reader. All the curves are provided in appendix \ref{sec:appendix_IV}.}
    \label{fig:i-v_curves}
\end{figure}
\par\end{center}

For a better representation of the temperature dependent improvement in dark current, it was decided to work with the current measured at 3~V overvoltage\footnote{The breakdown voltage is determined from the square root of the current versus the voltage. Two linear fits are performed, one for the region below the breakdown, and one for the region above. The breakdown is taken as the point of intersection of these two linear fits. This method is used all along this paper since it has been compared to other methods (first and second derivative of the current's logarithm) that gave similar breakdown values.} (standard operating point recommended by the manufacturer). The time evolution of the 3~V overvoltage current for different storage temperatures are shown in Figure \ref{fig:I_vs_time} for the three types of SiPMs. In this figure the y-axis shows the normalized current, meaning the dark current relative to that measured during the first measurement after irradiation.

\begin{center}
\begin{figure}[H]
\captionsetup{justification=centering}
\begin{centering}
\includegraphics[width=.6\textwidth]{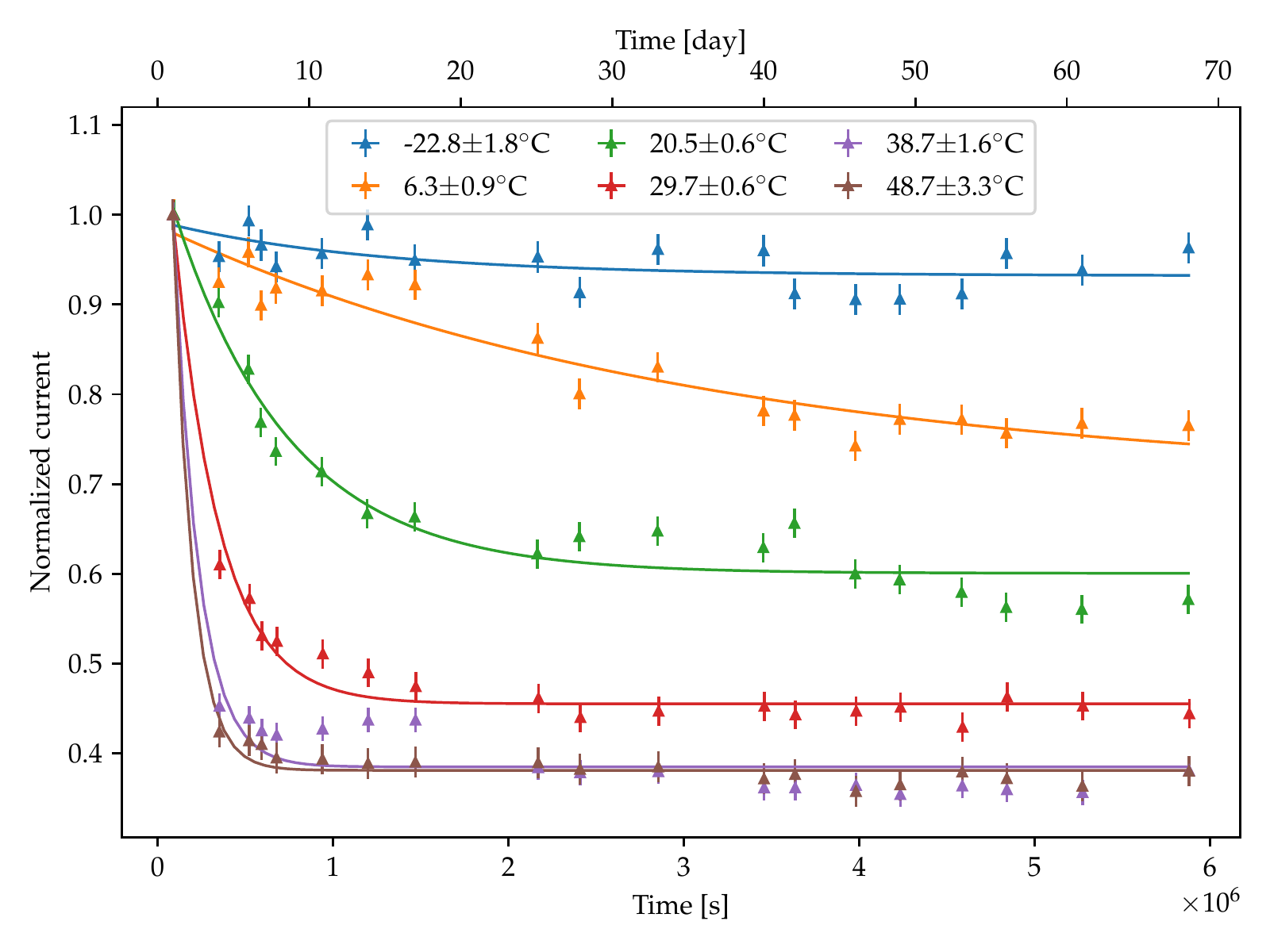}\\
\includegraphics[width=.6\textwidth]{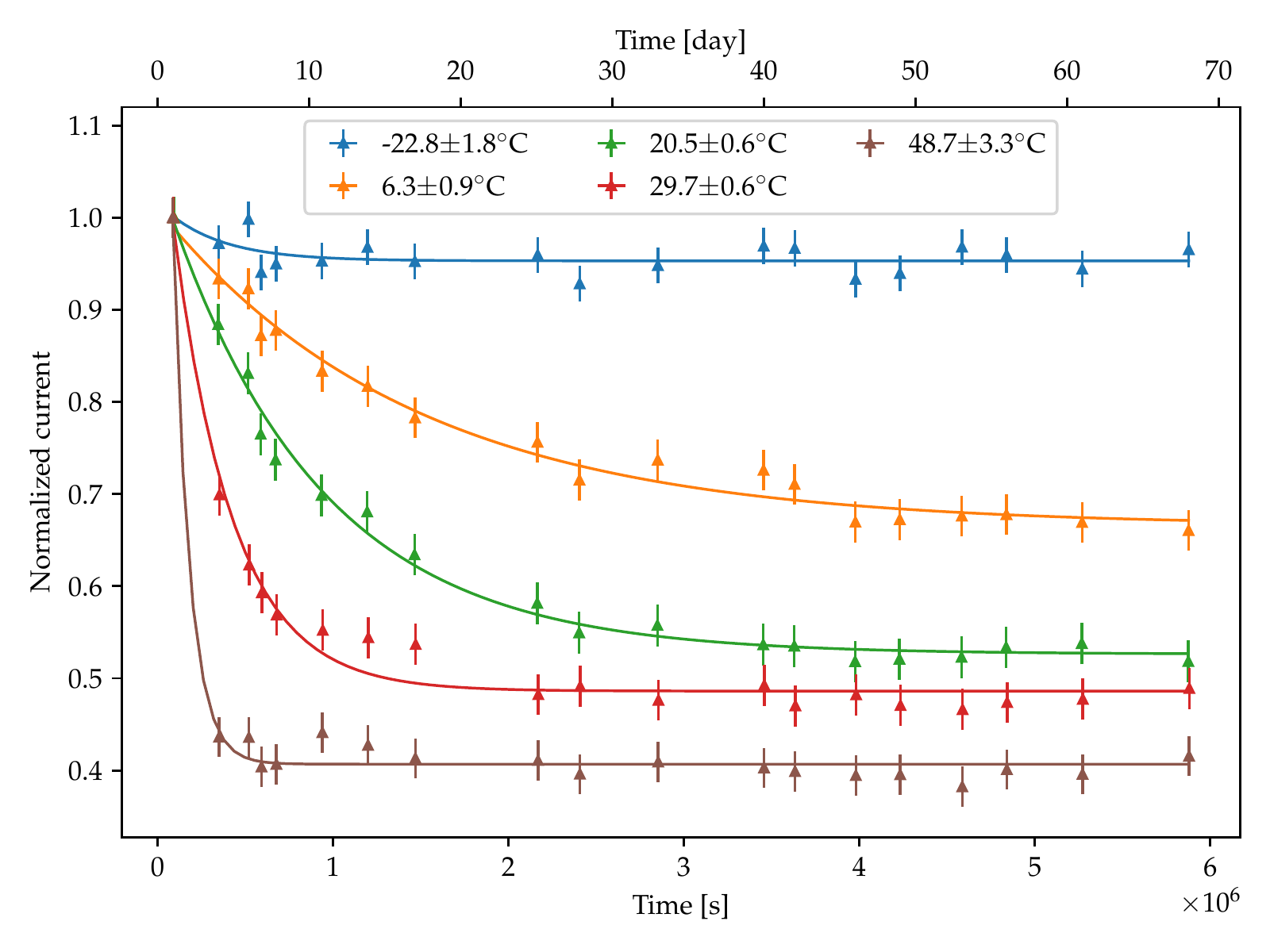}\\
\includegraphics[width=.6\textwidth]{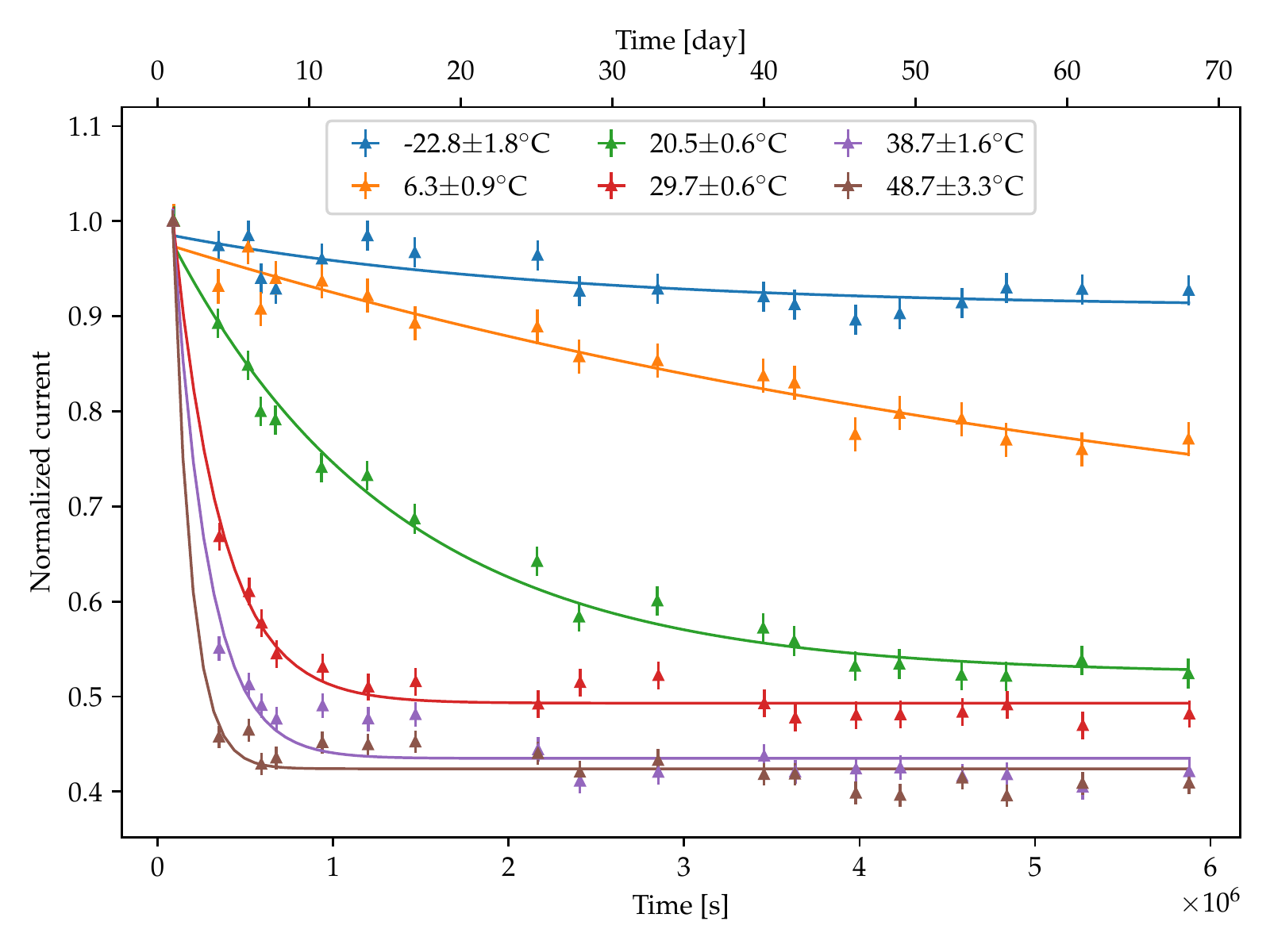}
  \par\end{centering}
  \protect\caption{Normalized current (to the first data point in time) measured at 3V over-voltage vs. time after irradiation for SiPMs stored at different temperatures (as explained in the legend). The curves are fitted with an exponential function, for the 25$\mu m$ (top), 50$\mu m$ (middle), and 75$\mu m$ (bottom) SiPMs.}  
    \label{fig:I_vs_time}
\end{figure}
\par\end{center}

It can be noticed that the current at this fixed operation point is following an exponential decay with time, whose amplitude, time constant, and offset vary with the storage temperature. The data points in Figure \ref{fig:I_vs_time} are therefore fitted with an exponential function, whose fit parameters are plotted versus the storage temperature in Figure \ref{fig:exp_fit_params}. The exponential offset is linked to the fraction of the post-irradiation dark current that can be recovered after an infinite amount of time, while the exponential slope can be seen as a dark current recovery rate. It can clearly be seen that the annealing process is both faster and more efficient (meaning that we recover a bigger part of the current) at higher temperatures. Furthermore, almost no annealing effect is observed at $-22.8\pm1.8^\circ$C.

\begin{center}
\begin{figure}[H]
\captionsetup{justification=centering}
\begin{centering}
\includegraphics[width=.5\textwidth]{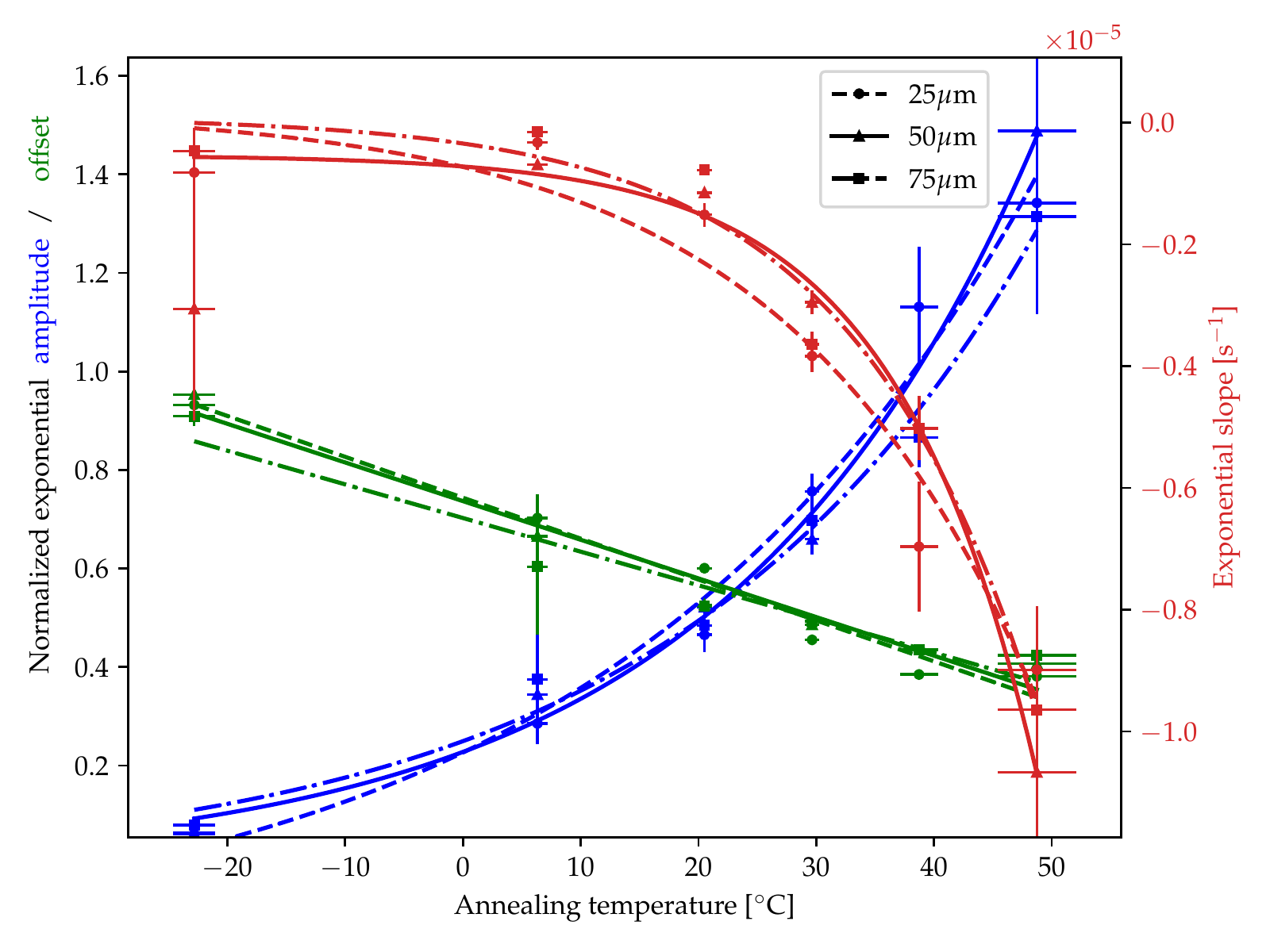}\includegraphics[width=.5\textwidth]{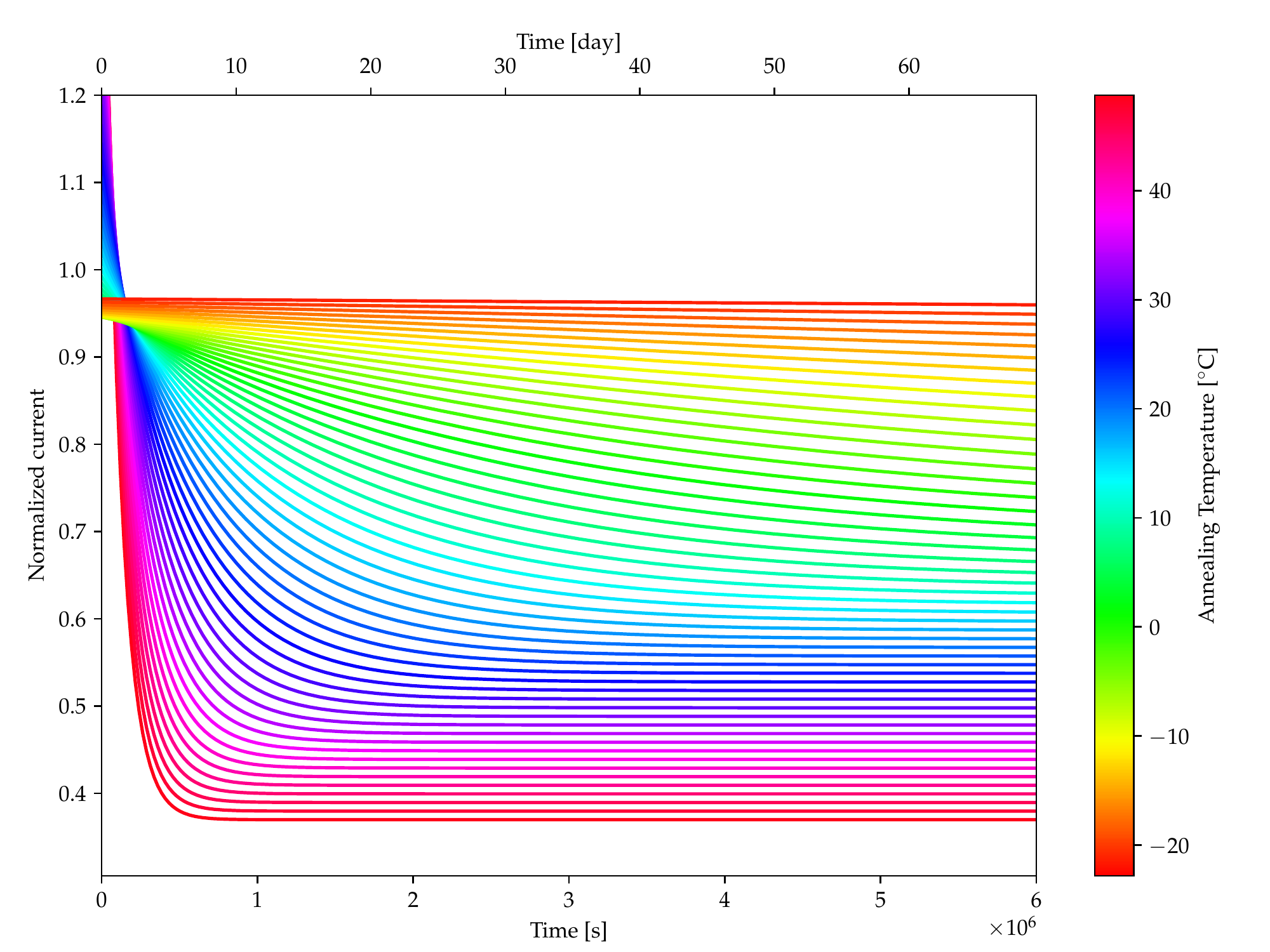}
  \par\end{centering}
  \protect\caption{Normalized exponential fit parameters for current vs. time after irradiation (left) and the temperature dependency of the exponential annealing effect for the 75~$\mu$m SiPM as predicted based on the fit results (right)}
    \label{fig:exp_fit_params}
\end{figure}
\par\end{center}

The exponential offset has a linear dependency in storage temperature, while the amplitude and exponential slope appear to have an exponential behavior. These exponential fit parameters are consequently fitted, and the fit results are shown in Table \ref{tab:fitting_expfit}. Using these fit parameters one can get the exponential decay shape of the current versus time for a wide range of temperatures. The right plot of Figure \ref{fig:exp_fit_params} shows the evolution of this theoretical exponential decay with temperature for the 75~$\mu$m SiPM.

\begin{table}[ht]
\centering
\hspace*{-0.5cm}\begin{tabular}{|l|c||c|c|c|}
\hline
Parameter & Pitch [$\mu m$] & Amplitude & Slope [$^\circ C^{-1}$] & Offset \\ \hline
\multirow{3}{*}{\color{blueplot}Amplitude} & 25 & (4.12$\pm$2.06$)\cdot 10^{-1}$ & (2.76$\pm$0.81$)\cdot 10^{-2}$ & (-1.86$\pm$2.17$)\cdot 10^{-1}$ \\ \cline{2-5}
 & 50 & (2.33$\pm$0.82$)\cdot 10^{-1}$ & (3.79$\pm$0.62$)\cdot 10^{-2}$ & (-5.73$\pm$97.05$)\cdot 10^{-3}$ \\ \cline{2-5}
 & 75 & (2.66$\pm$0.98$)\cdot 10^{-1}$ & (3.26$\pm$0.64$)\cdot 10^{-2}$ & (-1.60$\pm$10.99$)\cdot 10^{-2}$ \\ \hline
\multirow{3}{*}{\color{redplot}Slope} & 25 & (-9.59$\pm$8.83$)\cdot 10^{-7}$~$s^{-1}$ & (4.75$\pm$1.74$)\cdot 10^{-2}$ & (2.32$\pm$12.77$)\cdot 10^{-7}$~$s^{-1}$ \\ \cline{2-5}
 & 50 & (-1.84$\pm$1.17$)\cdot 10^{-7}$~$s^{-1}$ & (8.22$\pm$1.26$)\cdot 10^{-2}$ & (-5.36$\pm$3.44$)\cdot 10^{-7}$~$s^{-1}$ \\ \cline{2-5}
 & 75 & (-4.41$\pm$3.26$)\cdot 10^{-7}$~$s^{-1}$ & (6.36$\pm$1.45$)\cdot 10^{-2}$ & (9.79$\pm$69.20$)\cdot 10^{-8}$~$s^{-1}$ \\ \hline
\multirow{3}{*}{\color{greenplot}Offset} & 25 & -$^*$ & (-8.30$\pm$0.66$)\cdot 10^{-3}$ & (7.44$\pm$0.21$)\cdot 10^{-1}$ \\ \cline{2-5}
 & 50 & -$^*$ & (-7.84$\pm$0.93$)\cdot 10^{-3}$ & (7.37$\pm$0.27$)\cdot 10^{-1}$ \\ \cline{2-5}
 & 75 & -$^*$ & (-6.83$\pm$0.88$)\cdot 10^{-3}$ & (7.03$\pm$0.27$)\cdot 10^{-1}$ \\ \hline
\end{tabular}
\caption{Fit parameters of the normalized temperature-dependant exponential parameters shown in Figure \ref{fig:exp_fit_params}. The amplitude and slope are fitted with exponential functions of the form $a\exp(bx)+c$, while a linear fit$^*$ of the form $ax+b$ is performed on the temperature dependency of the offset.}
\label{tab:fitting_expfit}
\end{table}

Another way of analysing the impact of annealing on the dark current is to use the current related damage rate $\alpha$, as defined in \cite{Moll1999}:

\begin{equation}
\label{eq:alpha_def}
\alpha = \frac{\Delta I}{\Phi_{eq}V}
\end{equation}

where V is the sensitive volume in $cm^3$, $\Phi_{eq}$ the fluence in $cm^{-2}$, and $\Delta I$ the current increase due to irradiation\footnote{As discussed in the following talk: \href{https://indico.cern.ch/event/1093102/contributions/4813355/attachments/2430110/4160996/EG-SiPM-RadHard-CERNWorkshop.pdf}{"SiPMs in high radiation environment" by Erika Garutti}, one should use an extended version of \eqref{eq:alpha_def} for devices with gain like SiPMs: $\alpha_G = \frac{\Delta I}{\Phi_{eq}\cdot V\cdot G\cdot \textrm{ECF}\cdot \textrm{PDE}}$ where G is the gain, ECF the excess charge factor, and PDE the photo detection efficiency. Since a relatively low fluence was used in this study, the effect of radiation on the Gain can be neglected. Moreover, as the $\alpha$ parameter is extracted for a quite low overvoltage (3~V), the ECF can be expected to be quite low, so as its dependence on radiation damage for a small fluence. We therefore use the definition in \eqref{eq:alpha_def} for the current related damage rate.}.


The fluence with which the SiPMs have been irradiated is $10^8$ $p/cm^{2}$, and the effective surface for each type of sensor is derived using the numbers summarized in table \ref{tab:sipm_effective_surface}.

\begin{table}[ht]
\centering
\hspace*{-0.5cm}\begin{tabular}{|c|c|c|c|}
\hline
n\textsubscript{cells} & cell pitch [$\mu$m] & Fill factor [\%] & Effective channel surface [mm$^2$] \\ \hline\hline
57600 & 25 & 47 & 16.92 \\ \hline
14400 & 50 & 74 & 26.64 \\ \hline
6400 & 75 & 82 & 29.52 \\ \hline
\end{tabular}
\caption{SiPM channel effective surface for S13360-6025/50/75 SiPMs}
\label{tab:sipm_effective_surface}
\end{table}

In order to get the sensitive volume of each type of sensor, the effective surfaces in table \ref{tab:sipm_effective_surface} need to be multiplied by the thickness of the Silicon depletion layer, which is about 2~$\mu$m for Hamamatsu SiPMs \cite{GARUTTI201969}. The depletion width can be estimated using its relation to the micro-cell capacitance: $C_{\mu\textrm{cell}}=\varepsilon_0\cdot\varepsilon_{\textrm{Si}}\cdot\frac{A}{d}$, with $\varepsilon_0=9.95\cdot 10^{-14}F/cm$ the vacuum permittivity, $\varepsilon_{\textrm{Si}}$ the relative permittivity of Silicon, A the effective surface of the micro-cell, and $d$ the depletion width. Having the sensitive volume and the fluence, the dark current data from Figure \ref{fig:I_vs_time} can be converted into current related damage rate (see Figure \ref{fig:alpha_vs_time}).







The current related damage rate obeys the following relation with time \cite{Moll1999}:
\begin{equation}
\label{eq:alpha_expression}
\alpha (t) = \alpha_I \exp(-\frac{t}{\tau}) + \alpha_0 - \beta \ln(\frac{t}{t_0})
\end{equation}

where $\alpha_I$, $\alpha_0$, and $\beta$ are the amplitudes of each terms in A/cm, $t_0$ defines the time unit, and $\tau$ follows the Arrhenius relation \cite{CHILINGAROV1995432}:

\begin{equation}
\label{eq:arrhenius}
\frac{1}{\tau}\equiv k(T)=k_0\exp(-\frac{E_a}{k_B T})
\end{equation}

where $k_B=8.617\cdot 10^{-5}$~eV/K is the Boltzmann constant, $E_a$ is the activation energy of the defects, and T the annealing temperature.

The relation \eqref{eq:alpha_expression} is used to fit the current related damage rate for the three types of SiPMs at different annealing temperatures. The fits are shown in Figure \ref{fig:alpha_vs_time}, while the time dependent fit parameters are given in Table \ref{tab:alpha_fit_params} for the different amplitudes and the Arrhenius plots for the exponential decay time are provided in Figure \ref{fig:arrhenius_plot}. Note that there is almost no recovery for the coldest annealing temperature ($-22.8\pm1.8^\circ$C), resulting into big errors on the fit parameters due to the flat nature of the curve.

\begin{center}
\begin{figure}[H]
\captionsetup{justification=centering}
\begin{centering}
\includegraphics[width=.6\textwidth]{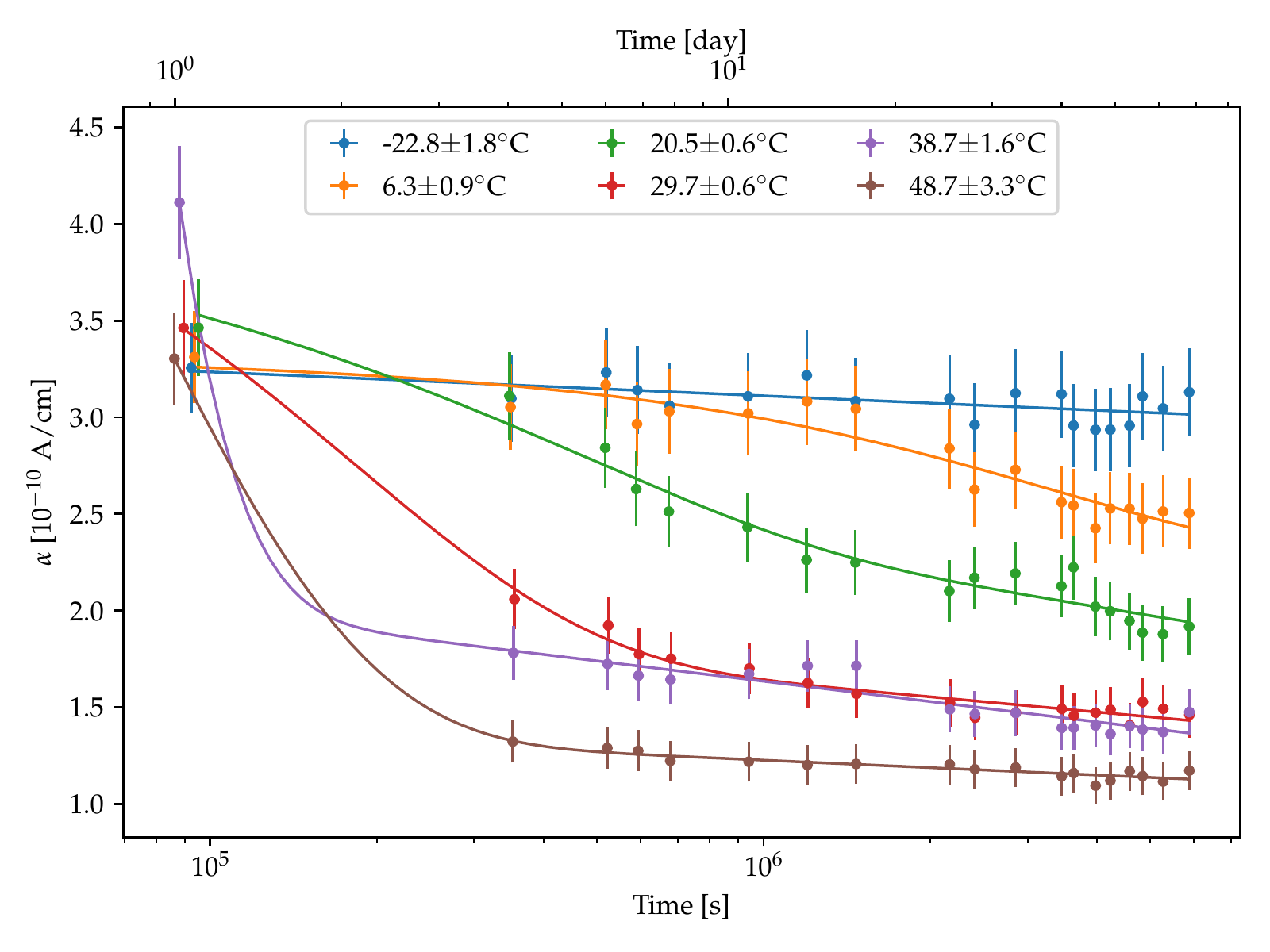}\\
\includegraphics[width=.6\textwidth]{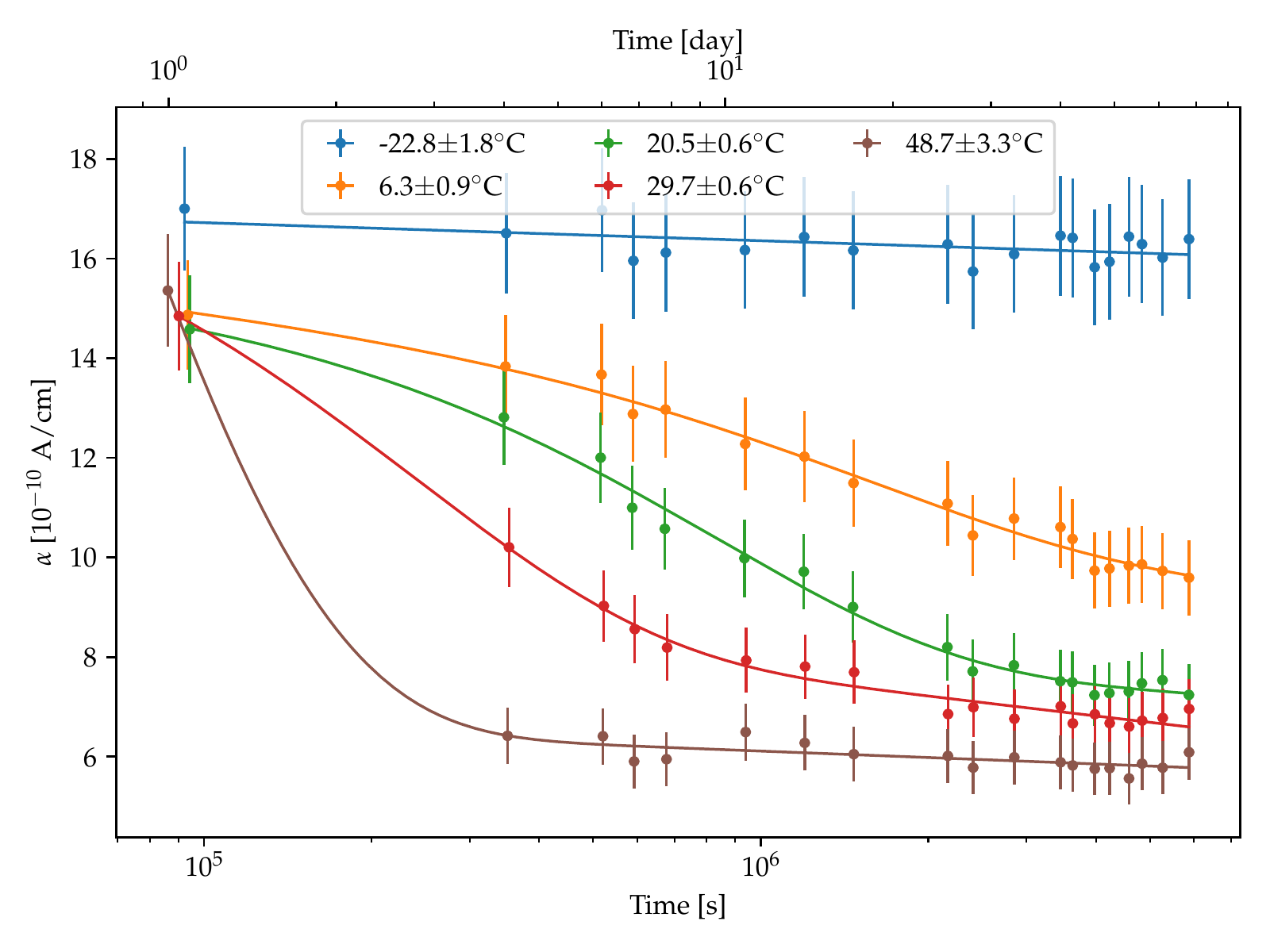}\\
\includegraphics[width=.6\textwidth]{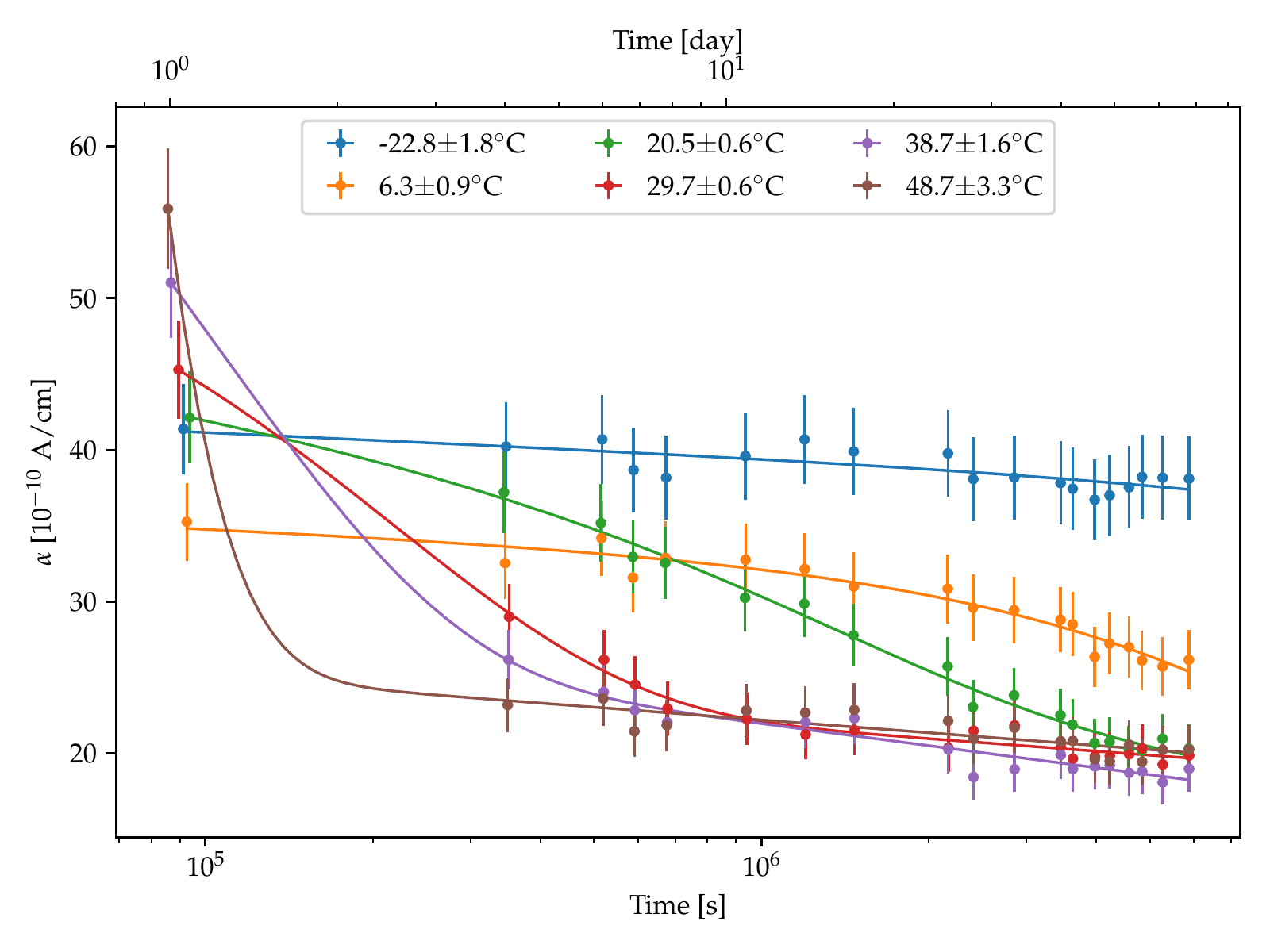}
  \par\end{centering}
  \protect\caption{Current related damage rate vs. time after irradiation, fitted with the function \eqref{eq:alpha_expression}, for the 25$\mu m$ (top), 50$\mu m$ (middle), and 75$\mu m$ (bottom) SiPMs.  Systematic effects coming both from the measurement procedure and from variations among samples contributes to the error on $\alpha$, whereas the error bars used in Figure \ref{fig:I_vs_time} only contained  a contribution from the former source systematic uncertainties, the latter being washed out by the current normalization. We do observe outlying curves due to a spread between samples, e.g. for the 75$\mu$m SiPM at 6.3$^\circ$C, but the trend of these curves behave as expected.}
    \label{fig:alpha_vs_time}
\end{figure}
\par\end{center}

\begin{table}[ht]
\centering
\hspace*{-0.5cm}\begin{tabular}{|l|c||c|c|c|}
\hline
Pitch & Temperature [$^\circ$C] & $\alpha_I$ [$10^{-10}$A/cm] & $\alpha_0$ [$10^{-10}$A/cm] & $\beta$ [$10^{-10}$A/cm] \\ \hline\hline
\multirow{6}{*}{25 $\mu m$} & $-22.8\pm1.8$ & (2.93$\pm$2204.00)$\cdot 10^{-1}$ & (3.57$\pm$220.80) & (5.46$\pm$4.26)$\cdot 10^{-2}$ \\ \cline{2-5}
 & $6.3\pm0.9$ & (8.80$\pm$3.28)$\cdot 10^{-1}$ & (2.69$\pm$1.57) & (2.43$\pm$10.48)$\cdot 10^{-2}$ \\ \cline{2-5}
 & $20.5\pm0.6$ & (8.37$\pm$3.01)$\cdot 10^{-1}$ & (5.56$\pm$1.06) & (2.33$\pm$0.70)$\cdot 10^{-1}$ \\ \cline{2-5}
 & $29.7\pm0.6$ & (2.50$\pm$0.21) & (3.06$\pm$0.35) & (1.05$\pm$0.24)$\cdot 10^{-1}$ \\ \cline{2-5}
 & $38.7\pm1.6$ & (1.30$\pm$4018.00)$\cdot 10^{2}$ & (3.73$\pm$0.28) & (1.52$\pm$0.19)$\cdot 10^{-1}$ \\ \cline{2-5}
 & $48.7\pm3.3$ & (6.65$\pm$1.66) & (1.99$\pm$0.13) & (5.55$\pm$0.85)$\cdot 10^{-2}$ \\ \hline
\multirow{5}{*}{50 $\mu m$} & $-22.8\pm1.8$ & (1.57$\pm$1566.00) & (1.69$\pm$156.80)$\cdot 10^{1}$ & (1.57$\pm$1.59)$\cdot 10^{-1}$ \\ \cline{2-5}
 & $6.3\pm0.9$ & (3.32$\pm$1.19) & (1.81$\pm$0.50)$\cdot 10^{1}$ & (5.52$\pm$3.26)$\cdot 10^{-1}$ \\ \cline{2-5}
 & $20.5\pm0.6$ & (6.79$\pm$1.05) & (1.19$\pm$0.43)$\cdot 10^{1}$ & (2.95$\pm$2.80)$\cdot 10^{-1}$ \\ \cline{2-5}
 & $29.7\pm0.6$ & (8.53$\pm$0.71) & (1.53$\pm$0.18)$\cdot 10^{1}$ & (5.56$\pm$1.17)$\cdot 10^{-1}$ \\ \cline{2-5}
 & $48.7\pm3.3$ & (3.48$\pm$2.12)$\cdot 10^{1}$ & (8.63$\pm$0.92) & (1.83$\pm$0.62)$\cdot 10^{-1}$ \\ \hline
\multirow{6}{*}{75 $\mu m$} & $-22.8\pm1.8$ & (3.67$\pm$169.20) & (4.61$\pm$16.27)$\cdot 10^{1}$ & (7.47$\pm$7.61)$\cdot 10^{-1}$ \\ \cline{2-5}
 & $6.3\pm0.9$ & (2.16$\pm$4.03)$\cdot 10^{1}$ & (1.97$\pm$3.51)$\cdot 10^{1}$ & (5.69$\pm$6.88)$\cdot 10^{-1}$ \\ \cline{2-5}
 & $20.5\pm0.6$ & (1.32$\pm$0.40)$\cdot 10^{1}$ & (5.72$\pm$1.71)$\cdot 10^{1}$ & (2.40$\pm$1.10) \\ \cline{2-5}
 & $29.7\pm0.6$ & (3.11$\pm$0.22)$\cdot 10^{1}$ & (3.82$\pm$0.45)$\cdot 10^{1}$ & (1.19$\pm$0.30) \\ \cline{2-5}
 & $38.7\pm1.6$ & (5.38$\pm$0.94)$\cdot 10^{1}$ & (5.04$\pm$0.41)$\cdot 10^{1}$ & (2.06$\pm$0.28) \\ \cline{2-5}
 & $48.7\pm3.3$ & (1.94$\pm$7203.00)$\cdot 10^{3}$ & (3.91$\pm$0.35)$\cdot 10^{1}$ & (1.22$\pm$0.24) \\ \hline
\end{tabular}
\caption{Fit parameters vs. annealing temperature obtained using the expression \eqref{eq:alpha_expression} for the 3 microcell pitches. The fit for $-22.8\pm1.8^\circ$C gives very big error due to the fact that $\alpha$ versus time is flat since there is almost not annealing at this temperature.}
\label{tab:alpha_fit_params}
\end{table}

\begin{center}
\begin{figure}[H]
\captionsetup{justification=centering}
\begin{centering}
\includegraphics[width=.7\textwidth]{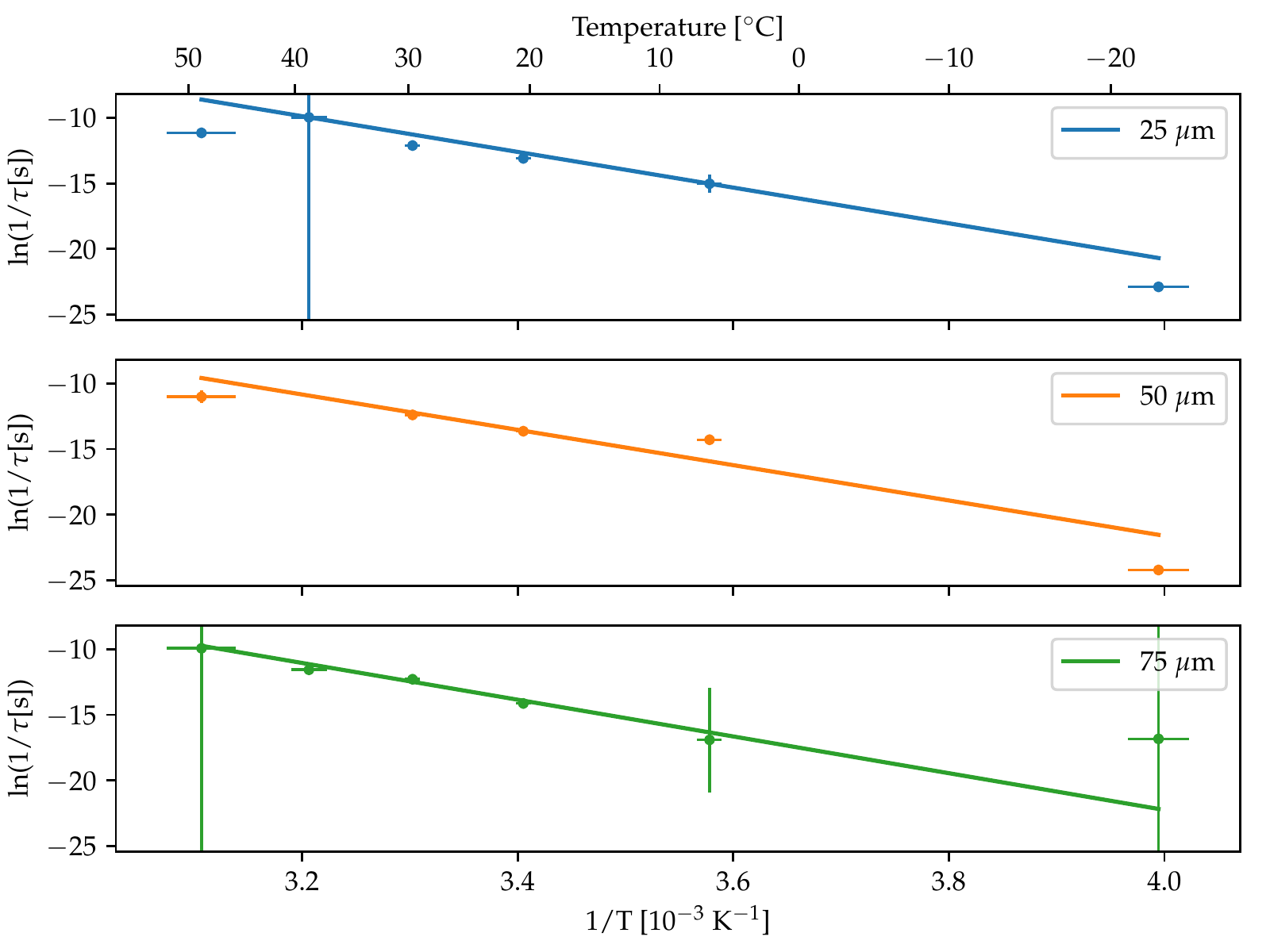}
  \par\end{centering}
  \protect\caption{Arrhenius plot for the decay time $\tau$ for the 25$\mu m$ (top), 50$\mu m$ (middle), and 75$\mu m$ (bottom) SiPMs}
    \label{fig:arrhenius_plot}
\end{figure}
\par\end{center}

The Arrhenius plots in Figure \ref{fig:arrhenius_plot}, which show the fit values for $\tau$ as a function of temperature, are fitted using the relation \eqref{eq:arrhenius}. The fit parameters and their corresponding physical quantities are provided in Table \ref{tab:arrhenius_fit_params}. $k_0$ corresponds to the frequency at which the defects try to escape \cite{Honniger2007}. A rate of the order $10^{12}-10^{13}$~Hz is expected from a dissociative process \cite{Moll1999,PhysRevB.75.155202,Corbett1966}. This is compatible with the values obtained from the offset of the fit. It is therefore likely that the exponential decay part in \eqref{eq:alpha_expression} is caused by a dissociation of the defects.

\begin{table}[H]
\centering
\bgroup
\def\arraystretch{1.5}
\hspace*{-0.5cm}\begin{tabular}{|l||c|c|c|c|}
\hline
Pitch & ln($k_0$ [$s^{-1}$]) & $k_0$ [$s^{-1}$] & Slope [K] & $E_a$ [eV] \\ \hline\hline
25 $\mu m$ & (3.36$\pm$0.37)$\cdot 10^{1}$ & $4.05\cdot 10^{14}\:\substack{+1.78\cdot 10^{15} \\ -4.05\cdot 10^{14}}$ & (-1.36$\pm$0.11)$\cdot 10^{4}$ & 1.17$\pm$0.09 \\ \hline
50 $\mu m$ & (3.22$\pm$1.08)$\cdot 10^{1}$ & $9.82\cdot 10^{13}\:\substack{+1.89\cdot 10^{16} \\ -9.82\cdot 10^{13}}$ & (-1.35$\pm$0.32)$\cdot 10^{4}$ & 1.16$\pm$0.28 \\ \hline
75 $\mu m$ & (3.37$\pm$0.75)$\cdot 10^{1}$ & $4.51\cdot 10^{14}\:\substack{+1.49\cdot 10^{16} \\ -4.51\cdot 10^{14}}$ & (-1.40$\pm$0.23)$\cdot 10^{4}$ & 1.21$\pm$0.20 \\ \hline
\end{tabular}
\egroup
\caption{Linear fit parameters obtained for the Arrhenius plots in Figure \ref{fig:arrhenius_plot}. The activation energy $E_a$ for defect migration or dissociation is computed from the slope ($E=k_BT$), while $k_0$ obtained from the offset of the fit corresponds to the frequency at which defects will try to escape \cite{Honniger2007}.}
\label{tab:arrhenius_fit_params}
\end{table}









The annealing effect has also been studied at higher temperatures, by storing at the end of the measurement campaign the SiPMs previously stored at 48.7$\pm$3.3$^\circ$C for a few days in a climatic chamber. The SiPMs were first annealed at a temperature of 75$^\circ$C, and then at 100$^\circ$C. Figure \ref{fig:annealing_hightemp} shows the I-V curve taken for a 25~$\mu$m SiPM after stabilization\footnote{The I-V curve shown in the figure were taken after a few days for the 75 and 100$^\circ$C annealing temperatures, where the annealing goes very fast. The I-V curves for the other annealing temperatures correspond to the last ones shown in Figure \ref{fig:all_IVs_25}, taken after about 2 months of annealing.} for each measured annealing temperature as well as the current at 3~V overvoltage extracted from each of these I-V curves for all three types of SiPMs.

\begin{center}
\begin{figure}[H]
\captionsetup{justification=centering}
\begin{centering}
\includegraphics[width=.5\textwidth]{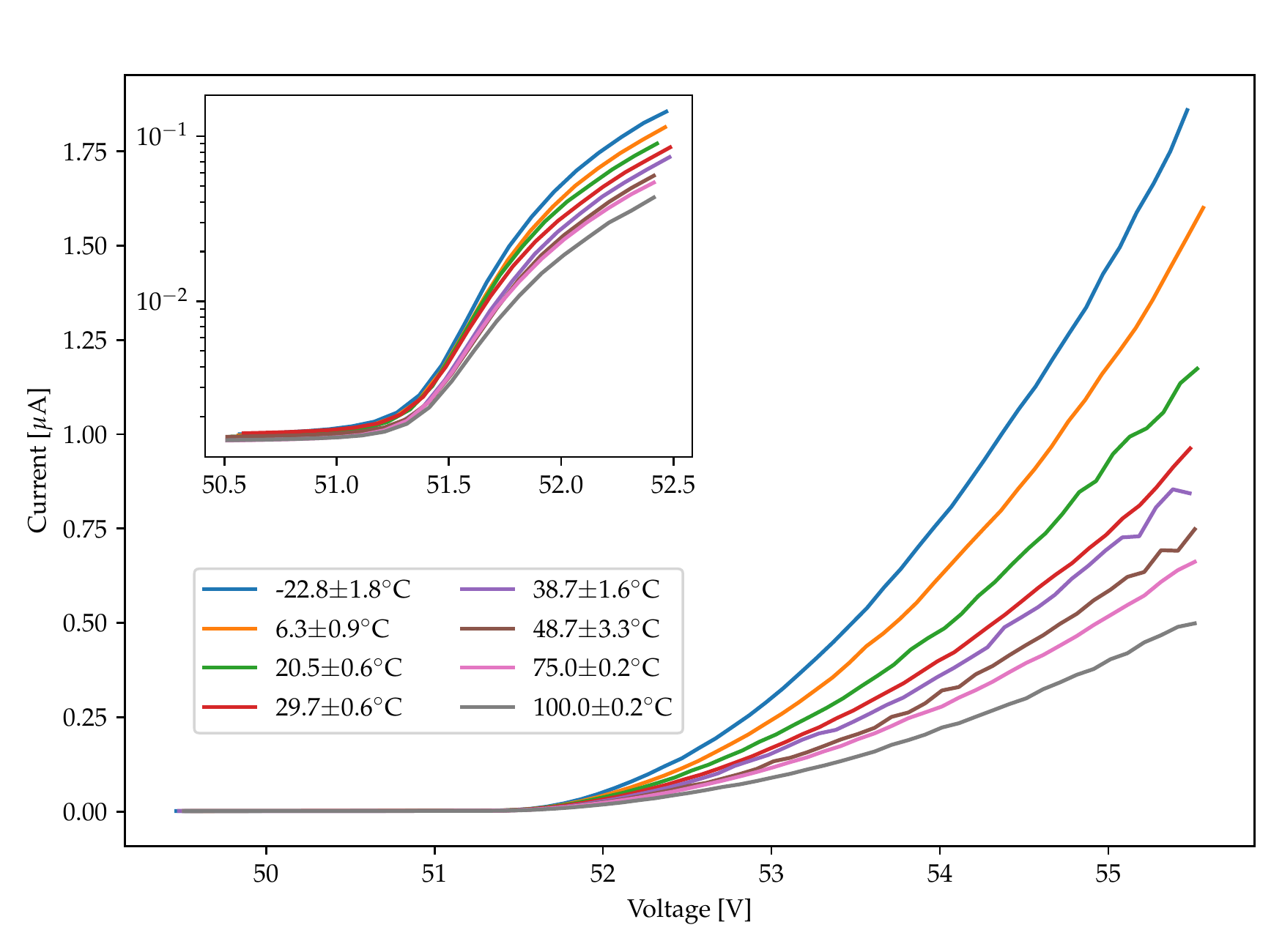}\includegraphics[width=.5\textwidth]{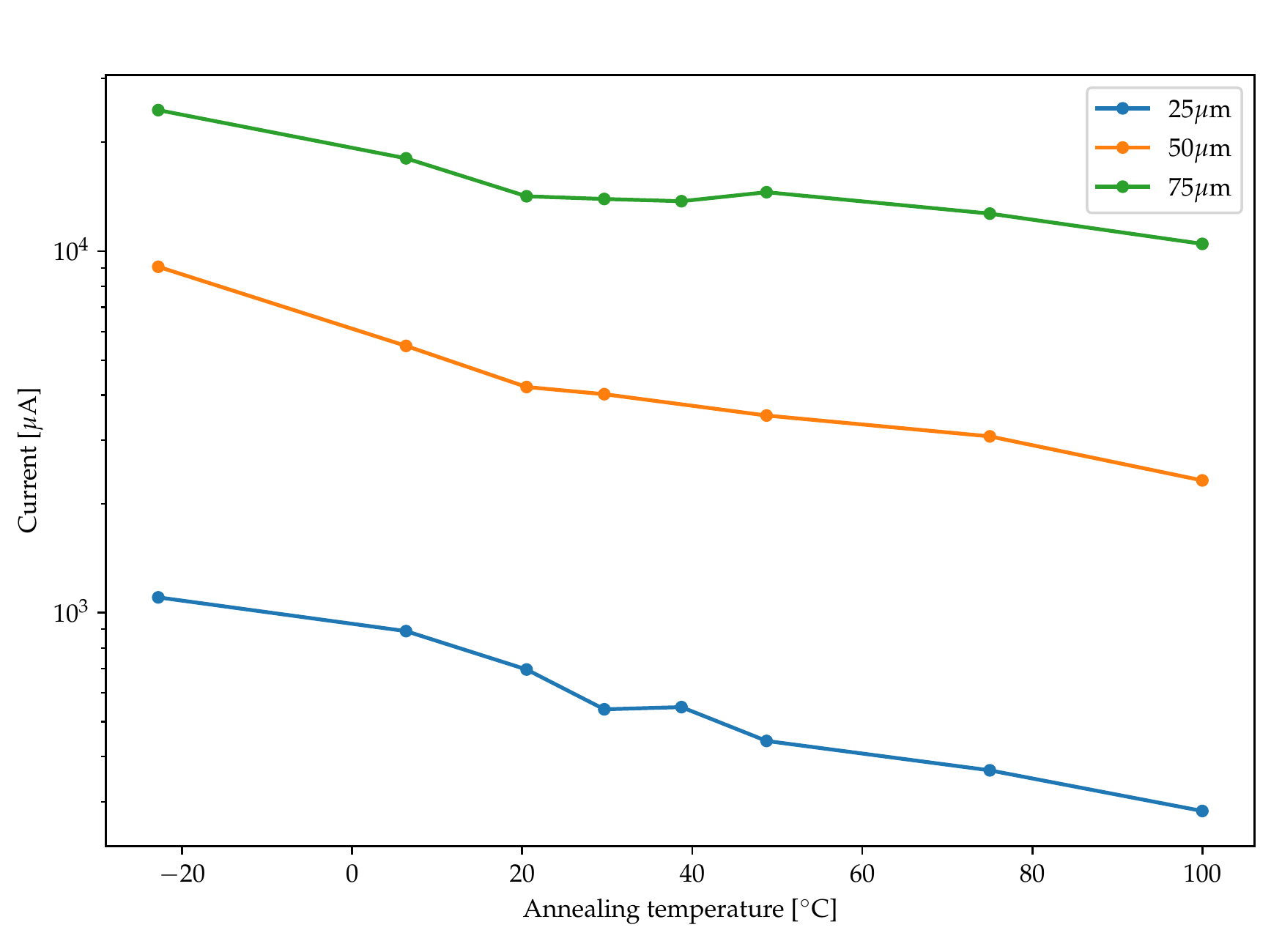}
  \par\end{centering}
  \protect\caption{I-V curve measured after stabilization for annealing temperatures up to 100$^\circ$C for the 25~$\mu$m SiPM (left) and current at 3~V overvoltage for 25, 50, and 75~$\mu$m SiPMs after stabilization including high annealing temperatures (right).}
    \label{fig:annealing_hightemp}
\end{figure}
\par\end{center}

Figure \ref{fig:annealing_hightemp_I3Vnorm} shows the current at 3~V overvoltage of Figure \ref{fig:annealing_hightemp} normalized by the current measured right after the irradiation. The curves therefore provides the recovery fraction for the 25, 50, and 75 $\mu$m SiPMs as a function of the annealing temperature. The solid lines given for all three kinds of SiPM are the normalized current values after an infinite time, calculated using the exponential fit parameters in Table \ref{tab:fitting_expfit}. Even though there is a quite good match up to 50$^\circ$C between the current measured after about 2~months and the value at infinity computed from the fit, the measured current at 75 and 100$^\circ$C are higher than the values from the extrapolation of the fit. As previously explained, the SiPMs were only stored for a couple of days in a climatic chambers at both 75 and 100$^\circ$C, but this is sufficient for the annealing to stabilize since the effect is significantly faster at high temperatures. The non-linearity observed in Figure \ref{fig:annealing_hightemp_I3Vnorm} is therefore due to a saturation effect of the annealing process at high temperatures, which is not accounted for in the solid lines derived from the fits since those fits were performed in the range -22.8$\pm$1.8 to 48.7$\pm$3.3$^\circ$C.

\begin{center}
\begin{figure}[H]
\captionsetup{justification=centering}
\begin{centering}
\includegraphics[width=.6\textwidth]{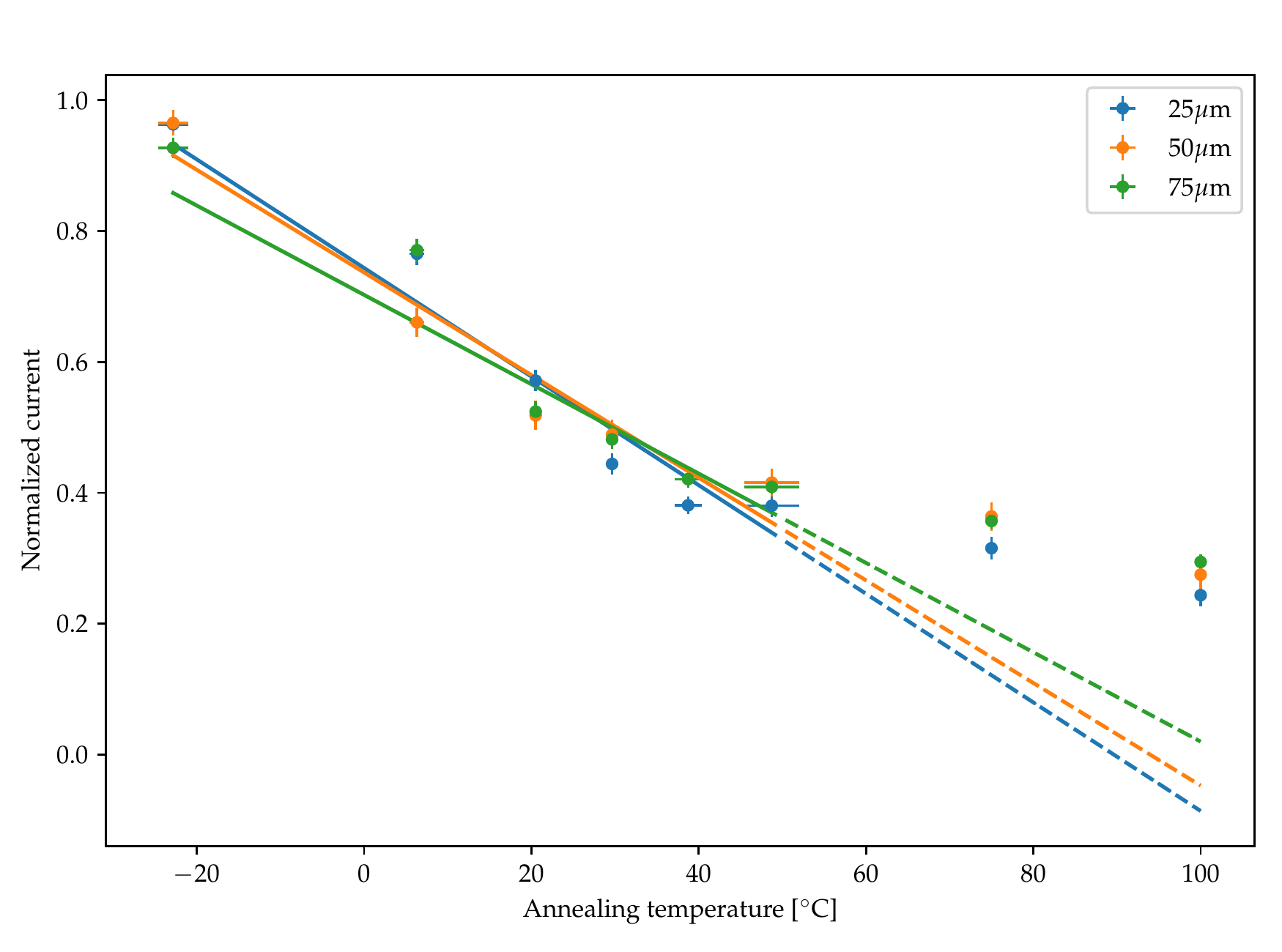}
  \par\end{centering}
  \protect\caption{Normalized current at 3~V overvoltage for 25, 50, and 75~$\mu$m SiPMs after stabilization including high annealing temperatures. A linear fit is performed on the data points below 50$^\circ$C. A saturation of the annealing effect is observed at higher temperatures.}
    \label{fig:annealing_hightemp_I3Vnorm}
\end{figure}
\par\end{center}

Although the above annealing studies were performed with breakdown correction to ensure the same overvoltage was used when comparing the current between different SiPMs, the breakdown stability has also been checked, as shown for the 50~$\mu$m SiPM in Figure \ref{fig:BD_vs_time}. It can be seen from this plot that the breakdown voltage\footnote{One should not that the breakdown voltages in this analysis are computed by performing linear fits on the regions below and above breakdown of the $\sqrt{I}$ vs. V curve, and by taking the intersection point between the two lines. Other methods using first and second voltage derivative of the current logarithm were found to give similar breakdown values} variations for the different annealing temperature are all within 0.1~V. A correlation between the breakdown variations of the different SiPMs can clearly be observed from these curves, implying the variation is a result of local temperature variations in the laboratory\footnote{The breakdown voltages are corrected for the temperature measured in the clean room where the I-V characterizations are made. The observed fluctuations with time are therefore likely due to local temperature variations in the laboratory, but they do not affect the analysis since the breakdown is computed for every measurement to ensure that the current is always extracted for the same overvoltage.}. A similar behavior is observed for the 25 and 75~$\mu$m SiPMs.

\begin{center}
\begin{figure}[H]
\captionsetup{justification=centering}
\begin{centering}
\includegraphics[width=.6\textwidth]{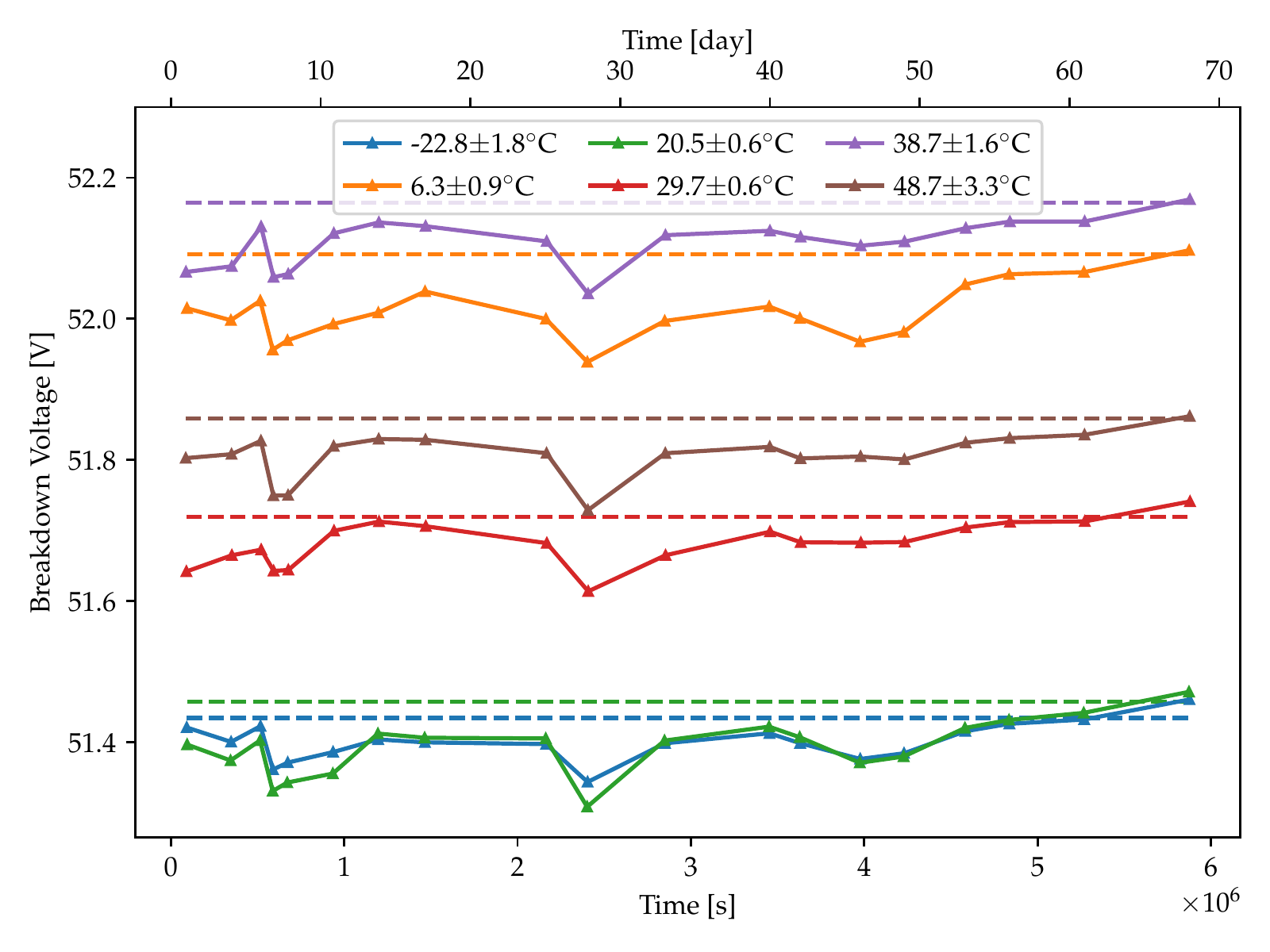}
  \par\end{centering}
  \protect\caption{Temperature corrected breakdown voltage vs. time after irradiation for the 75$\mu m$ SiPMs. A similar behavior is observed for the 25 and 50$\mu m$ SiPMs. The common variations of the breakdowns is due to local temperature variations in the measurement room. Horizontal dashed lines corresponds to the breakdowns measured before irradiation.}
    \label{fig:BD_vs_time}
\end{figure}
\par\end{center}

\subsection{Temperature dependence of the annealing effect on the SiPM dark spectrum}

Analysing the effect of radiation damage and the related recovery on the SiPM's dark current and breakdown voltage does not give a complete picture of the evolution of the performances of the sensor. In order to get an idea of the photon resolution degradation after irradiation and behavior with annealing, the dark spectra of the detectors should also be studied. An example of fitted spectra for a non-irradiated 50$\mu$m SiPM measured at 0$^\circ$C is shown in Figure \ref{fig:initial_darkspectrum}, where the applied thresholds mentioned in section \ref{subsec:char_setup} are indicated with the red lines on the top plot.

\begin{center}
\begin{figure}[H]
\captionsetup{justification=centering}
\begin{centering}
\includegraphics[width=.6\textwidth]{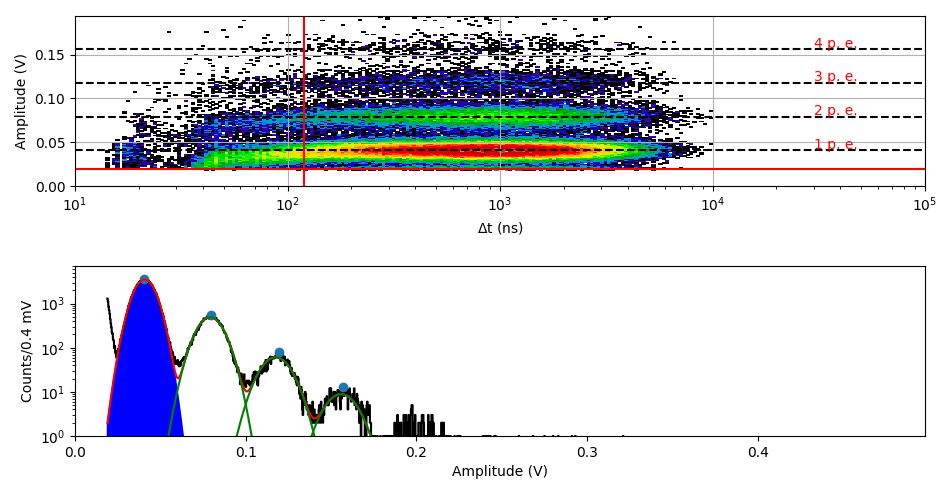}
  \par\end{centering}
  \protect\caption{Trigger count map in the amplitude-time space, where an amplitude threshold of 20~mV and an inter-pulse time threshold of 120~ns are applied (top) and corresponding dark spectra fitted with a sum of Gaussians (bottom) for a non-irradiated 50$\mu$m SiPM as measured at 0$^\circ$C}
    \label{fig:initial_darkspectrum}
\end{figure}
\par\end{center}

As can be seen in Figure \ref{fig:SPE_50}, the performance recovery associated with the annealing effect does not only apply to dark current, but also to the photon resolution.  Indeed, the peaks in the dark spectra gets narrower and more recognizable at higher annealing temperatures both at 0 and 20$^\circ$C. We only show here the spectra for 50~$\mu$m SiPMs, since similar behaviors are observed for the 75~$\mu$m ones. The dark spectrum for a non irradiated SiPM is also given as a reference.

\begin{center}
\begin{figure}[H]
\captionsetup{justification=centering}
\begin{centering}
\includegraphics[width=.5\textwidth]{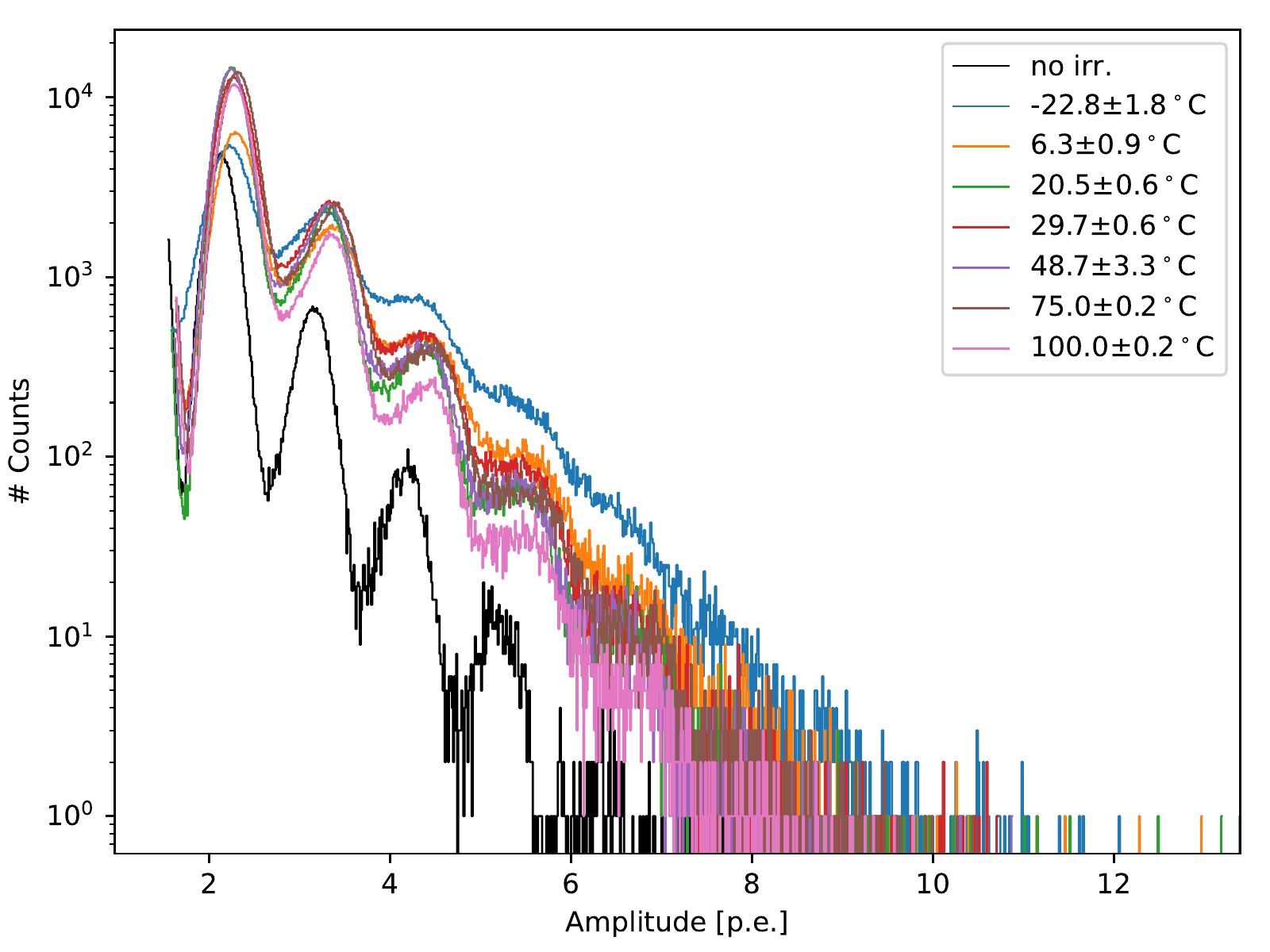}\includegraphics[width=.5\textwidth]{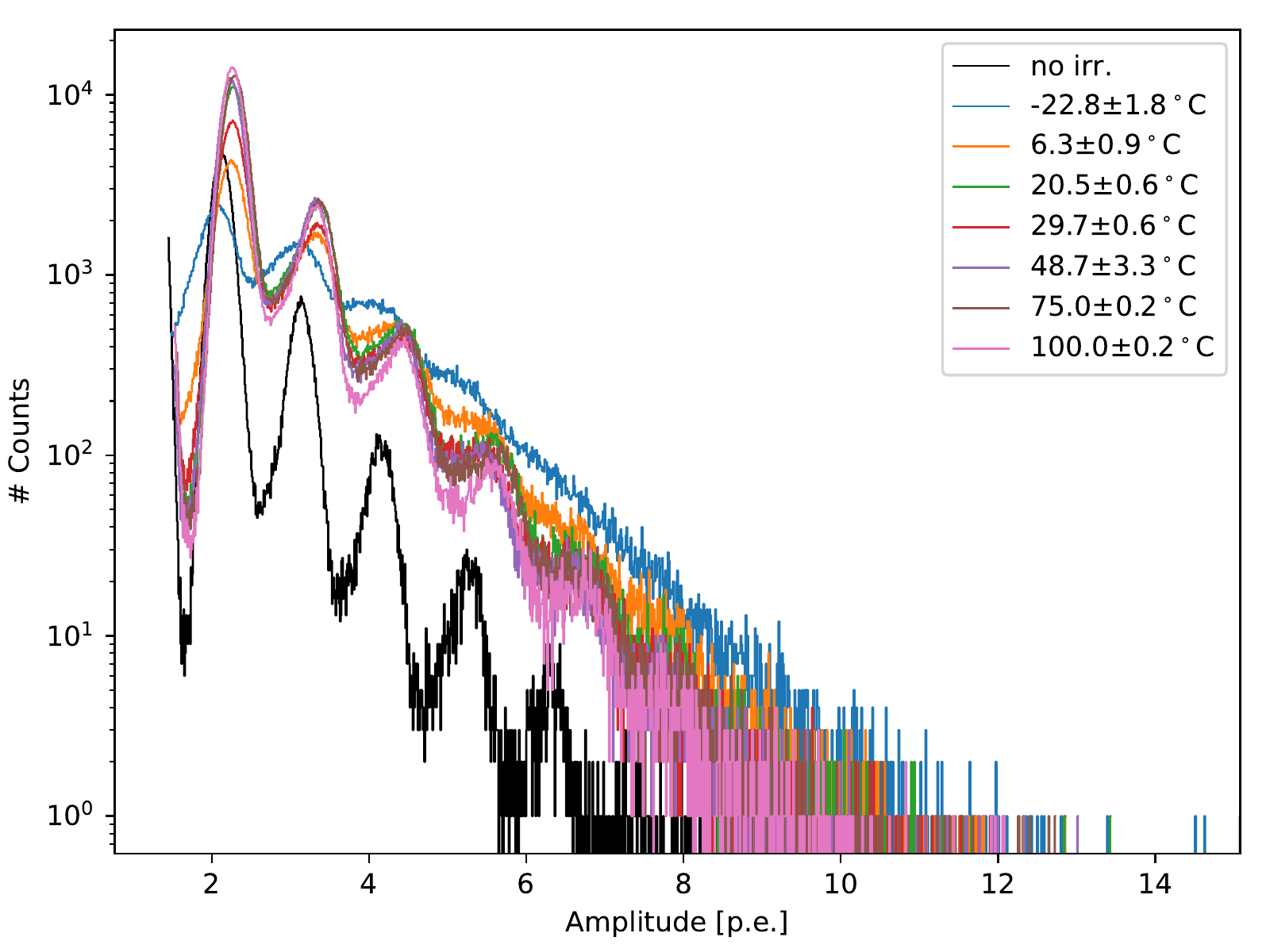}
  \par\end{centering}
  \protect\caption{Dark spectrum for a 50~$\mu$m SiPM as measured at 0 (left) and 20$^\circ$C (right) for different annealing temperatures}
    \label{fig:SPE_50}
\end{figure}
\par\end{center}


Another interesting feature to analyse is the effect of annealing on the DCR as a function of the applied threshold. From these curves, provided in Figure \ref{fig:DCR_50} for the 50~$\mu$m, we first see the improvement of resolution with annealing, but also a change in slope, implying that the main contribution to the rate is shifting towards low amplitudes. 

\begin{center}
\begin{figure}[H]
\captionsetup{justification=centering}
\begin{centering}
\includegraphics[width=.5\textwidth]{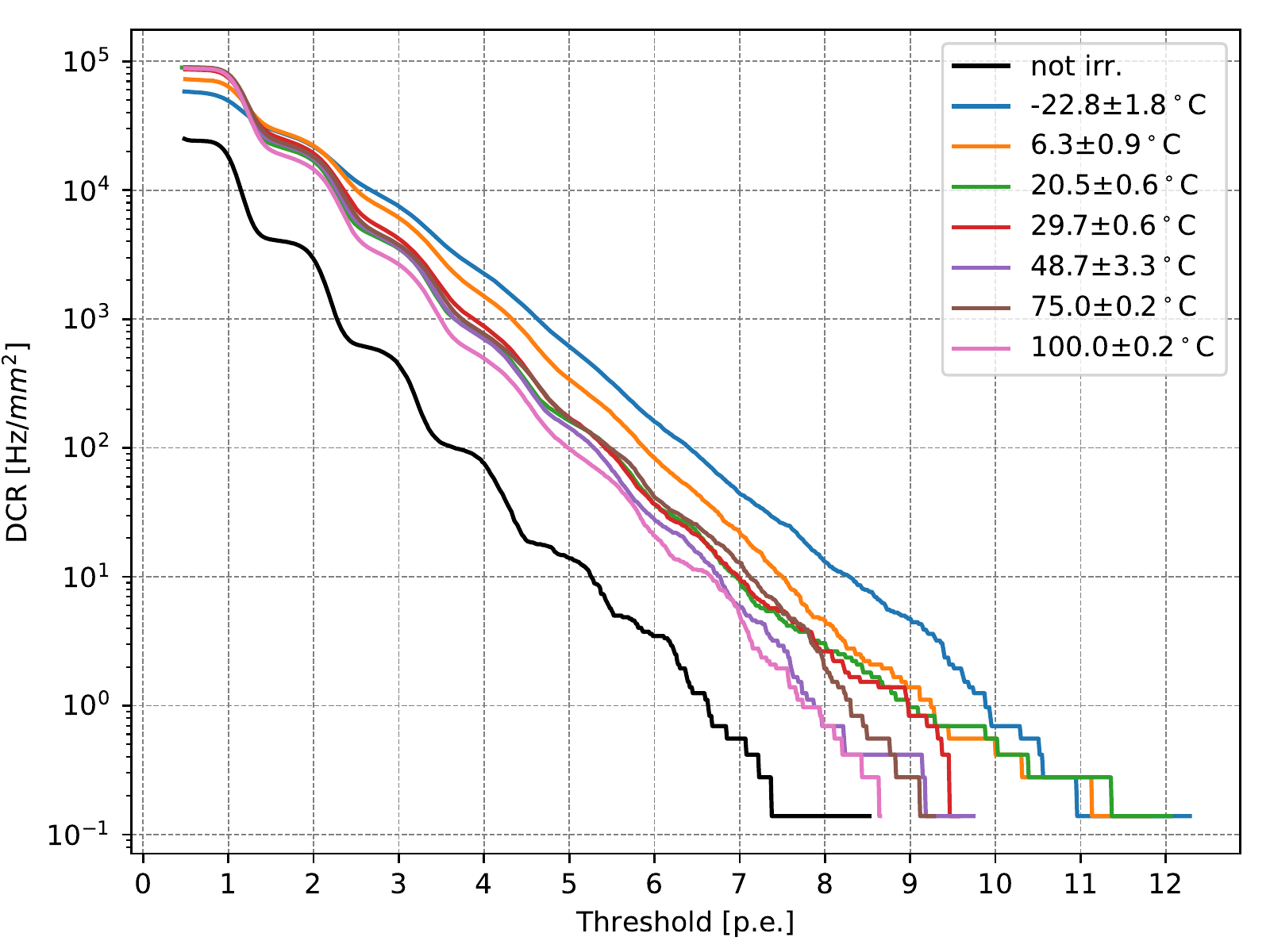}\includegraphics[width=.5\textwidth]{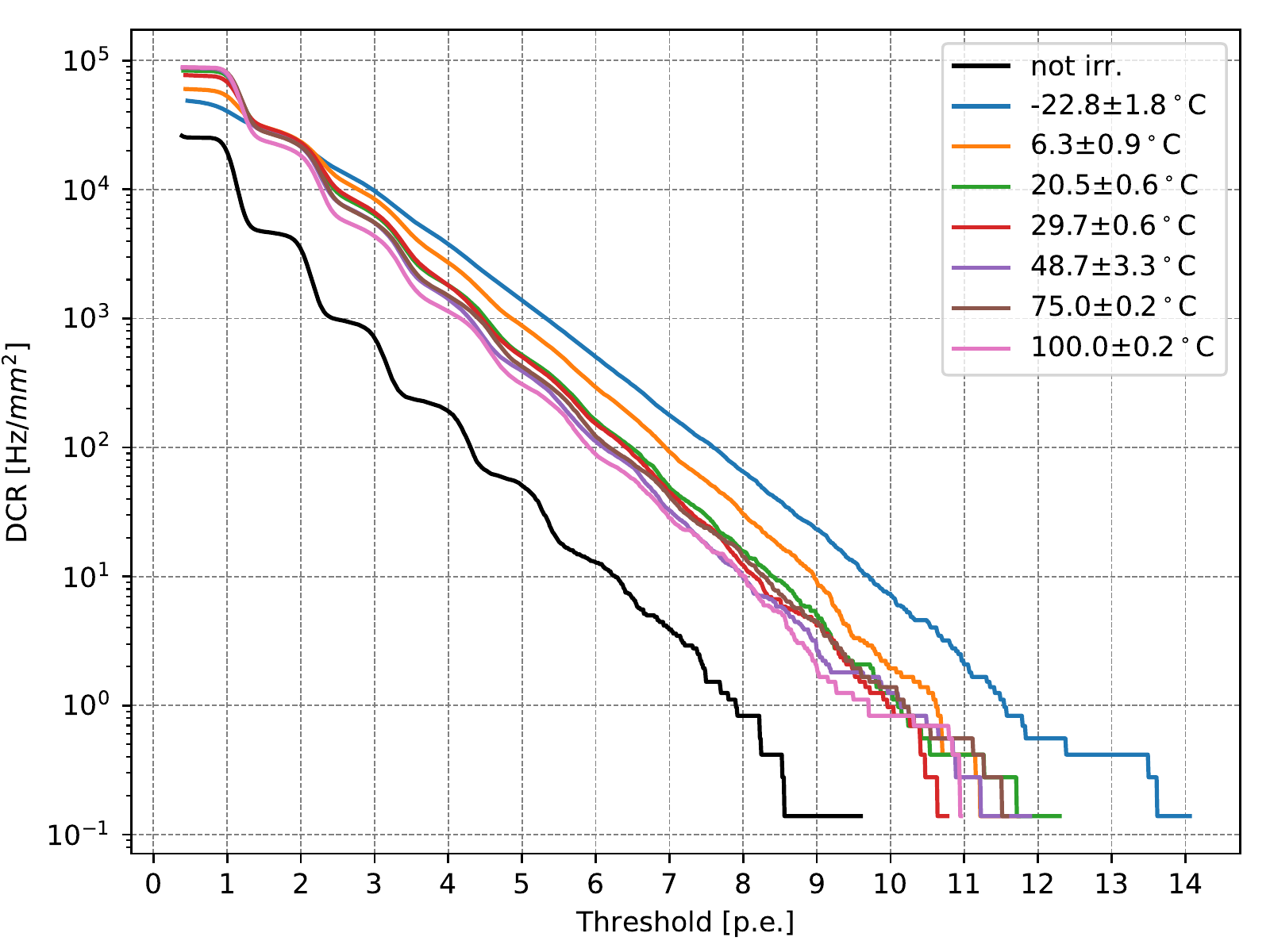}
  \par\end{centering}
  \protect\caption{DCR as a function of the threshold for a 50~$\mu$m SiPM as measured at 0 (left) and 20$^\circ$C (right) for different annealing temperatures}
    \label{fig:DCR_50}
\end{figure}
\par\end{center}




\begin{center}
\begin{figure}[H]
\captionsetup{justification=centering}
\begin{centering}
\includegraphics[width=.5\textwidth]{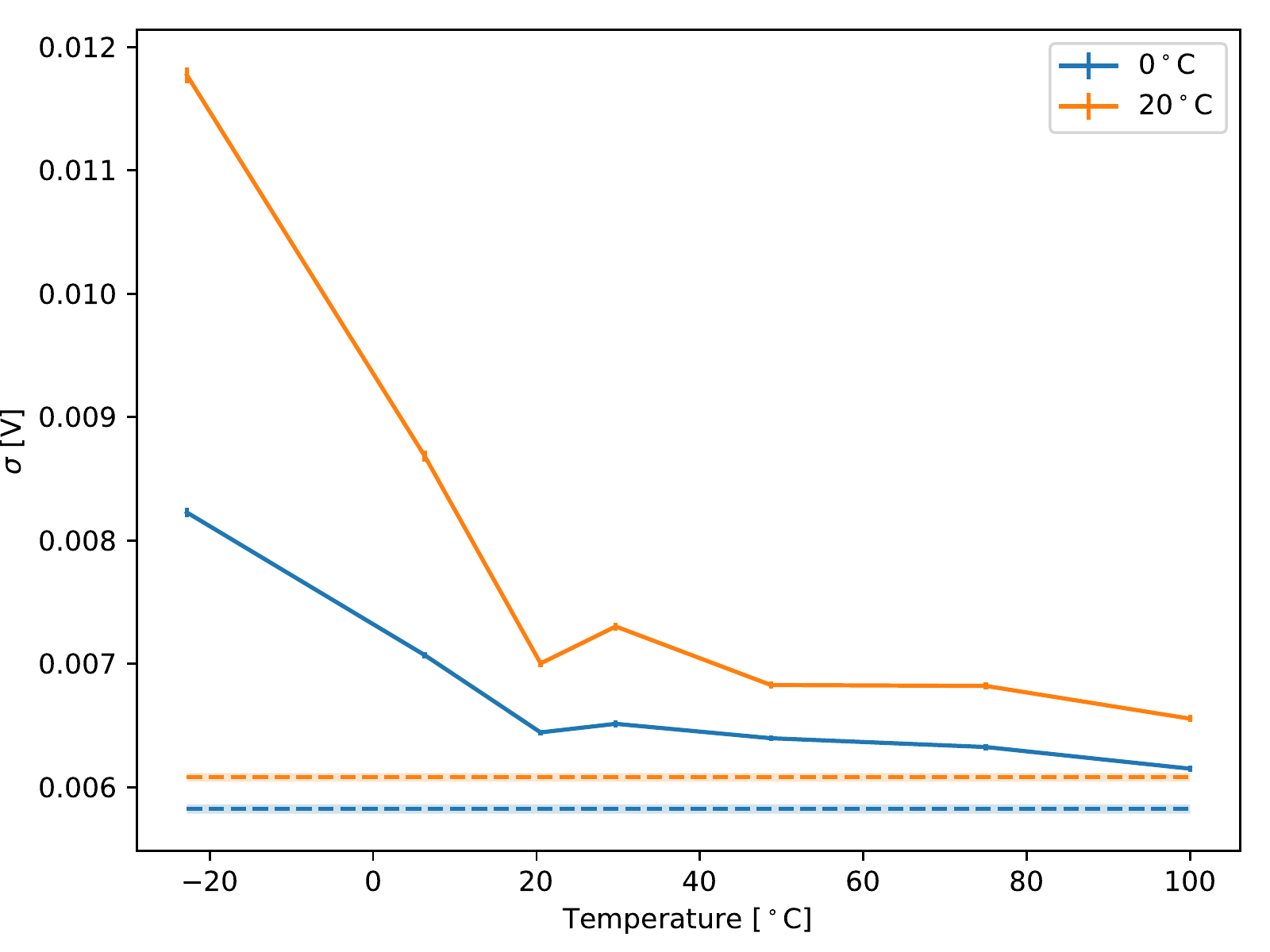}
  \par\end{centering}
  \protect\caption{First photo-electron peak width vs. annealing temperature for the 50~$\mu$m SiPMs as measured at 0 and 20$^\circ$C. The horizontal dashed lines correspond to the values before the irradiation.}
    \label{fig:Pxt_sigma1stpeak_vs_T}
\end{figure}
\par\end{center}

The width of the first peak in the dark spectra, corresponding to single photoelectron events, is shown as a function of annealing temperature in Figure \ref{fig:Pxt_sigma1stpeak_vs_T}, where we clearly see the improvement at high temperatures. A linear fit of the peak width versus peak number in each spectrum as been performed, the offset of the linear function corresponding to the first peak width, and the slope to the increase in the width from a peak $n$ to a peak $n+1$. While the former as shown a degradation with irradiation, and as discussed above an improvement with annealing, the latter was not affected by the irradiation.

%

\subsection{Bias voltage dependence of the annealing effect}

All the above results were obtained for SiPMs passively stored at different annealing temperatures. When used in an experiment, the sensors will be biased, which implies the need of a comparison of the annealing effect between biased and unbiased SiPMs. Figure \ref{fig:I_BD_vs_time_HV} shows the 3~V overvoltage current and breakdown voltage time evolution for different 50~$\mu$m SiPMs stored at $20.5\pm0.6^\circ$C with different bias voltage. Two sensors were stored with a 3~V overvoltage, which corresponds to the recommended operation point by the manufacturer, and two other chips were stored with higher overvoltages of 8 and 12~V. The data for an unbiased SiPM stored at the same temperature, which corresponds to the one already presented in section \ref{subsec:annealing_current}, is also plotted for comparison.

\begin{center}
\begin{figure}[H]
\captionsetup{justification=centering}
\begin{centering}
\includegraphics[width=.5\textwidth]{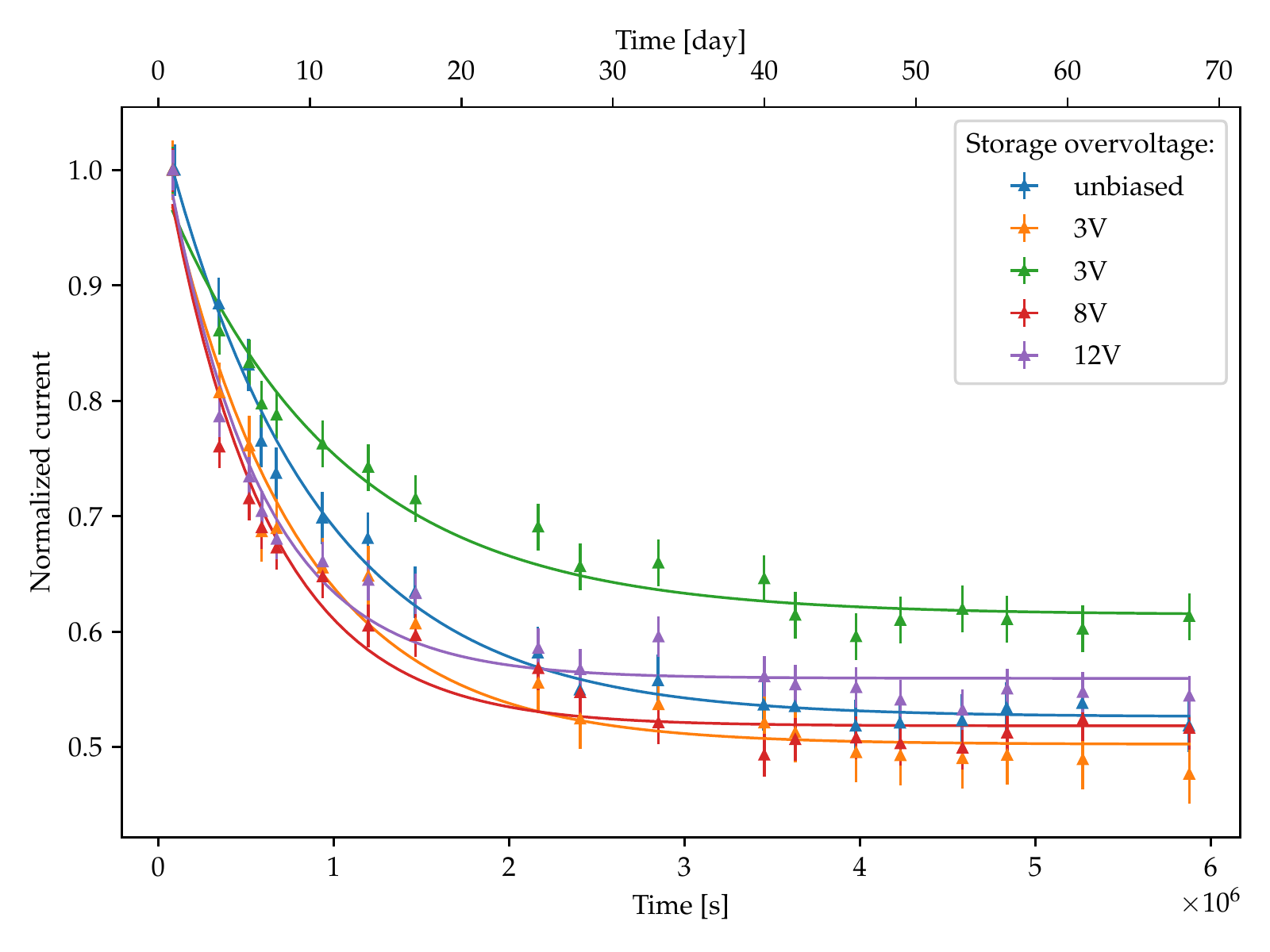}\includegraphics[width=.5\textwidth]{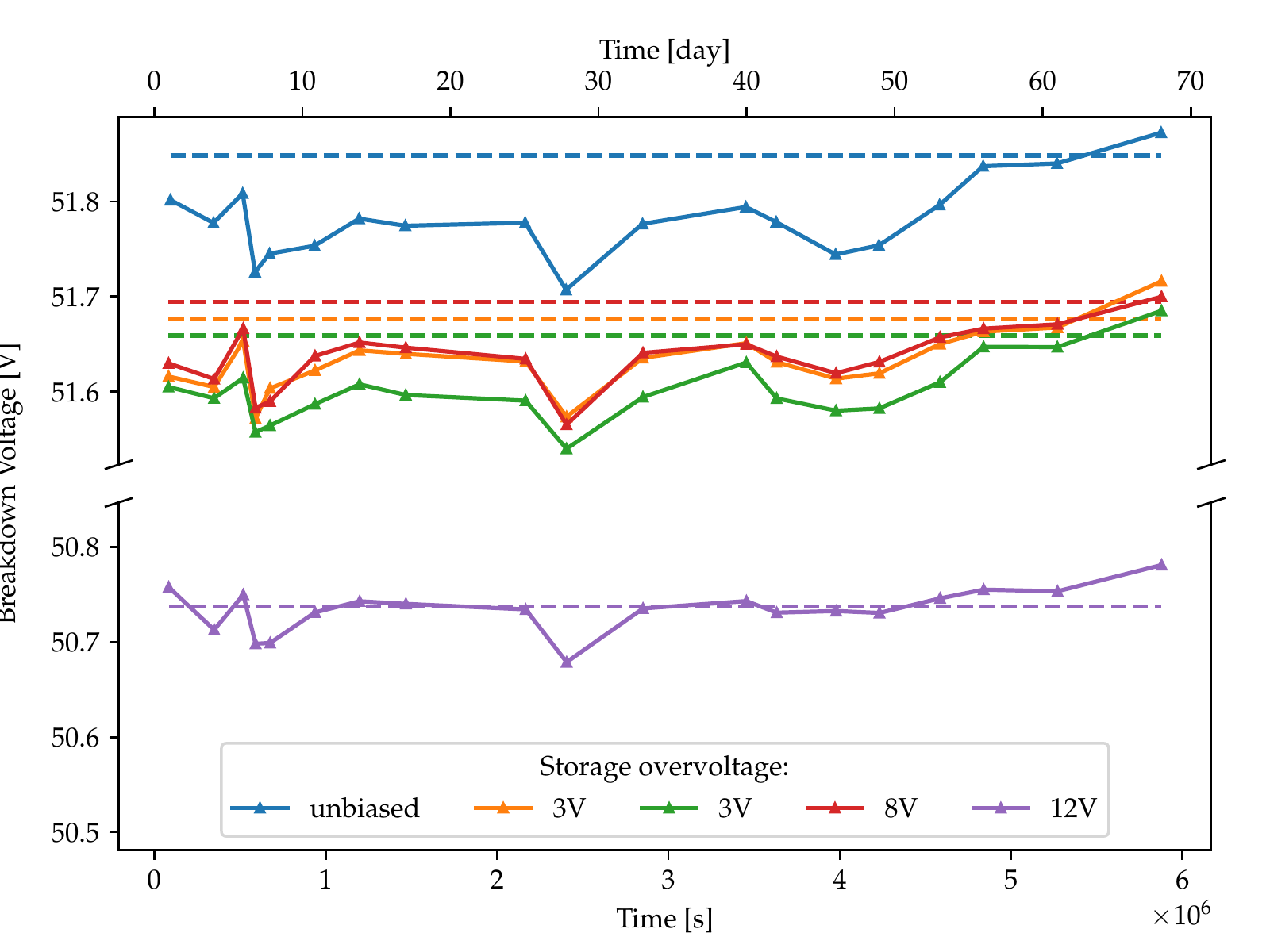}
  \par\end{centering}
  \protect\caption{Normalized current measured at 3V over-voltage (left) and breakdown voltage (right) vs. time after irradiation for the 50$\mu m$ SiPM stored at room temperature ($20.5\pm0.6^\circ$C)}
    \label{fig:I_BD_vs_time_HV}
\end{figure}
\par\end{center}

Both the dark current and the breakdown voltage show a similar behavior for unbiased and biased detectors. One of the current versus voltage curves is outlying for one of the chips stored at 3~V. This is due to a spread among SiPM chips. The trends for the different biased SiPMs are compatible with the unbiased one, making the results presented in the previous sections still valid for experiments operating at a normal overvoltage. The exponential fit parameters of the normalized current provided in Table \ref{tab:exp_fit_HV}.

\begin{table}[H]
\centering
\hspace*{-0.5cm}\begin{tabular}{|l||c|c|c|}
\hline
Overvoltage [V] & Normalized amplitude & Slope [$s^{-1}$] & Normalized offset \\ \hline\hline
Unbiased & $(5.22\pm0.16)\cdot 10^{-1}$ & $(-1.15\pm0.07)\cdot 10^{-6}$ & $(5.26\pm0.05)\cdot 10^{-1}$ \\ \hline
3 & $(5.20\pm0.28)\cdot 10^{-1}$ & $(-1.34\pm0.13)\cdot 10^{-6}$ & $(5.02\pm0.09)\cdot 10^{-1}$ \\ \hline
3 & $(3.80\pm0.19)\cdot 10^{-1}$ & $(-1.00\pm0.11)\cdot 10^{-6}$ & $(6.15\pm0.08)\cdot 10^{-1}$ \\ \hline
8 & $(5.21\pm0.30)\cdot 10^{-1}$ & $(-1.73\pm0.16)\cdot 10^{-6}$ & $(5.18\pm0.08)\cdot 10^{-1}$ \\ \hline
12 & $(4.90\pm0.29)\cdot 10^{-1}$ & $(-1.88\pm0.18)\cdot 10^{-6}$ & $(5.60\pm0.07)\cdot 10^{-1}$ \\ \hline
\end{tabular}
\caption{Exponential fit parameters for the normalized 3~V overvoltage current versus time shown on the left plot of Figure \ref{fig:I_BD_vs_time_HV} for 50$\mu$m SiPMs stored at an annealing temperature of 20$^\circ$C with different bias voltages.}
\label{tab:exp_fit_HV}
\end{table}

The dark spectra and DCR versus threshold were also measured for different annealing bias voltages (see Figure \ref{fig:SPE_DCR_HV}). The unbiased (blue curve) and first 3~V biased (orange curve) SiPMs show a similar photoelectron resolution, but the biased SiPM has a higher dark count rate. Furthermore, no significant difference in photoelectron resolution is observed for different bias voltages. Therefore, if a good thermal contact to the sensor is ensured, no electrical annealing effects \cite{electrical_annealing} due to local self-heating are observed, which makes the passive study performed in the previous sections still relevant.

\begin{center}
\begin{figure}[H]
\captionsetup{justification=centering}
\begin{centering}
\includegraphics[width=.5\textwidth]{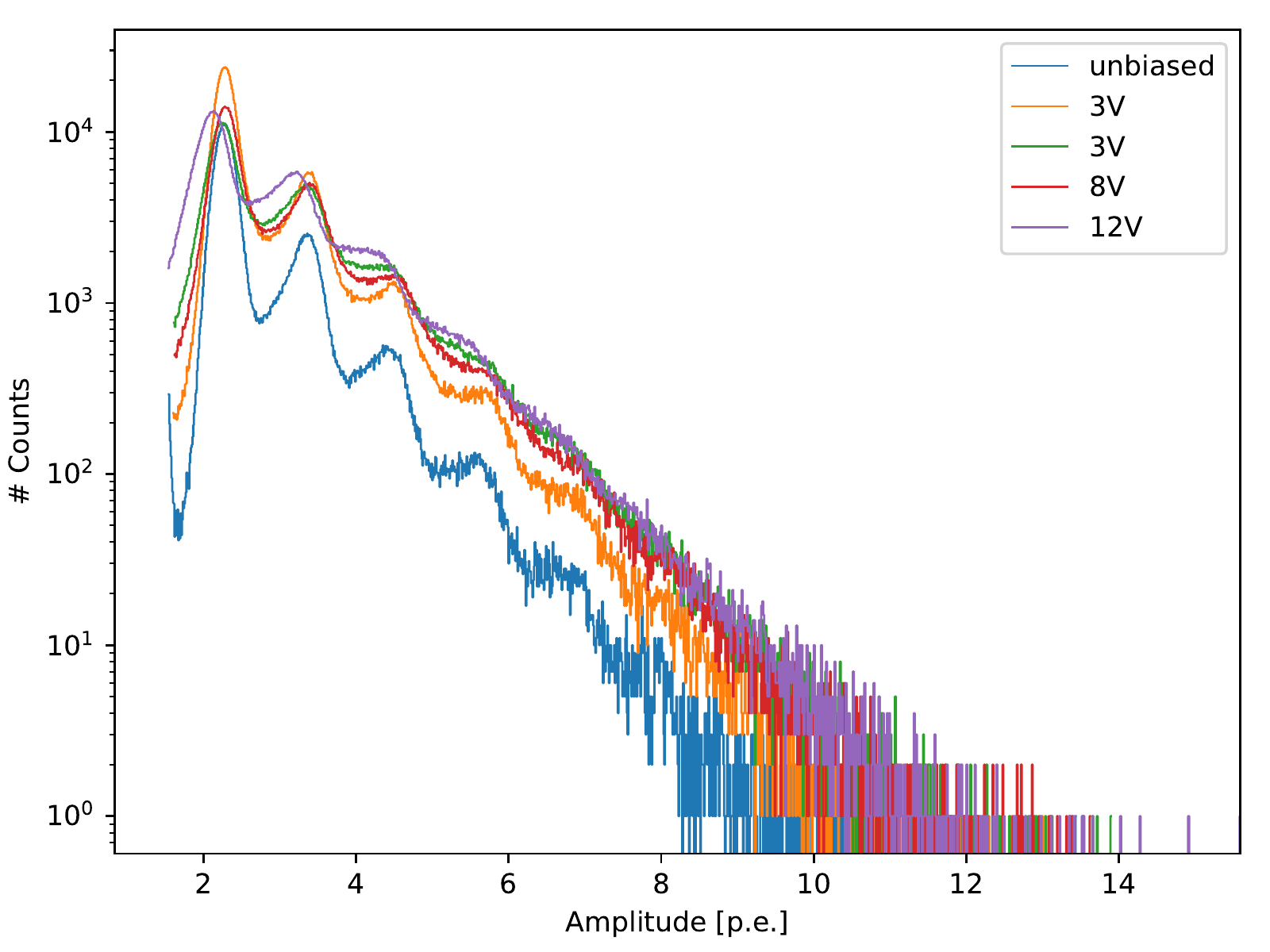}\includegraphics[width=.5\textwidth]{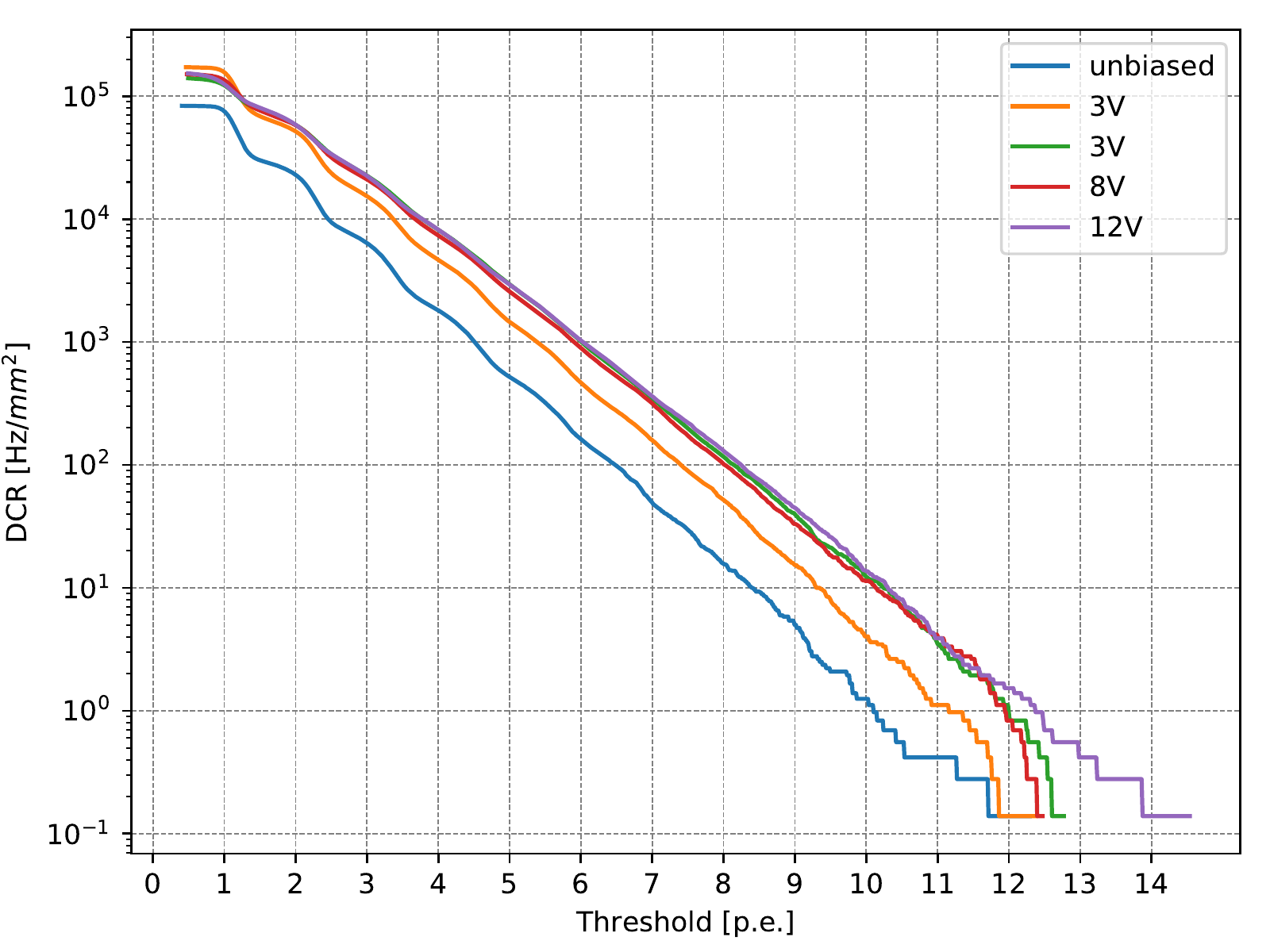}
  \par\end{centering}
  \protect\caption{Dark spectrum (left) and DCR versus threshold (right) for 50$\mu$m SiPMs with different annealing bias voltage, measured at 20$^\circ$C}
    \label{fig:SPE_DCR_HV}
\end{figure}
\par\end{center}

\newpage
\section{Implications for the POLAR-2 experiment and other SiPM-based space instruments}

We discuss in this section an application of the annealing effect characterized in the previous sections to a space-borne experiment: POLAR-2. POLAR-2 \cite{ICRC21_POLAR-2} is a Compton polarimeter dedicated to the measurement of the polarization of Gamma-Ray Bursts' (GRBs) prompt emission. It consists of an array of 6400 elongated plastic scintillators each readout by Hamamatsu SiPMs of the S13360-6075 series. POLAR-2 is planned for a minimal operation time of 2 years with a launch early 2025 to the China Space Station (CSS, orbit: 383~km average altitude and 42$^\circ$ inclination). Since the major drawback of SiPMs over PMTs (which were used in the predecessor experiment of POLAR-2, POLAR \cite{PRODUIT2018259}) is the dark noise, which is temperature related, we opted for an active cooling system based on Peltier elements and copper bars to extract the heat and operate the sensors as cold as possible (see Figure \ref{fig:cooling_POLAR-2}). Those thermo-electric cooling units will be directly placed in the back of the SiPMs, in order to cool them down as much as possible to reduce the noise and reduce the detection threshold.

\begin{center}
\begin{figure}[H]
\captionsetup{justification=centering}
\begin{centering}
\hspace*{0.5cm}\includegraphics[height=.35\textwidth]{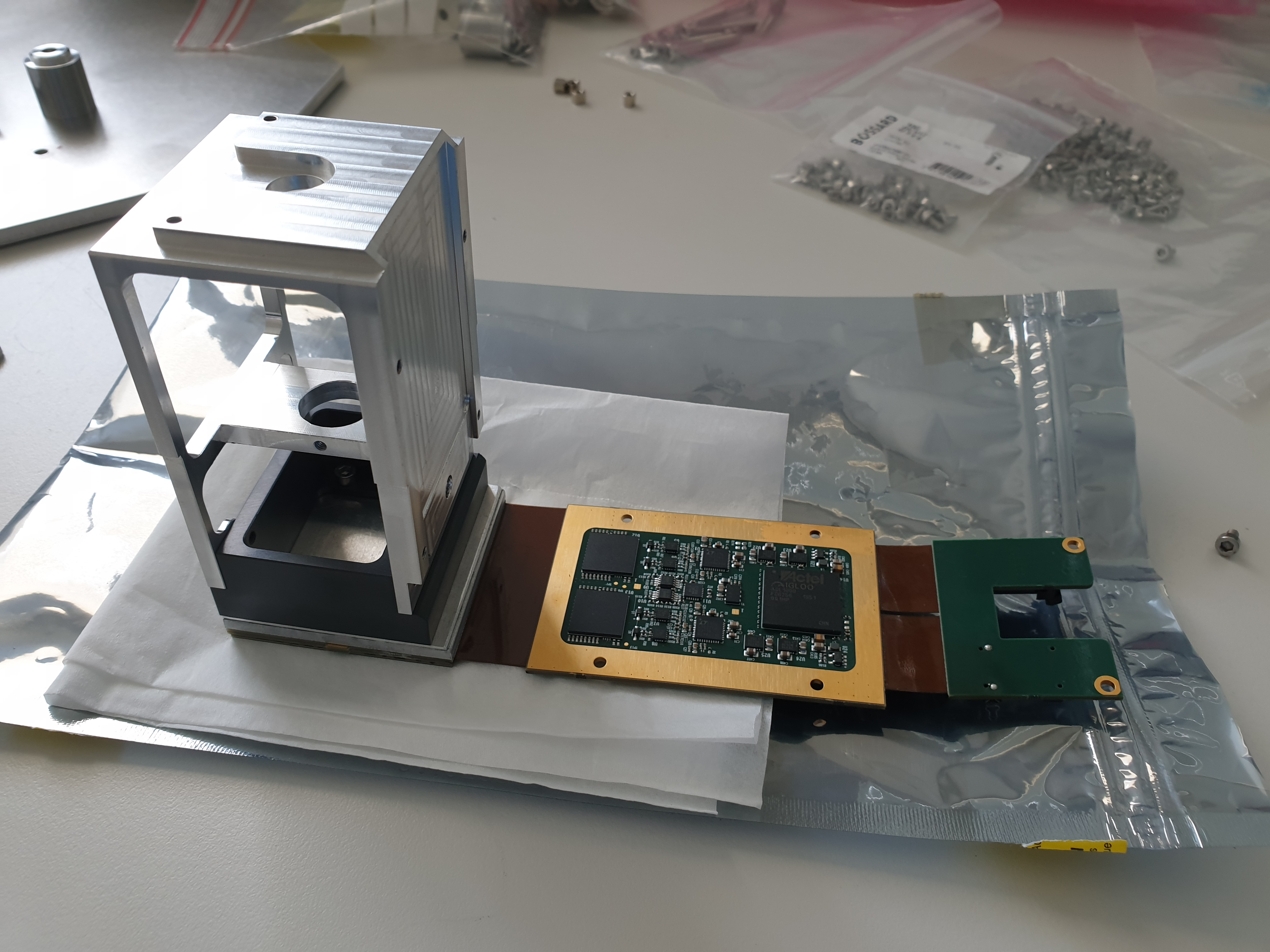}\hspace*{0.1cm}\includegraphics[height=.35\textwidth]{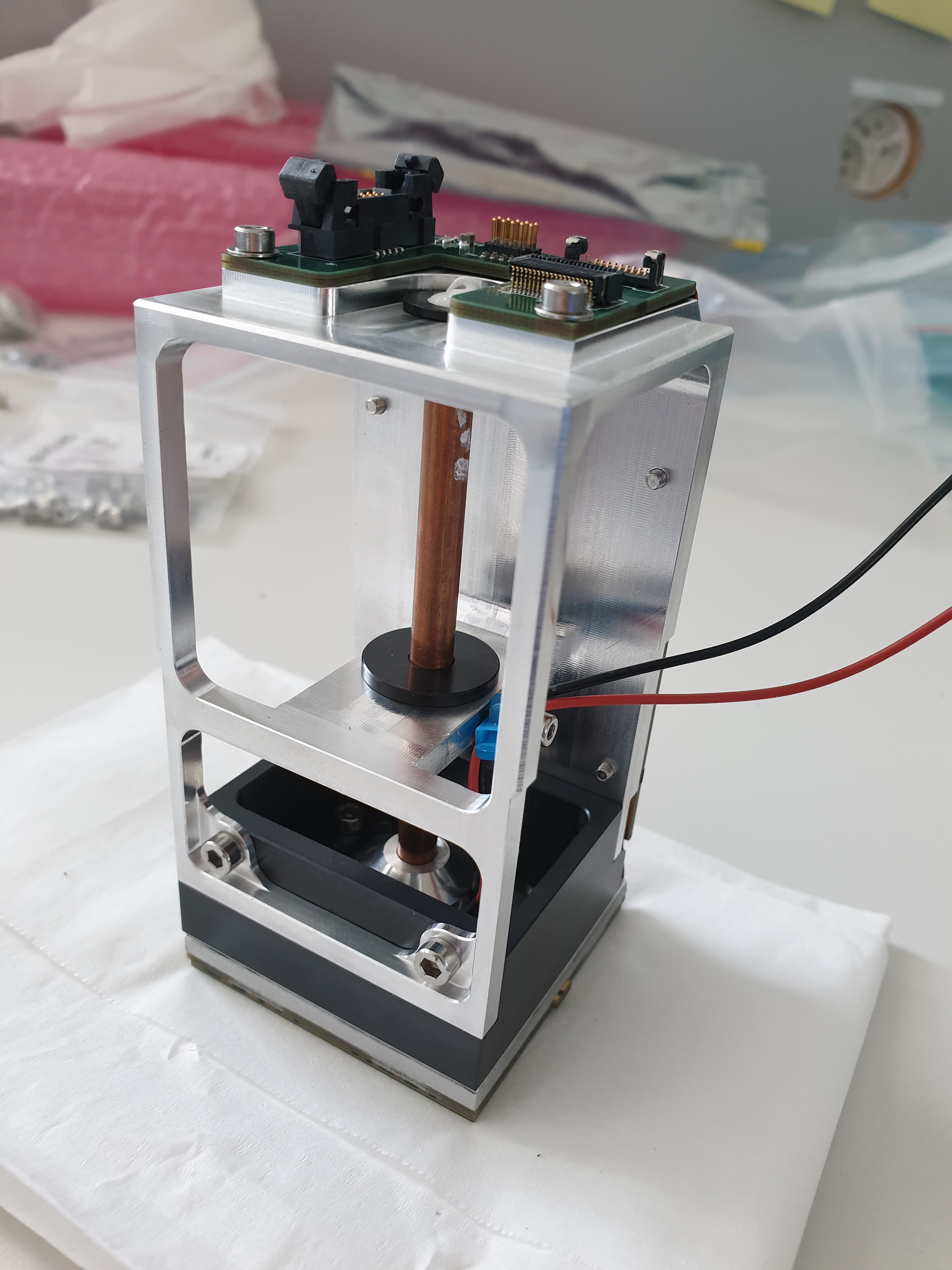}\hspace*{-0.8cm}\includegraphics[height=.35\textwidth]{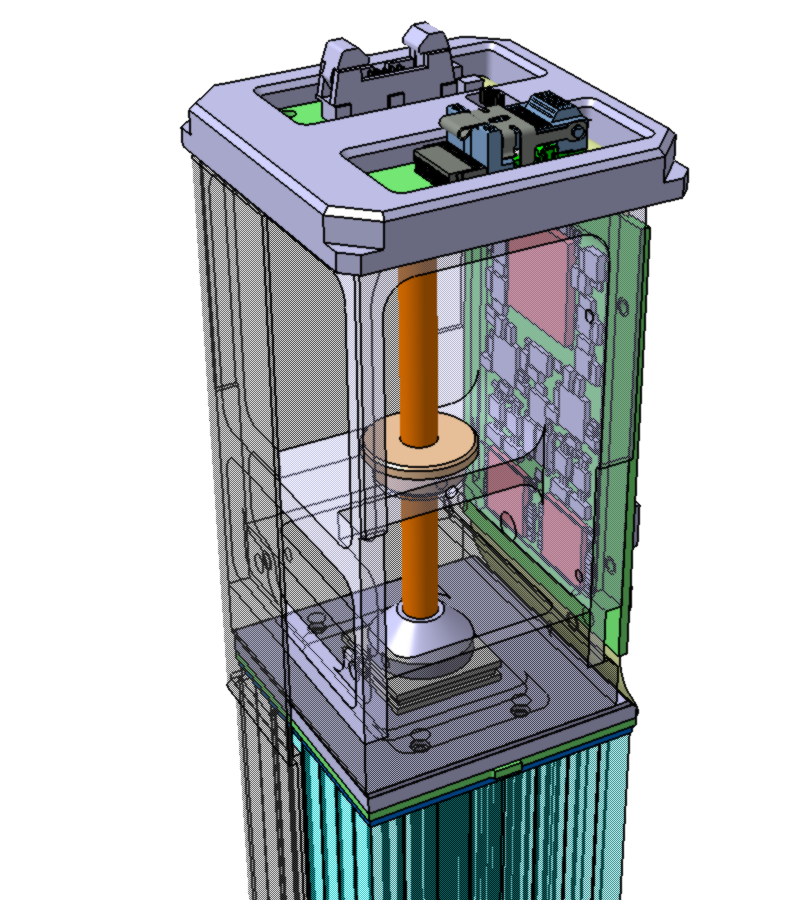}
  \par\end{centering}
  \protect\caption{Picture (left/middle) and CAD model (right) of the module level cooling system of POLAR-2}
    \label{fig:cooling_POLAR-2}
\end{figure}
\par\end{center}

As POLAR-2 will be operating for at least two years in LEO, its SiPMs will be subject to radiation damage. The yearly radiation dose at which the POLAR-2 SiPMs will be exposed to has been determined as being 0.0789~Gy/yr (at an altitude of 383~km, which corresponds to the average altitude of the CSS) using Geant4 simulations of the entire instrument, as described in \cite{SiPMpaper} where the corresponding radiation damage has been studied.\\

The POLAR-2 6400 scintillators are divided into 100 polarimeter modules of 8$\times$8 channels, each of which is readout with 4 S13361-6075PE-04 arrays from Hamamatsu. A S13361-6075PE-04 consist of 16 channels of the 6~mm S13360-6075PE SiPM type studied in this paper. The array have been irradiated with different doses to study the radiation damage of the SiPM channels \cite{SiPMpaper}. The annealing of the array is compared to the one of the single channel SiPM, which consist of the same technology, in Figure \ref{fig:annealing_array}. This figure shows the I-V curve at different time after proton irradiation with a dose of 0.267~Gy, as well as the I-V before irradiation. The arrays were stored at an approximate temperature of 25$^\circ$C (not controlled like in the case of the single channel studies presented in the rest of the paper).

\begin{center}
\begin{figure}[H]
\captionsetup{justification=centering}
\begin{centering}
\includegraphics[height=.36\textwidth]{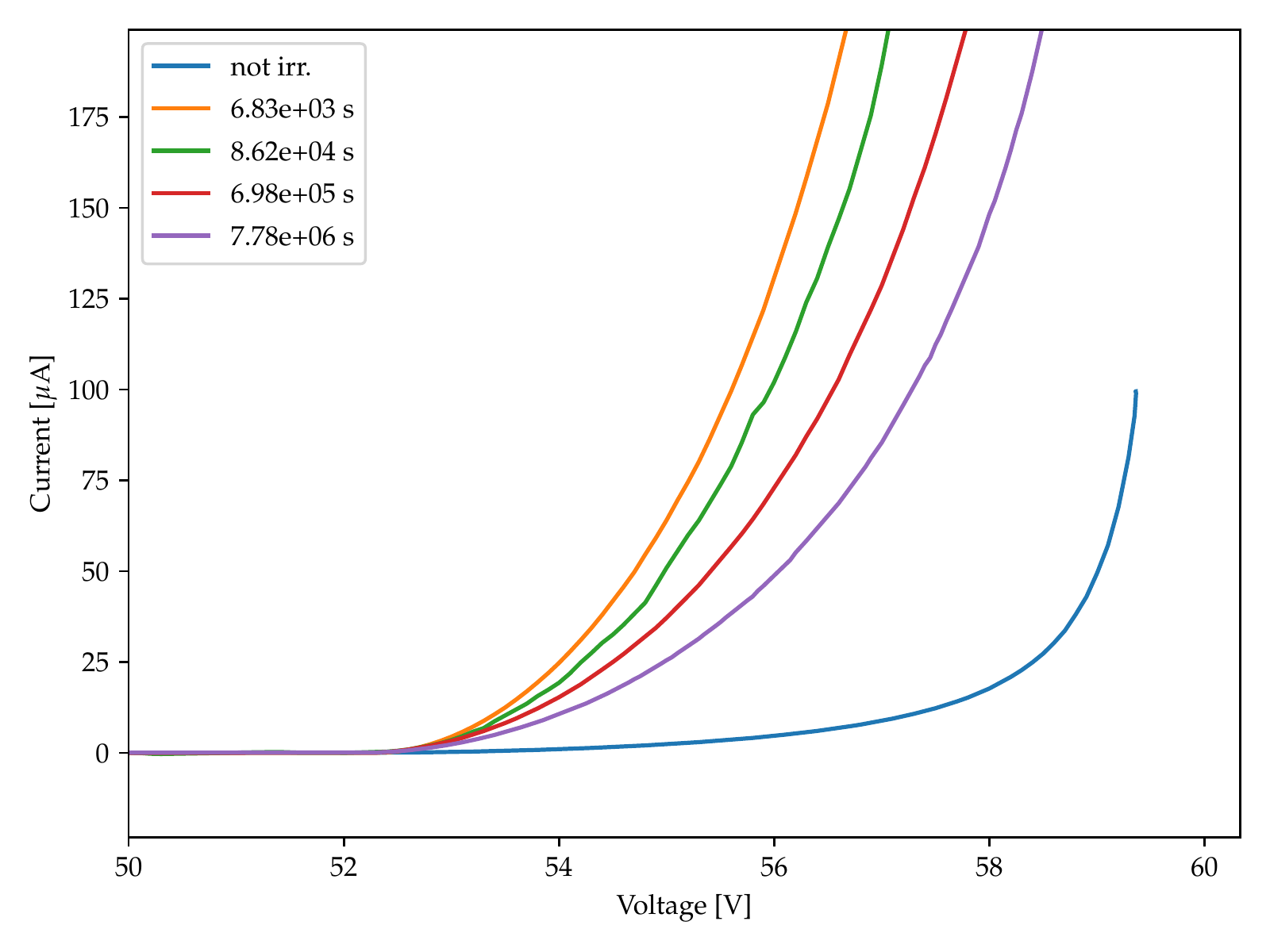}\includegraphics[height=.37\textwidth]{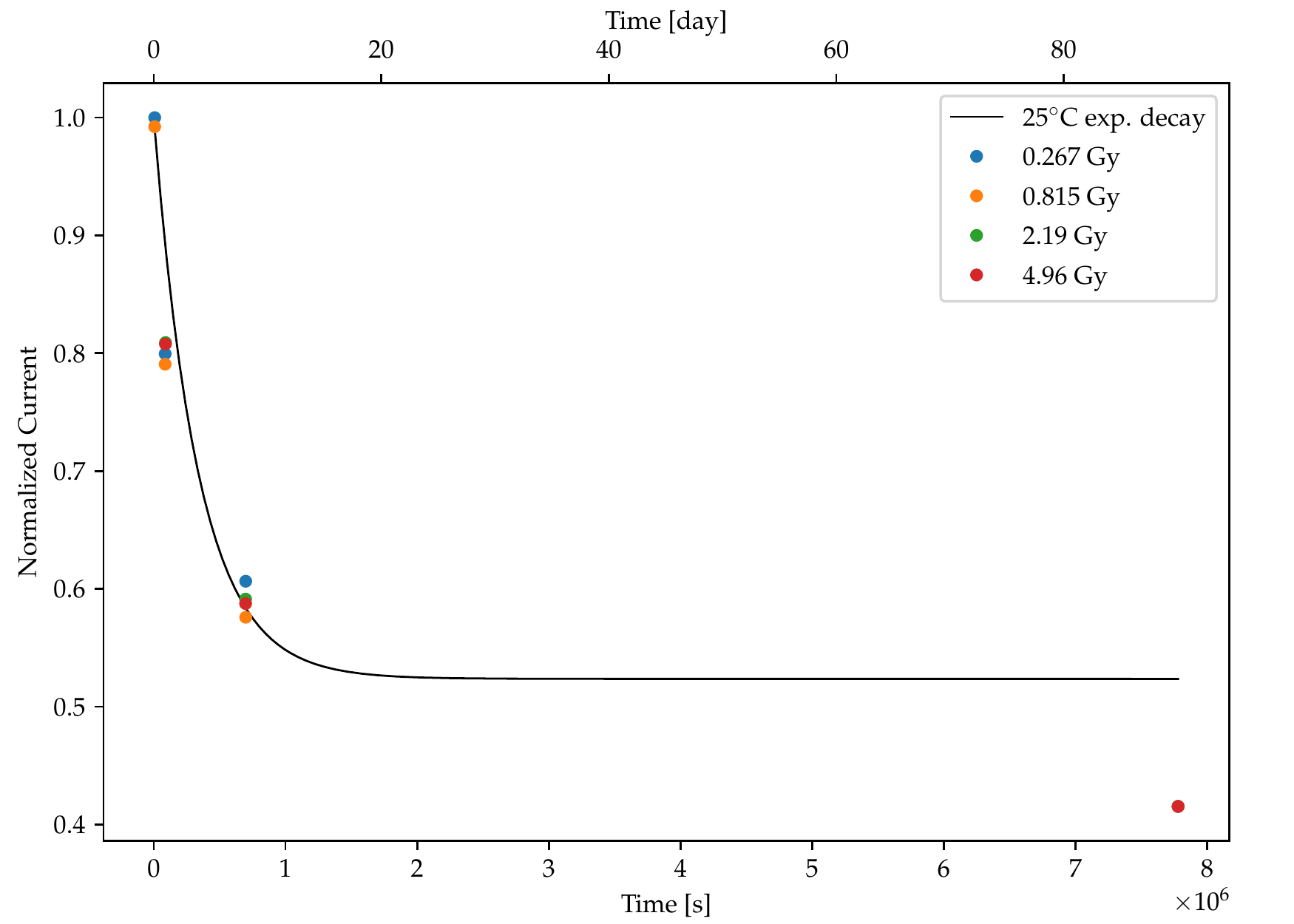}
  \par\end{centering}
  \protect\caption{I-V curve time evolution for the S13361-6075PE SiPM stored at 25$^\circ$C and irradiated with 0.267~Gy (left) and time evolution of the normalized current at 2~V overvoltage for an S13361-6075PE SiPM exposed to 0.267, 0.815, 2.19, and 4.96~Gy (right). The black curve corresponds to the exponential decay at an annealing temperature of 25$^\circ$C interpolated from the study on single channel SiPMs (see Figure \ref{fig:exp_fit_params}).}
    \label{fig:annealing_array}
\end{figure}
\par\end{center}

The 3~V overvoltage current is also plotted as a function of time in Figure \ref{fig:annealing_array} for 4 different doses. A black line gives the exponential decay expected for a 25$^\circ$C annealing temperature extracted from the fits of Figure \ref{fig:exp_fit_params}. It can be noticed that the first data points match well the curve extracted from the single channel SiPM study, while the last data point for each dose is lower than expected for the array. This can easily be explained by the fact that the arrays were not stored in a controlled temperature environment (since the purpose of those sensor were to study their radiation damage right after the irradiation campaign, not for annealing studies). The arrays were therefore annealed more than when being stored continuously at 25$^\circ$C, probably because of an increase of temperature in the storage room sometimes during the 2 months due to the weather. Based on this and the fact that the first 3 points match our expectations, we will assume that the arrays, as used in POLAR-2, have the same behavior than the single channel SiPMs regarding annealing. This is to be expected since the technology is the same and the SiPMs only differ in the mechanics which couples them to the neighbouring channels.\\

As already mentioned, POLAR-2 will be equipped with an active cooling system based on Peltier elements directly placed on the back of the SiPMs to reduce their operating temperature. The plan is therefore to invert the current polarity of the Peltier elements after some operating time in order to heat up the SiPMs and partially recover their original performances. Nevertheless,  thermal simulations are still ongoing to optimize the thermal design of the instrument. A passive cooling, where the Peltier elements are removed, is also being investigated. In this option Kapton heaters would be used to heat up the SiPMs for annealing purposes.

\begin{table}[H]
\centering
\hspace*{-0.5cm}\begin{tabular}{|l|c||c|c|c|}
\hline
\multicolumn{2}{|l||}{Operating temperature [$^\circ$C]} & -10 & 0 & 20 \\ \hline\hline
\multirow{3}{4cm}{\textbf{Scenario~1} "No~Annealing"} & 25$\mu$m & \multicolumn{3}{c|}{$15.7\pm0.8$}\\ \cline{2-5}
 & 50$\mu$m & \multicolumn{3}{c|}{$15.4\pm2.2$}\\ \cline{2-5}
 & 75$\mu$m & \multicolumn{3}{c|}{$7.2\pm0.9$}\\ \hline\hline
\multirow{3}{4cm}{\textbf{Scenario~2} "Continuous~annealing"} & 25$\mu$m & $13.15\pm0.73$ & $11.93\pm0.67$ & $9.49\pm0.59$ \\ \cline{2-5}
 & 50$\mu$m & $12.74\pm1.84$ & $11.61\pm1.67$ & $9.35\pm1.36$ \\ \cline{2-5}
 & 75$\mu$m & $5.80\pm0.72$ & $5.36\pm0.65$ & $4.51\pm0.55$ \\ \hline\hline
\multirow{3}{6.5cm}{\textbf{Scenario~3a}~(1~day) "Continuous~+~Stimulated~annealing"} & 25$\mu$m & $8.02\pm0.45$ & $7.27\pm0.41$ & $5.79\pm0.36$ \\ \cline{2-5}
 & 50$\mu$m & $7.40\pm1.07$ & $6.75\pm0.97$ & $5.43\pm0.79$ \\ \cline{2-5}
 & 75$\mu$m & $3.59\pm0.45$ & $3.31\pm0.40$ & $2.79\pm0.34$ \\ \hline
\multirow{3}{6.5cm}{\textbf{Scenario~3b}~(2~days) "Continuous~+~Stimulated~annealing"} & 25$\mu$m & $5.87\pm0.33$ & $5.32\pm0.30$ & $4.24\pm0.26$ \\ \cline{2-5}
 & 50$\mu$m & $5.48\pm0.79$ & $4.99\pm0.72$ & $4.02\pm0.58$ \\ \cline{2-5}
 & 75$\mu$m & $2.69\pm0.34$ & $2.49\pm0.30$ & $2.10\pm0.25$ \\ \hline
\multirow{3}{6.5cm}{\textbf{Scenario~3c}~(10~days) "Continuous~+~Stimulated~annealing"} & 25$\mu$m & $4.32\pm0.24$ & $3.92\pm0.22$ & $3.12\pm0.19$ \\ \cline{2-5}
 & 50$\mu$m & $4.39\pm0.63$ & $4.00\pm0.57$ & $3.22\pm0.47$ \\ \cline{2-5}
 & 75$\mu$m & $2.09\pm0.26$ & $1.93\pm0.24$ & $1.63\pm0.20$ \\ \hline
\end{tabular}
\caption{Increase of the dark current at 3~V overvoltage after 1 year of operation for different annealing scenarios. The numbers are also provided for the 25 and 50$\mu$m SiPMs for other experiments using these sensors. The different annealing scenarios are defined as follow: \textbf{Scenario 1.} Increase of dark current after 1 year of operation in LEO neglecting the annealing effect during the operation. No annealing is considered here and the radiation damage is assumed to be independent of the temperature. \textbf{Scenario 2.} Increase of dark current after 1 year of operation in LEO considering the continuous annealing happening all along the operation, given the operating temperature. \textbf{Scenario 3.} Same as \textit{Scenario 2} but followed by a 1 (\textbf{a}), 2 (\textbf{b}) or 10 (\textbf{c}) days stimulated annealing at 50$^\circ$C after 1 year of operation in order to recover part of the original performances. The annealing is stimulated via an active heating of the sensors.}
\label{tab:annealing_strategies}
\end{table}

An increase of the dark current at 3~V overvoltage by a factor 7.2$\pm$0.9 is expected every year for the orbit and shielding conditions of POLAR-2 if operating at low enough temperatures where the annealing effect is not significant. This yearly dark current increase, also provided in Table \ref{tab:annealing_strategies} for the three types of sensors studied in this paper, is an upper value for an instrument exposed to a dose rate of 0.0789~Gy/yr. The current increase of SiPMs can be considered as proportional to the dose rate for LEO experiments receiving relatively low levels of radiation. The factor provided for \textit{Scenario 1} can therefore be scaled up or down for other instruments with different amounts of shielding. Depending on the operating temperature of the instrument, some continuous annealing effect could dampen this current increase. We therefore provide the factor corrected for the annealing effect for the three types of SiPMs and for three different operating temperatures, namely -10, 0 and $20^\circ$C. This is shown in Table \ref{tab:annealing_strategies} as \textit{Scenario 2}. The effect of continuous annealing at operating temperature on the linearly increasing current is described by the following expression:

\begin{equation}
    F_I(t)=1+(K-1)\cdot t[yr]\cdot\qty[A(T)\exp(B(T)\cdot \tilde{t}[s])+C(T)] 
\end{equation}
where K is the yearly increase in current without annealing (Scenario 1) and A(T), B(T) and C(T) are the exponential amplitude, slope and offset obtained from the exponential fits of Figure \ref{fig:I_vs_time} and reported as a function of temperature in Figure \ref{fig:exp_fit_params}. $\tilde{t}$ is the time corrected for the duration of the SiPMs transportation from Krakow to Geneva after the irradiation session, and is calculated as $\tilde{t}=t+t_0=t+\frac{1}{B(T)}\ln(\frac{1-C(T)}{A(T)})$ where $t_0$ is the time of the first post-irradiation measurement. The values for $t_0$ are 76616~s, 72935~s and 70432~s, respectively for the 25, 50, and 75~$\mu$m SiPMs. This time correction allows to use the amplitude and offset obtained from the fits in Figure \ref{fig:I_vs_time} without having to renormalize them.\\

Finally, \textit{Scenario 3} of Table \ref{tab:annealing_strategies} provides the increase in current for the case of \textit{Scenario 2} followed by an active heating of the instrument performed after 1 year of operations at 50$^\circ$C in order to partially recover the dark current. The numbers are provided for a stimulated annealing of 1, 2 and 10 days. Assuming an operating point of 0$^\circ$C for POLAR-2, an annual increase of dark current by a factor $5.36\pm0.65$ can be expected. The effect of radiation on the polarimeter performances can be mitigated by heating up the sensors to a temperature of 50$^\circ$C for 1 (or 2) days after a year of operation. After this procedure, the 1 year increase of current would be reduced to $3.31\pm0.40$ (or $2.49\pm0.30$). Table \ref{tab:annealing_strategies} can be used to estimate dark current increase of S13360-6025/50/5 SiPMs related to radiation damage. It can also be used for estimating the required thermal conditions for any space borne experiment using SiPMs as photosensors, and planning an annealing strategy to extend the life time of the mission.

\newpage
\section{Conclusions}

Several S13360-60(25/50/75)PE SiPMs from Hamamatsu have been irradiated with 58~MeV protons in order to study the recovery of their performances due to annealing effects in the silicon lattice at different temperatures and bias voltages. The SiPMs have been exposed to a dose of 0.134~Gy, which was computed from Geant4 simulations as being a 2 years equivalent dose for the POLAR-2 experiment. POLAR-2 is a Compton polarimeter that will be operating on the China Space Station for an initial life time of 2 years starting 2025. This work can also be applied to other instruments using SiPMs and operating in low Earth orbit (LEO).

Once irradiated, the SiPMs have been passively stored at different temperatures ranging from -22.8$\pm$1.8 to 48.7$\pm$3.3$^\circ$C. I-V characteristics were measured regularly all along the 70~days of storage. The sensors stored at the coldest conditions saw no significant evolution of there dark current with time after irradiation, while the dark current of the other sensors improved with time thanks to thermal annealing of the silicon lattice. The exponentially decaying dark current has been parameterized with annealing temperature, showing that the effect depends significantly on the thermal conditions, with SiPMs stored at high temperatures recovering faster and closer to the pre-irradiation performances. The time evolving dark current has also been studied through the current related damage rate $\alpha$, which allows to better appreciate the behavior of the decay after long times. SiPMs were also stored at higher temperatures, namely 75 and 100$^\circ$C. Studying the annealing effect at higher temperatures showed a saturation of the annealing effect with a significant loss of linearity of the dark current above 60$^\circ$C. The breakdown voltage did not show any significant change, either due to irradiation or recovery.

In order to get a more complete picture of the thermally related recovery of the sensors, the dark spectra for different annealing temperatures have been studied. Dark spectra have been measured in a climatic chamber to ensure stable thermal conditions at 0 and 20$^\circ$C. An improvement of photon resolution was observed, especially at high annealing temperatures, both in the dark spectrum and by looking at the dark count rate as a function of the threshold. Furthermore, the width of the first photoelectron peak in the dark spectrum is decreasing with annealing temperatures. Therefore, not only the dark current but the entire SiPM performances are partially recovering through the annealing process.

Several sensors have also been stored with different bias condition, to ensure that the analysis performed with passive SiPMs is also valid when operating them in an experiment. Although a spread between samples was observed, the difference in behavior between passive and active sensors for different bias voltages was not significant, both for current annealing and photoelectron resolution after annealing. Ultimately, since applying a bias voltage to the SiPMs had no significant effect on the temporal behavior of the annealing, the passive study performed earlier is still valid for any real-life experiment, where the sensors are biased.

Finally, a concrete application to this analysis have been discussed with the POLAR-2 experiment. This instrument will consist of a matrix of 6400 elongated plastic scintillators, each readout by an S13360-6075PE channel. As any other instrument in low-Earth orbit (LEO), for which this annealing study also apply, POLAR-2 will be subject to radiation damage due to the harsh in-orbit environment, especially when crossing the South Atlantic Anomaly. In order to reduce the dark noise of the SiPMs as much as possible, a cooling system using Peltier elements has been designed. The sensors will therefore be operated as cold as possible to improve their sensitivity to low light levels, and therefore the sensitivity of the instrument to low energy events. Assuming an operating temperature of 0$^\circ$C, a yearly increase of dark current by a factor $5.36\pm0.65$ due to radiation damage is expected. The polarity of the Peltier elements can then be inverted for a few days whenever needed to anneal the SiPMs. Heating up the sensors for 1 or 2 days after 1 year of operation for POLAR-2 would suffice to reduce the current increase by a factor 2. SiPM based instruments having simpler thermal design based on passive cooling might consider the use of Kapton heaters to recover part of the original performances of the sensors after some operating time. This solution is also being investigated for the POLAR-2 mission.

\newpage
\section*{Acknowledgements}

We thank Gabriel Pelleriti and Javier Mesa for their help with the SiPMs storage PCBs and Fiona Hubert Martrou for her help with the development of the storage units. We gratefully acknowledge the Swiss Space Office of the State Secretariat for Education, Research and Innovation (ESA PRODEX Programme) which supported the development and production of the POLAR-2 detector. M.K. and N.D.A. acknowledge the support of the Swiss National Science Foundation. National Centre for Nuclear Research acknowledges support from Polish National Science Center under the grant UMO-2018/30/M/ST9/00757.

\normalem
\printbibliography

\appendix
\newpage
\section{Annealing related time evolution of the I-V characteristics for all the storage temperatures}
\label{sec:appendix_IV}

In this section are provided all the I-V curves measured for the 25, 50, and 75$\mu m$ SiPMs at each storage temperatures.

\begin{center}
\begin{figure}[H]
\captionsetup{justification=centering}
\begin{centering}
\includegraphics[width=.5\textwidth]{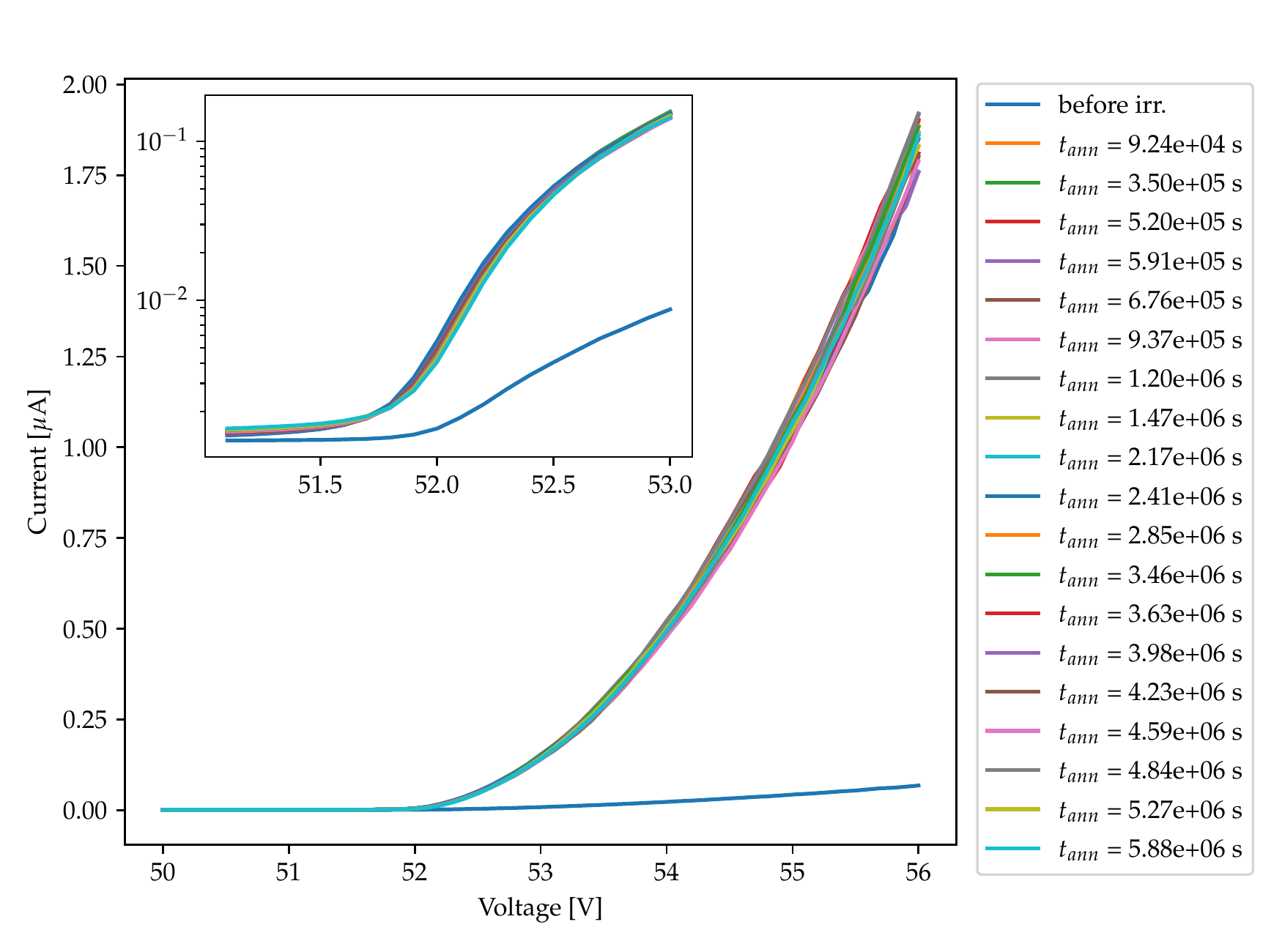}\includegraphics[width=.5\textwidth]{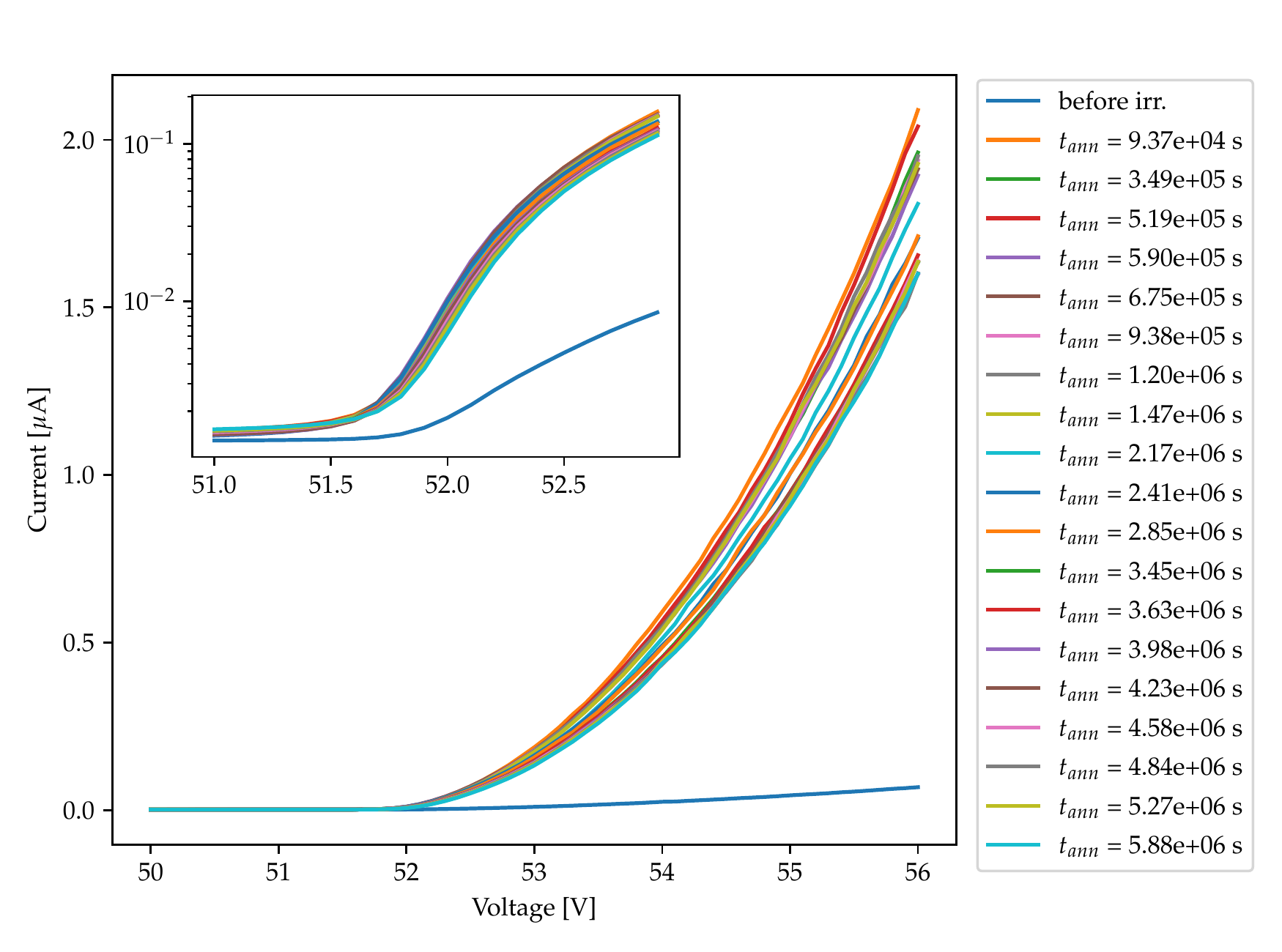}\\
\includegraphics[width=.5\textwidth]{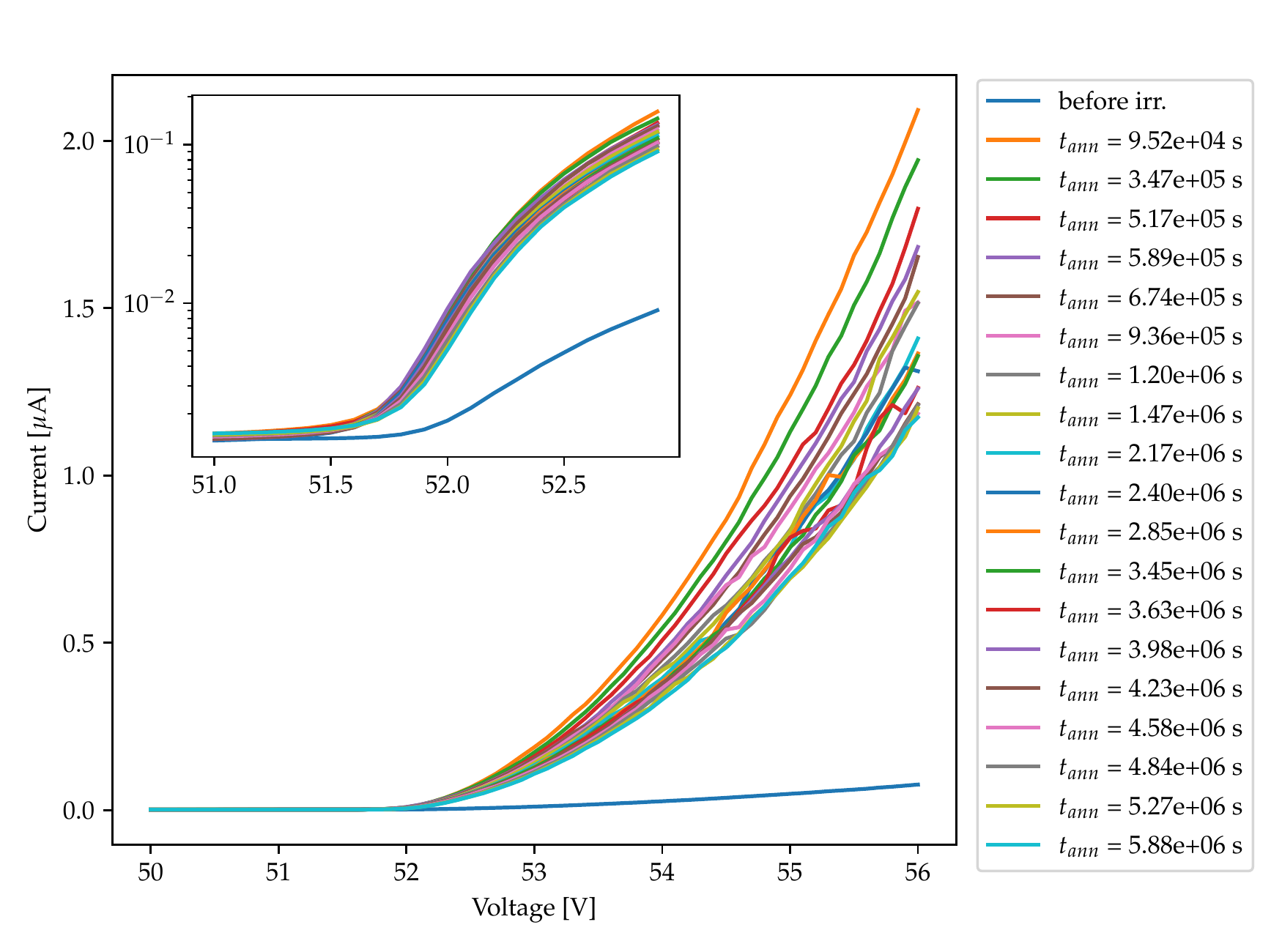}\includegraphics[width=.5\textwidth]{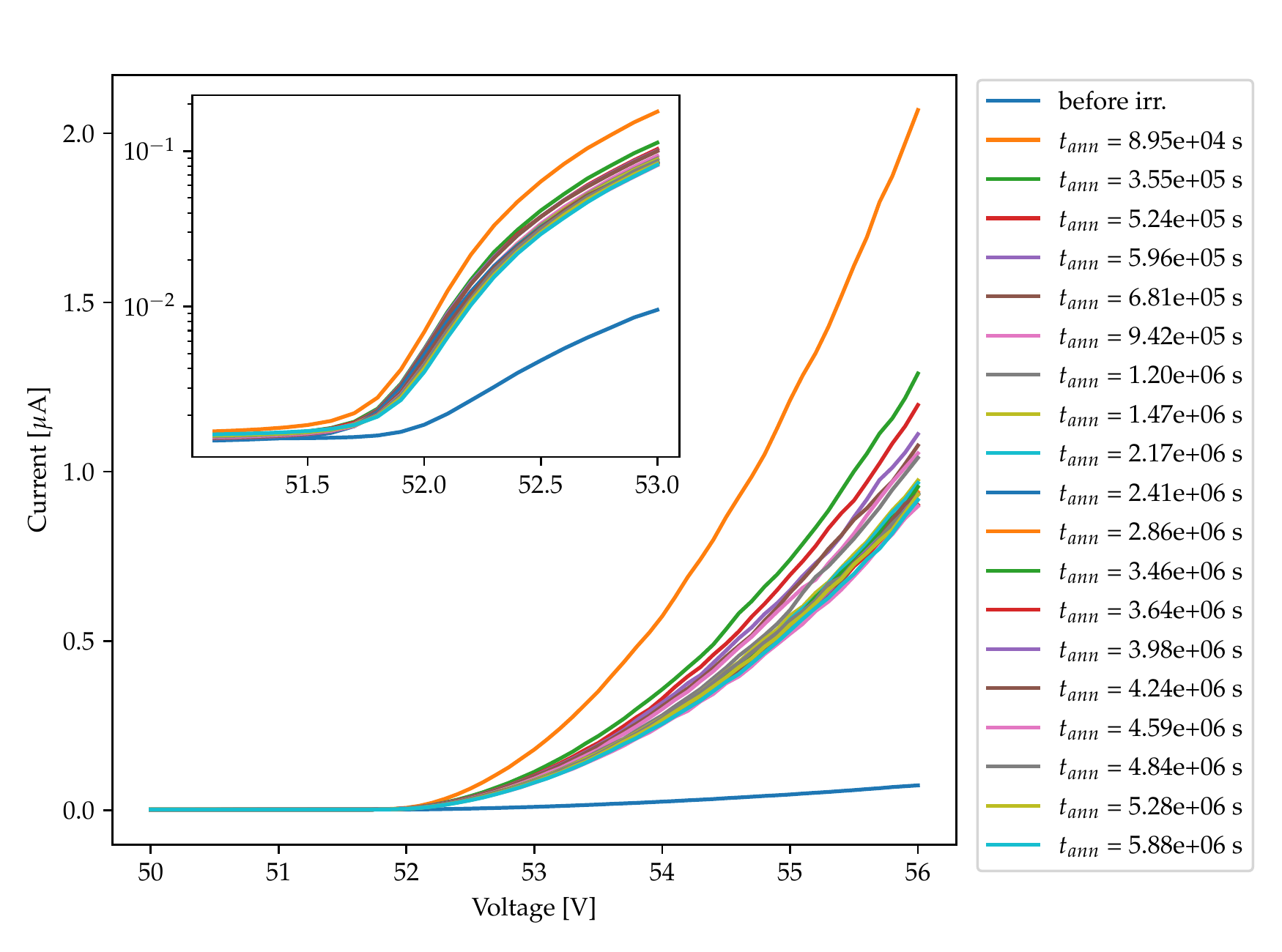}\\
\includegraphics[width=.5\textwidth]{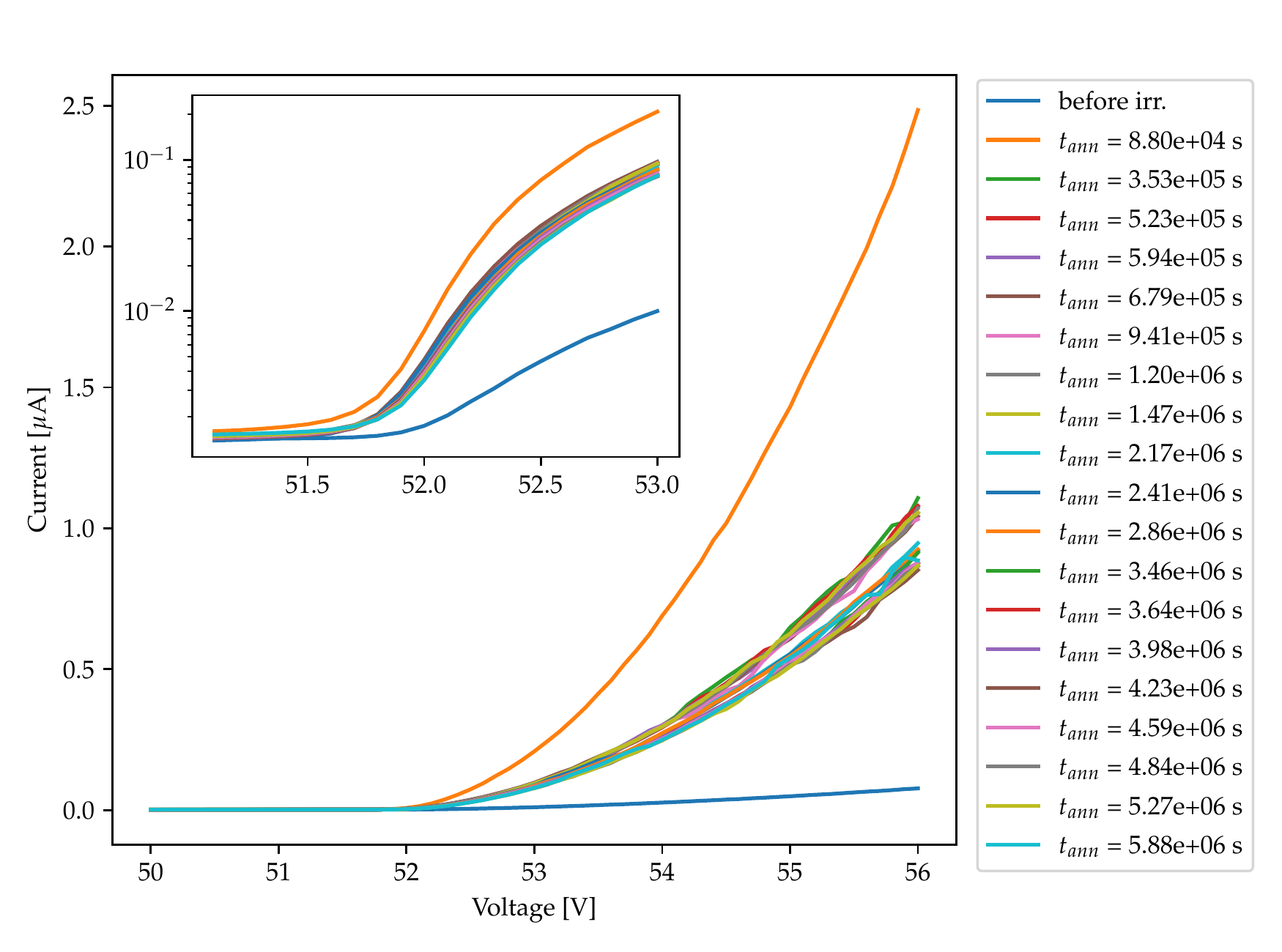}\includegraphics[width=.5\textwidth]{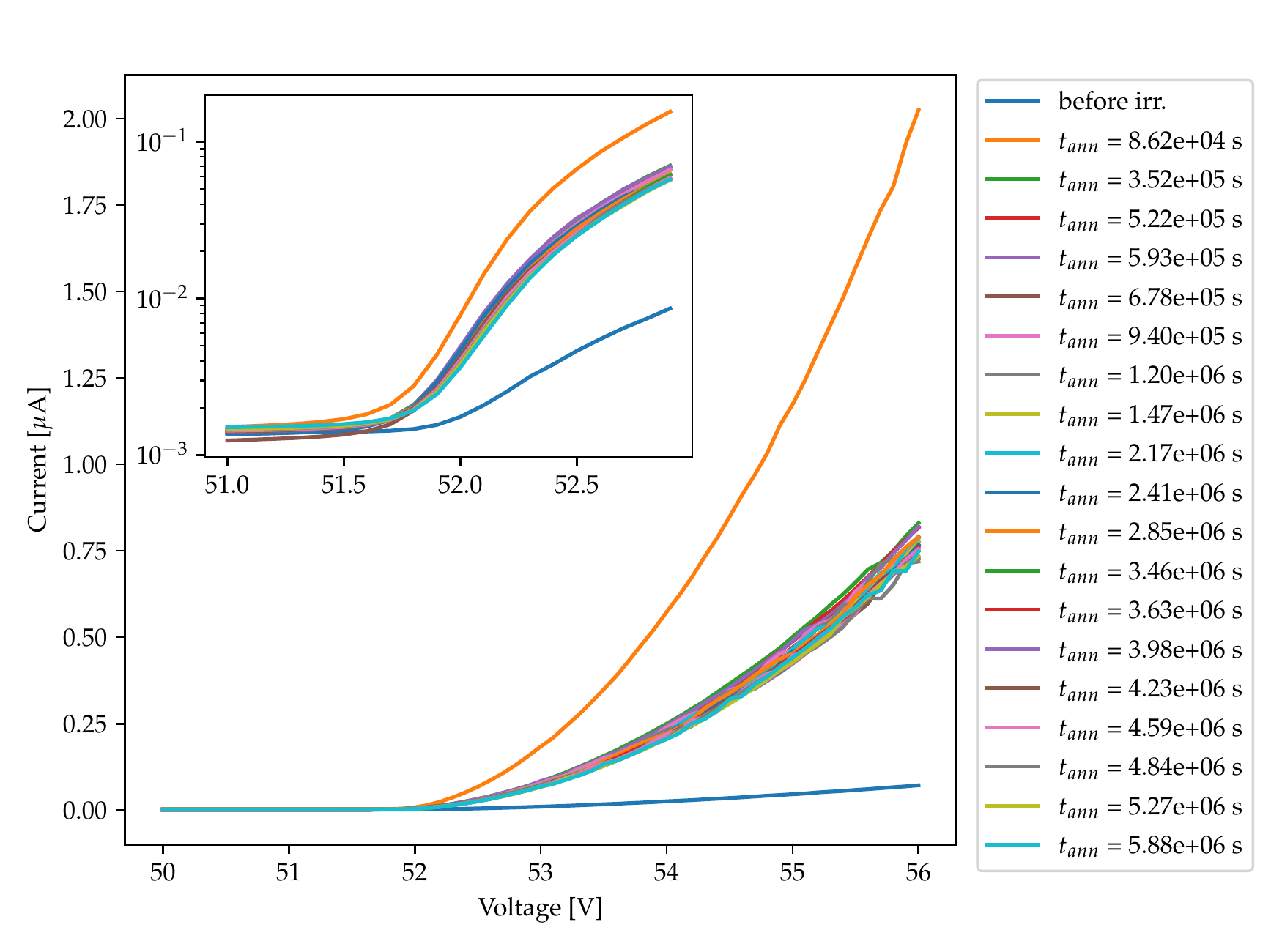}
  \par\end{centering}
  \protect\caption{Post-irradiation time evolution of the I-V characteristics of 25$\mu m$ at $-22.8\pm1.8^\circ$C, $6.3\pm0.9^\circ$C, $20.5\pm0.6^\circ$C, $29.7\pm0.6^\circ$C, $38.7\pm1.6^\circ$C and $48.7\pm3.3^\circ$C (from left to right, top to bottom)}
    \label{fig:all_IVs_25}
\end{figure}
\par\end{center}

\begin{center}
\begin{figure}[H]
\captionsetup{justification=centering}
\begin{centering}
\includegraphics[width=.5\textwidth]{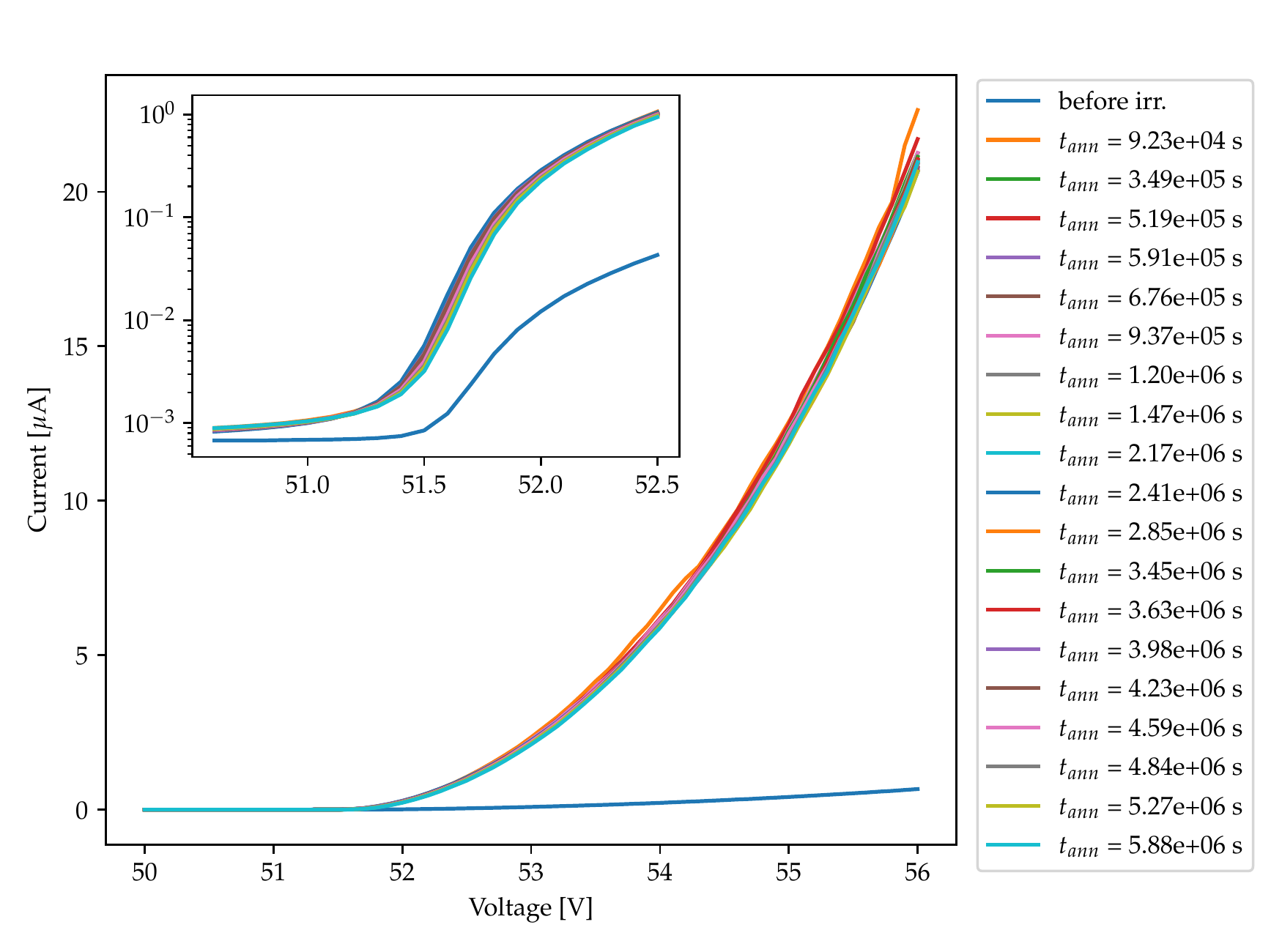}\includegraphics[width=.5\textwidth]{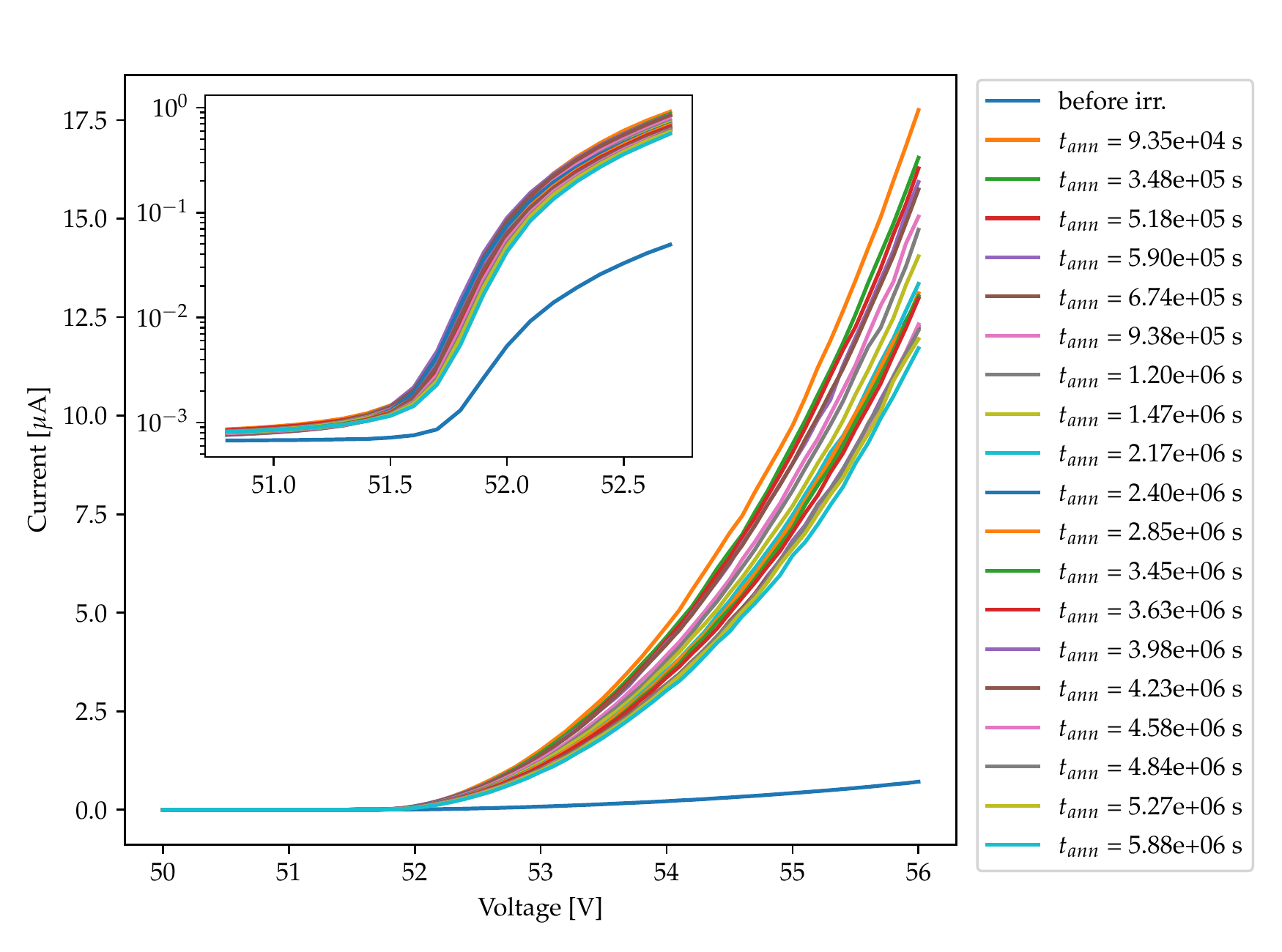}\\
\includegraphics[width=.5\textwidth]{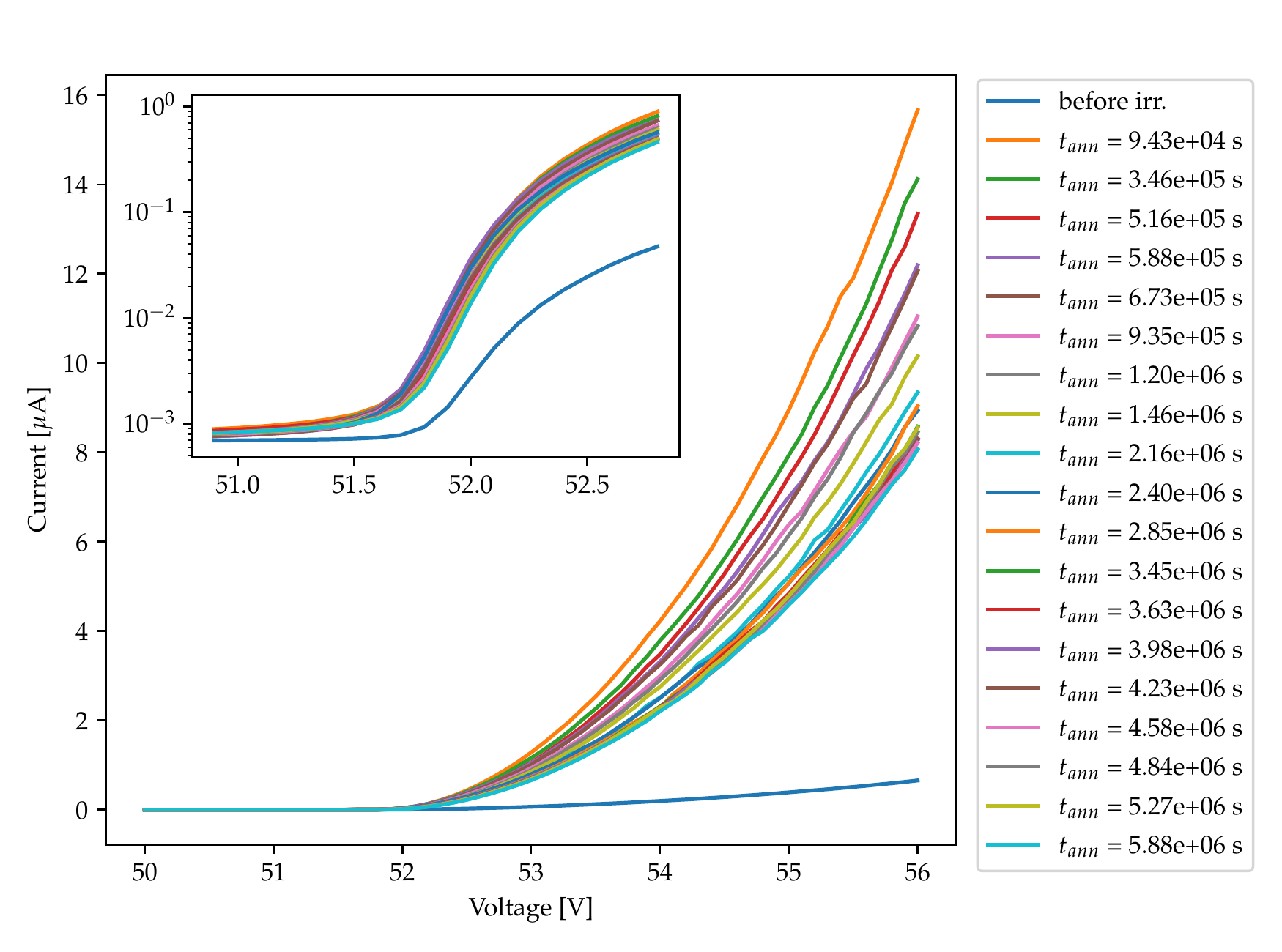}\\
\includegraphics[width=.5\textwidth]{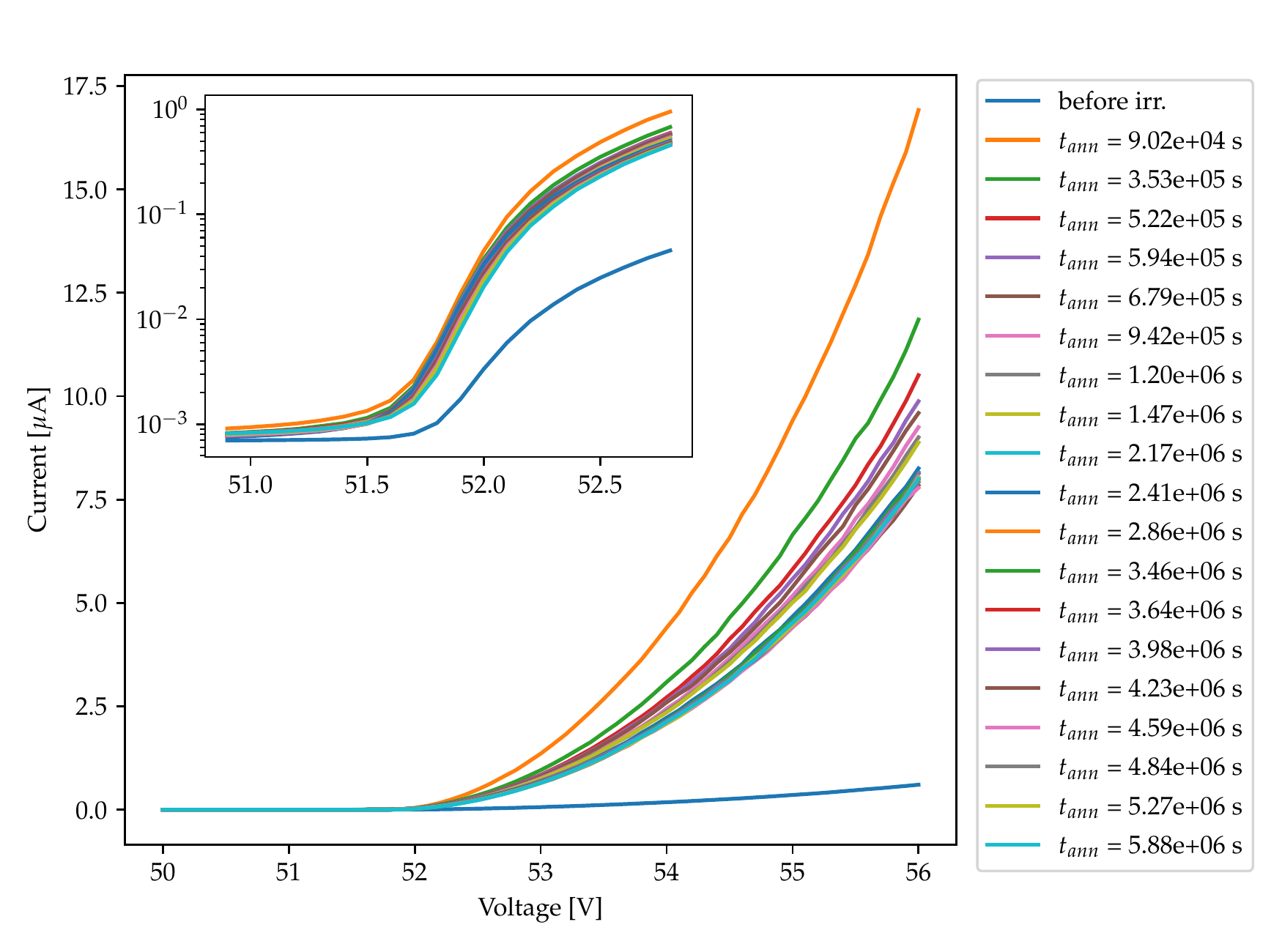}\includegraphics[width=.5\textwidth]{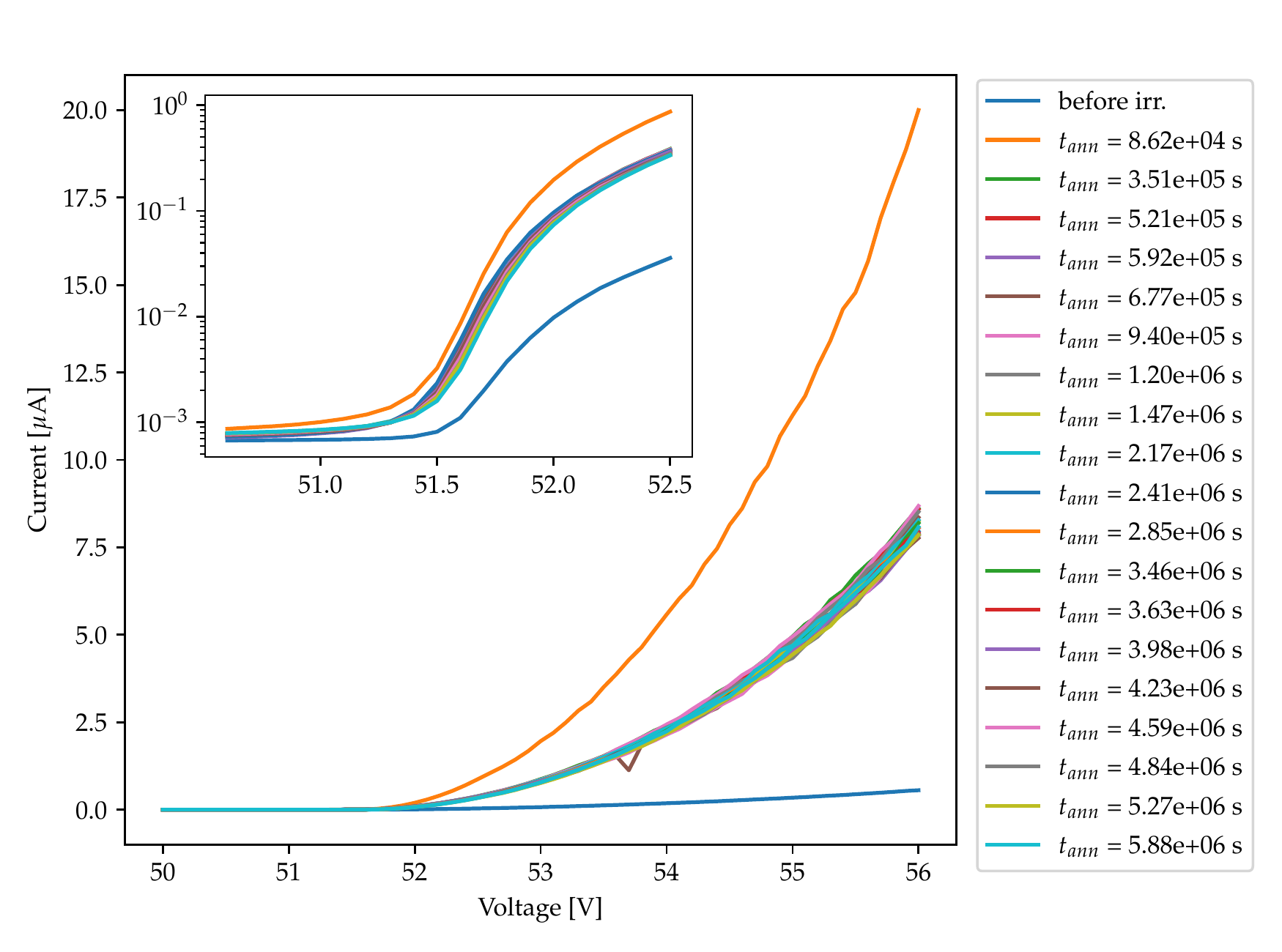}
  \par\end{centering}
  \protect\caption{Post-irradiation time evolution of the I-V characteristics of 50$\mu m$ at $-22.8\pm1.8^\circ$C, $6.3\pm0.9^\circ$C, $20.5\pm0.6^\circ$C, $29.7\pm0.6^\circ$C and $48.7\pm3.3^\circ$C (from left to right, top to bottom)}
    \label{fig:all_IVs_50}
\end{figure}
\par\end{center}

\begin{center}
\begin{figure}[H]
\captionsetup{justification=centering}
\begin{centering}
\includegraphics[width=.5\textwidth]{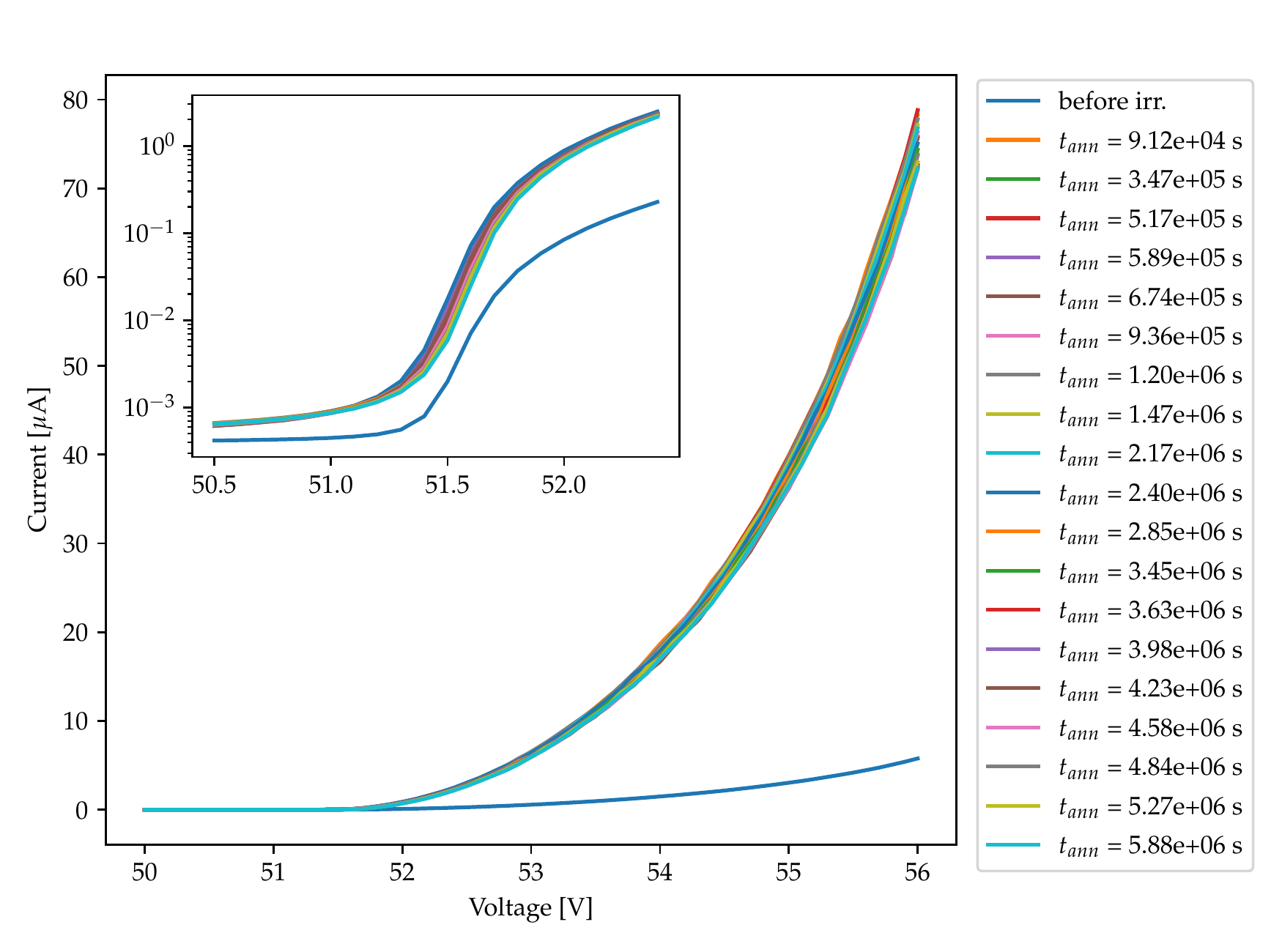}\includegraphics[width=.5\textwidth]{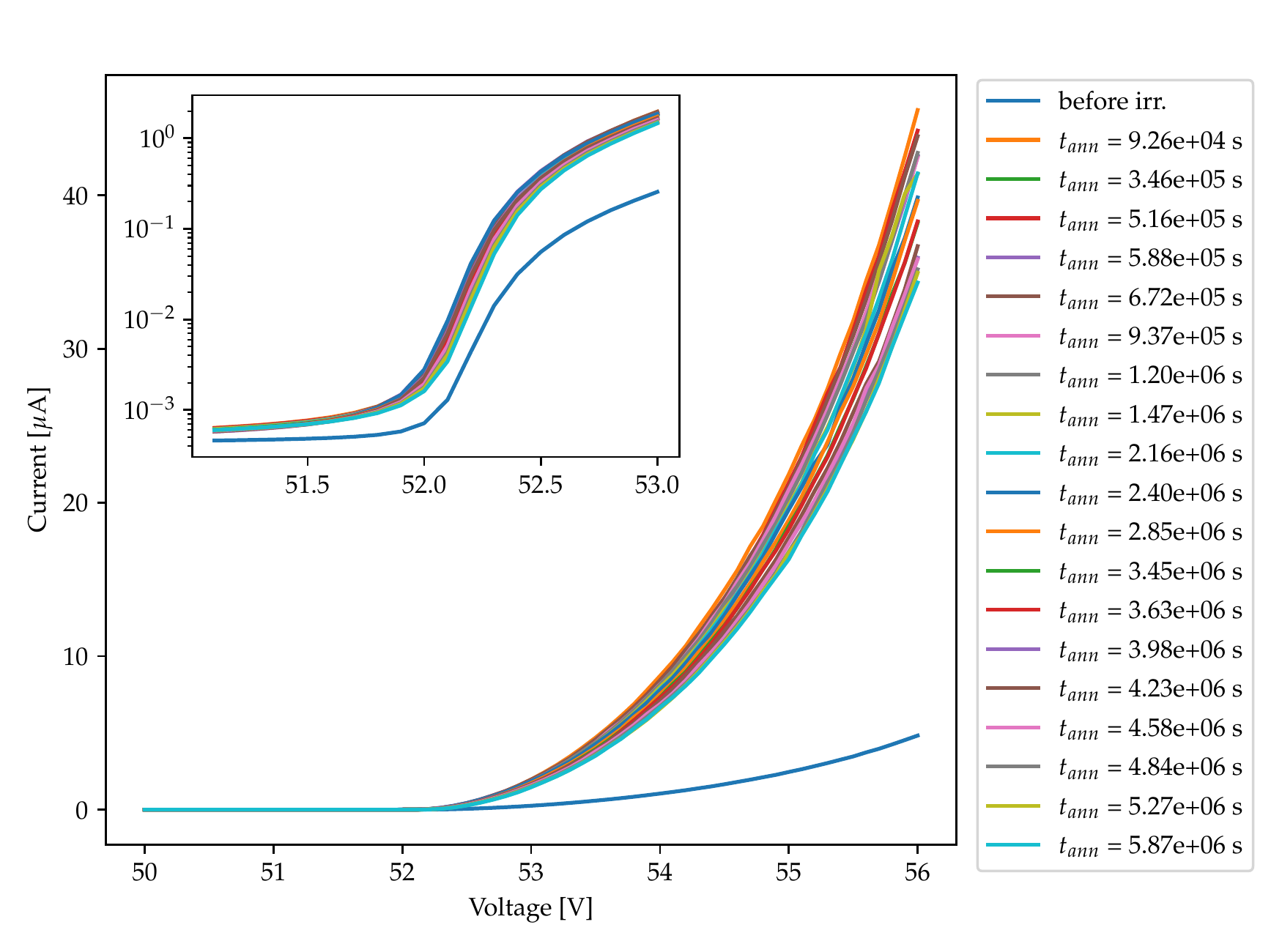}\\
\includegraphics[width=.5\textwidth]{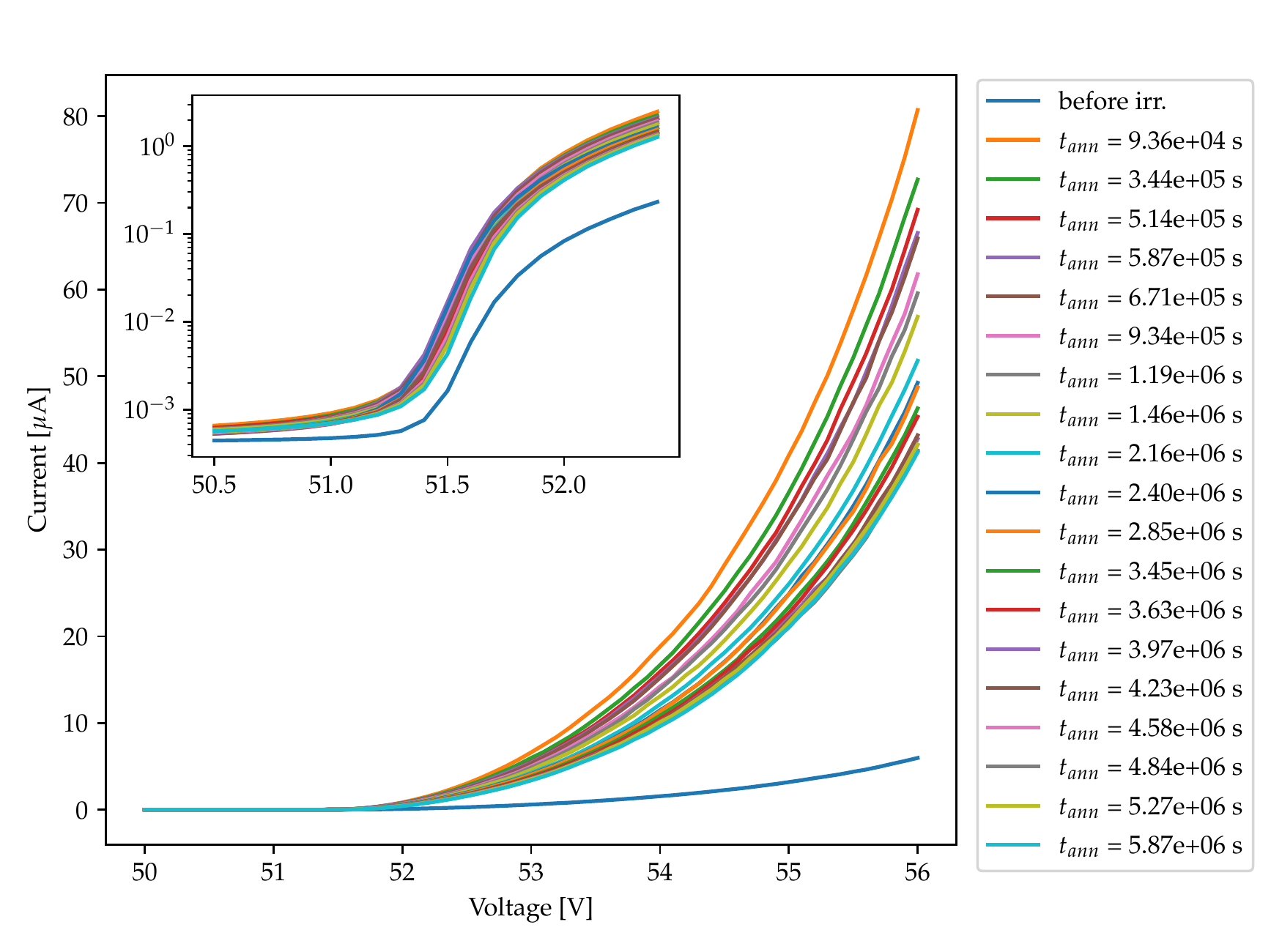}\includegraphics[width=.5\textwidth]{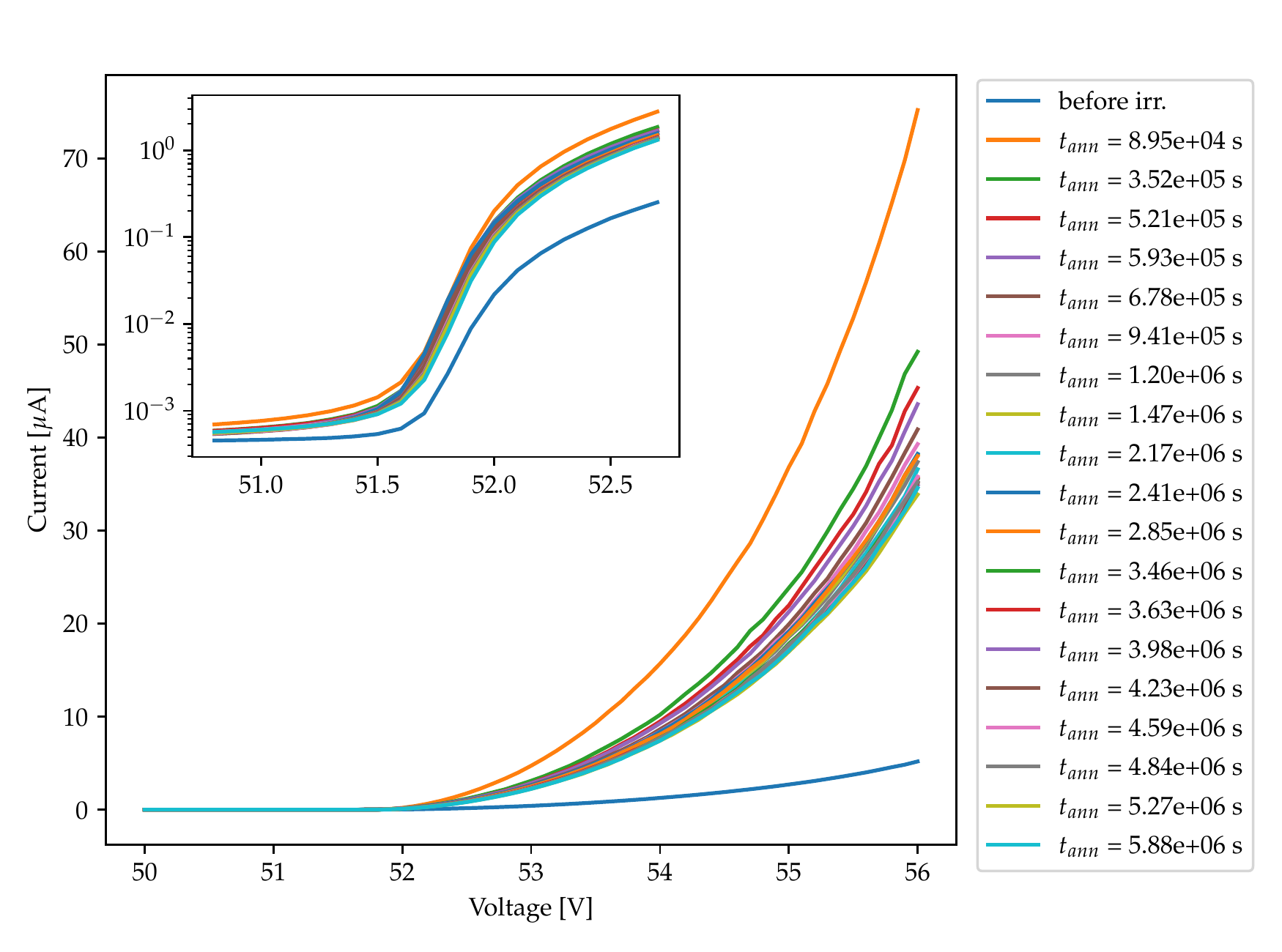}\\
\includegraphics[width=.5\textwidth]{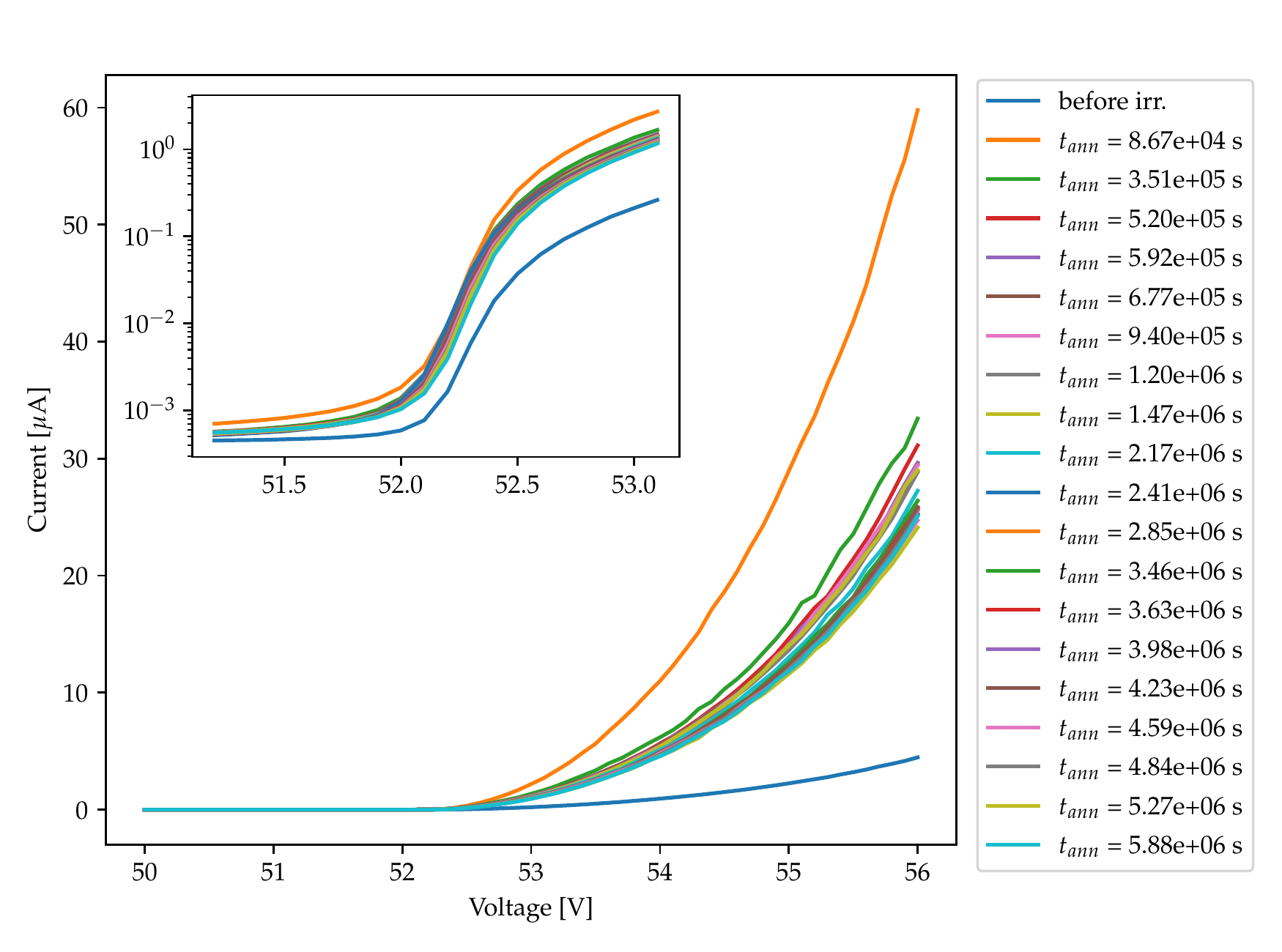}\includegraphics[width=.5\textwidth]{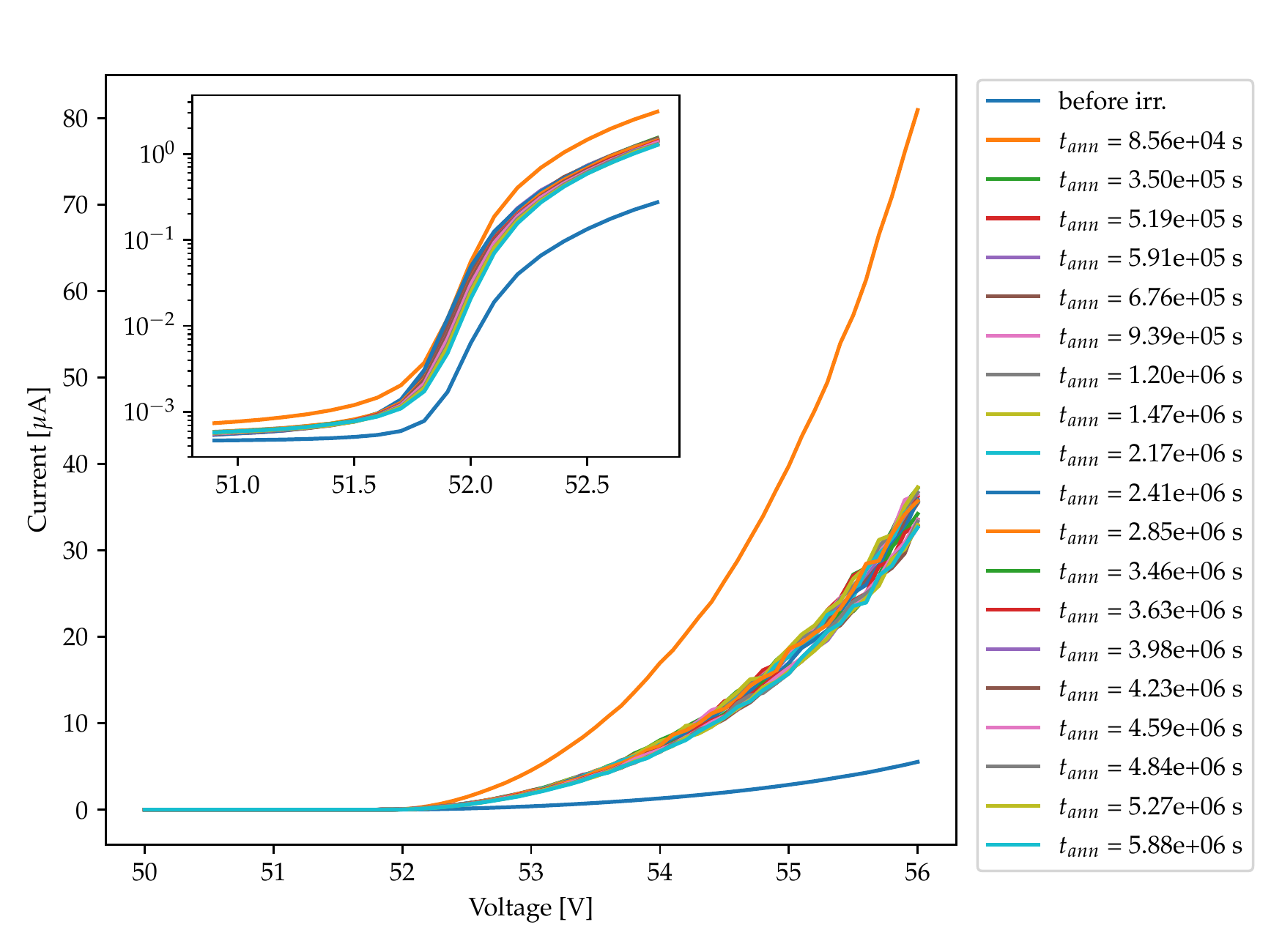}
  \par\end{centering}
  \protect\caption{Post-irradiation time evolution of the I-V characteristics of 75$\mu m$ at $-22.8\pm1.8^\circ$C, $6.3\pm0.9^\circ$C, $20.5\pm0.6^\circ$C, $29.7\pm0.6^\circ$C, $38.7\pm1.6^\circ$C and $48.7\pm3.3^\circ$C (from left to right, top to bottom)}
    \label{fig:all_IVs_75}
\end{figure}
\par\end{center}

\end{document}

%% file: authors.tex
\author[a]{N.~De~Angelis\footnote{Corresponding author. \newline \href{mailto:nicolas.deangelis@unige.ch}{nicolas.deangelis@unige.ch}}}
\author[a]{M.~Kole}  
\author[a]{F.~Cadoux}  
\author[a]{J.~Hulsman}  
\author[b]{T.~Kowalski}  
\author[b]{S.~Kusyk}  
\author[c]{S.~Mianowski}  
\author[c]{D.~Rybka}  
\author[a]{J.~Stauffer} 
\author[b]{J.~Swakon}  
\author[b]{D.~Wrobel}  
\author[a]{X.~Wu} 
\affil[a]{DPNC, University of Geneva, 24 Quai Ernest-Ansermet, CH-1205 Geneva, Switzerland}
\affil[b]{Institute of Nuclear Physics, Polish Academy of Sciences, PL-31342 Krakow, Poland}
\affil[c]{National Centre for Nuclear Research, ul. A. Soltana 7, 05-400 Otwock, Swierk, Poland}